\documentclass[final,3p,times]{elsarticle}

\usepackage[utf8]{inputenc}
\usepackage{graphicx}
\usepackage{pstricks}
\usepackage{array}
\usepackage{natbib}
\usepackage{amsmath}
\usepackage{pifont}
\usepackage{dsfont}
\usepackage{multirow}
\usepackage{amssymb}
\usepackage{wasysym}
\usepackage{ upgreek }

\def\be{\begin{equation}}
\def\ee{\end{equation}}

\journal{Physics Report}
\begin{document}
\begin{frontmatter}
\title{Numerical simulations of time-resolved quantum electronics}
\newcommand{\spsms}{CEA-INAC/UJF Grenoble 1, SPSMS UMR-E 9001, Grenoble F-38054, France}
\author{Benoit Gaury }
\author{Joseph Weston }
\author{Matthieu Santin }
\author{Manuel Houzet }
\author{Christoph Groth}
\author{Xavier Waintal\footnote{To whom correspondance should be sent, xavier.waintal@cea.fr}}
\address{\spsms}
\date{\today}

\begin{abstract}
Numerical simulation has become a major tool in quantum electronics both for
fundamental and applied purposes.  While for a long time those simulations
focused on stationary properties (e.g. DC currents), the recent experimental
trend toward GHz frequencies and beyond has triggered a new interest for
handling time-dependent perturbations. As the experimental frequencies get
higher, it becomes possible to conceive experiments which are both time-resolved
and fast enough to probe the internal quantum dynamics of the system.
This paper discusses the technical aspects -- mathematical and numerical --
associated with the numerical simulations of such a setup in the time domain
(i.e. beyond the single-frequency AC limit).  After a short review of the state
of the art, we develop a theoretical framework for the calculation of time-resolved
observables in a general  multiterminal system subject to an
arbitrary time-dependent perturbation (oscillating electrostatic gates, voltage pulses, time-varying magnetic fields, etc.) The approach is mathematically equivalent to (i)
the time-dependent scattering formalism, (ii) the time-resolved non-equilibrium
Green's function (NEGF) formalism and (iii) the partition-free approach.
The central object of our theory is a wave function that
obeys a simple Schrödinger equation with an additional source term that
accounts for the electrons injected from the electrodes. The time-resolved
observables (current, density, etc.) and the (inelastic) scattering matrix are simply expressed in term
of this wave function.  We use our approach to develop a numerical technique
for simulating time-resolved quantum transport.  We find that the use of this
wave function is advantageous for numerical simulations resulting in a speed up
of many orders of magnitude with respect to the direct integration of NEGF
equations. Our technique allows one to simulate realistic situations beyond
simple models, a subject that was until now beyond the simulation capabilities
of available approaches.
\end{abstract}

\end{frontmatter}
\tableofcontents
%\maketitle

%%%%%%%%%%%%%%%%%%%%%%%%%%%%%%%%%%%%%%%%%%%%%%%%%%%%%%%%%%%%%%%%%%%%%%%%%%
\section{Introduction}
%%%%%%%%%%%%%%%%%%%%%%%%%%%%%%%%%%%%%%%%%%%%%%%%%%%%%%%%%%%%%%%%%%%%%%%%%

The last ten years have seen the development of an increasing number of finite
frequency low temperature nanoelectronic experiments in the GHz range and
above. As the frequencies get higher, they become larger than the thermal background
(1 GHz corresponds to 50 mK) and eventually reach the internal characteristic frequencies of the system. Conceptually new types of experiments become possible where one probes directly the internal quantum dynamics of the devices. Beside a large interest for setups with very few
degrees of freedom (e.g. Qubits), open systems with continuous spectra -- the
subject of the present paper -- have been studied comparatively little.
Experiments performed in the time domain are even more recent.

This article is devoted to the development of the theoretical apparatus needed
to build efficient numerical techniques capable of handling the simulation of
nanoelectronic systems in the time domain.  While the theoretical basis for
such a  project are, by now, well established,
neither of the two (equivalent) standard approaches -- the time-dependent Non-equilibrium Green's function \cite{Wingreen-Meir_1993} (NEGF) and the time-dependent
scattering formalism \cite{Blanter2000} -- are, in their usual form,
well suited for a numerical treatment. Here we show that a third -
wave-function based -- approach, mathematically equivalent to both NEGF and
scattering theory, naturally leads to efficient numerical algorithms.

\begin{figure}[h]
    \centering
    \includegraphics[width=8cm]{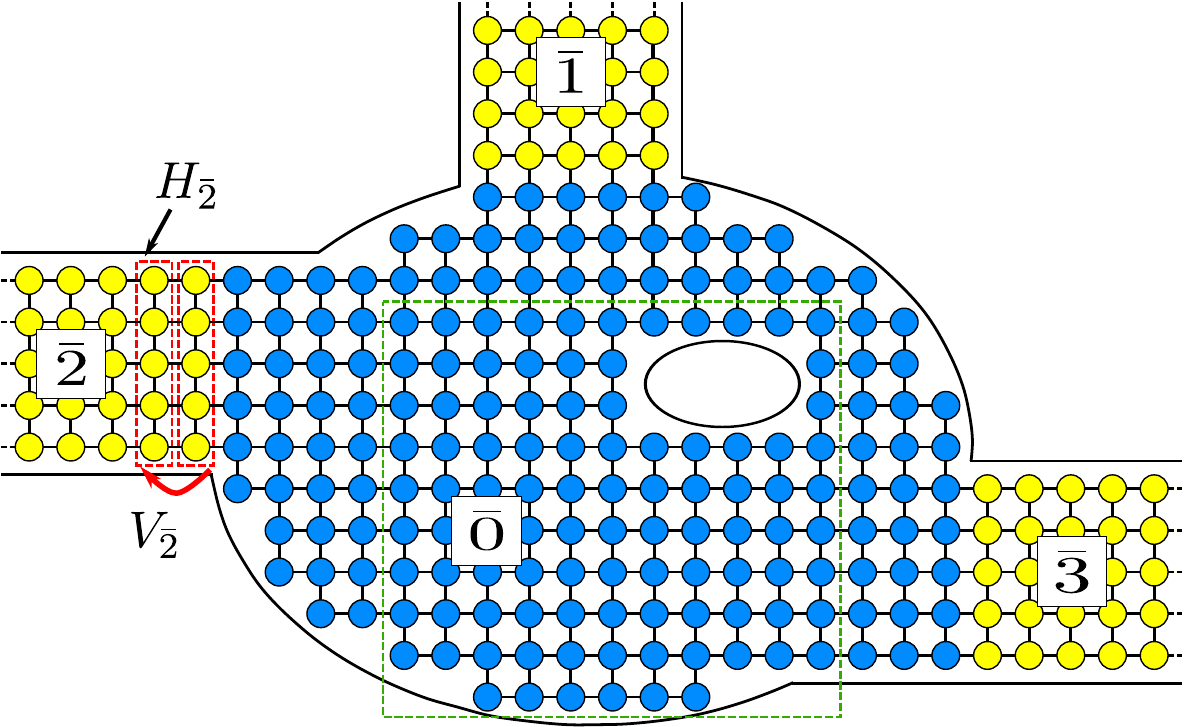}
    \caption{\label{fig: system}
     Sketch of a generic multiterminal system where the central part
     $\bar{0}$ (blue circles) is connected to three semi-infinite leads
     $\bar{1}$, $\bar{2}$, $\bar{3}$ (yellow circles).  The leads are kept at
     equilibrium with temperature $T_{\bar{m}}$ and chemical potential
     $\mu_{\bar{m}}$. The dashed green line indicates a region that will be
     integrated out in Fig.~\ref{fig: integrate_out} }
\end{figure}

The wave function approach followed here has a simple mathematical structure:
we consider a generic infinite system made of several semi-infinite electrodes
and a central mesoscopic region as sketched in Fig.~\ref{fig: system}.
Introducing the wave function $\Psi_{\alpha E}(\vec r,t)$ which depends on
space $\vec r$ and time $t$ as well as on the injection energy $E$ and mode
$\alpha $, we find that it obeys a Schrödinger equation with an additional
source term:
\be
\label{1} i\hbar \frac{\partial}{\partial t} \Psi_{\alpha
E}(\vec r,t) = H (t) \Psi_{\alpha E}(\vec r,t) + \sqrt{v_\alpha} \xi_{\alpha
E}(\vec r) e^{-iEt/\hbar},
\ee
where $H(t)$ is the time-dependent (one-particle) Hamiltonian of
the system while $\xi_{\alpha E}(\vec r)$ corresponds to the transverse wave
function of the conducting channel $\alpha $ at the electrode-device interface
(the number $\alpha$ is labeling both the different channels and the electrodes
to which they are associated) and $v_\alpha$ is the associated mode velocity.
The various observables are then easily expressed in terms of this wave
function. For instance, the particle current density (without electro-magnetic
field) reads:
\be
\label{2} \vec I(\vec r,t)={\rm Im} \sum_\alpha  \int
\frac{dE}{2\pi}\ \Psi^*_{\alpha E}(\vec r,t)\vec\nabla \Psi_{\alpha
E}(\vec r,t) f_\alpha (E)
\ee
where $f_\alpha (E)$ is the Fermi function in the
electrode of channel $\alpha$. The source term and mode velocities in
Eq.(\ref{1}) are standard objects of the theory of stationary quantum transport
and are readily obtained while Eq.(\ref{1}) itself can be integrated
numerically (as a function of time) without difficulty. The observables are
obtained in a second step by integrating Eq.~(\ref{2}) over the incident energy.
Eq. (\ref{1}) and Eq. (\ref{2}) thus form a simple set of equations in the mixed
time-energy domain, easily amenable to numerical evaluations. Note that a
treatment of the electron-electron interactions at the mean field level implies
that  these two equations become linked:  for instance the Hartree potential is
a function of the time-resolved local electron density. Hence, by introducing self-consistent potentials, these techniques 
can be easily extended to include electron-electron interactions at the mean field level (including time-dependent
density functional theory). As such a treatment is essentially independent from the non-interacting aspects discussed here, 
the link with the electrostatic, or more generally electron-electron interactions, will be mostly ignored
in the remaining of this article (except in the discussion of section \ref{electrostatic}).

This article is structured as follows. After a short survey of the literature
on time-resolved quantum transport with a focus on numerical work, this
article starts with the construction of  the theory of time-dependent
transport with an emphasis on drawing connections between the
various possible approaches. We discuss our general model
(Section~\ref{sec: model}) and the basic equations of the NEGF formalism
(Section \ref{sec: NEGF}) and then proceed in Section \ref{sec: wf} with the
introduction of the time-dependent wave function as a mathematical artifact to
reformulate the NEGF formalism. Section \ref{sec:S} is devoted to a
constructive presentation of the scattering approach. We show that it is
strictly identical to the wave function of Section \ref{sec: wf}. We also find
that the NEGF approach is equivalent to the partition-free approach introduced
in \cite{Cini1980} and further developed in Ref.\cite{Gross_DFT}.
This concludes the formalism part of this work. The next sections leverage on
the previous ones to build efficient numerical techniques for the simulation of
time-resolved transport. Section \ref{numerics} discusses (seven) different
practical algorithms that can be used to simulate time-dependent transport.
Section \ref{sec: applications} presents some numerical results that illustrate
the strengths and weaknesses of the approaches of Section \ref{numerics}.
We eventually converge toward a very simple algorithm (later referred to
as WF-D) which is many orders of magnitude faster than the direct NEGF approach
(yet mathematically equivalent). The reader is referred to Table
\ref{fig:benchmark} for a quick glance at the relative speeds of the different
approaches. In the last part of this article, we restrict the theory (so far
valid for an arbitrary form of the time-dependent perturbation) to the
particular case where a voltage pulse is applied to one of the electrodes.
Section \ref{landauer} is devoted to a derivation of the voltage pulse analogue to the
Landauer formula for DC transport. In Section \ref{sec:1d} we
apply the formalism to a very simple situation: the propagation and
spreading of a voltage pulse in a one dimensional wire. This application, for
which both analytical and numerical results can be obtained, serves to build
our physical understanding of the physics of fast pulses. We end this article
with Section \ref{sec:2d} where we present some simulations of a flying Qubit
inspired from recent experiments performed in a two-dimensional electron gas
\cite{Yamamoto2012}.

We end this introduction with a short review of the literature on the simulations of time-resolved
quantum transport.  Although we  briefly mention AC transport, the focus will
be on situations where the perturbation is localized in time.  Also, the
emphasis is on techniques that are suited to numerical treatments.

\subsection{From AC quantum transport to voltage pulses and
            ``electronic quantum optics''}

The history of AC quantum transport probably starts in the 60s with the
prediction and measurement of the photo assisted tunneling
\cite{TienGordon1963}, an effect that has attracted some renewed attention
recently in the context of noise measurements \cite{Reydellet2003}. Around the same time
was the discovery of the AC Josephson effect \cite{Likharev1986} between two superconductors.
Other early experiments showed that it was possible to pump current in a mesoscopic device
using the Coulomb blockade effect\cite{Pothier1992} or, more recently, the
quantum modulation of the wave function \cite{Brouwer1998,Giazotto2011} .

An important point that  was recognized early by B\"uttiker and his collaborators
is that a proper treatment of the electrostatics of the system was crucial when
dealing with finite frequency quantum transport \cite{Buttiker_capacitors,
Buttiker_admittances,
Buttiker_LC,Buttiker_dynamic_conductance,Buttiker_TDcurrent_partition}.
Indeed, in non-interacting AC theory, the electronic density fluctuates with
time and space and as a result the current is no longer a conserved quantity.
Allowing for the extra charges to be screened by nearby conductors restores the
neutrality of the global system, as well as current conservation once the
displacement currents are properly included. One finds that it is difficult to
observe the internal time scales of a device as they are often hidden by the
classical RC time. One of the most spectacular predictions of this
theory \cite{Buttiker_capacitors,Buttiker_LC} is that the resistance R that
sets the RC time of a ``quantum capacitor'' (a quantum dot
connected to one electrode and capacitively coupled to a gate) is given by half
the resistance quantum $h/(2e^2)$  irrespective of the actual
resistance of the constriction that forms the ``R'' part of the
device. The experimental verification of this prediction \cite{RC_Kirchhoff} was
a salient feature in the field of AC quantum physics.
More recent experiments include the measurement of a quantum LC circuit
\cite{Gabelli2007}, the statistics of the emitted photons
\cite{Portier2007,Portier2010} and the control of shot noise using multiple
harmonics \cite{gabelli2013}.

The theory of those AC quantum transport effects has now evolved into a field
in itself which lies outside the scope of this short discussion. We refer to
\cite{Moskalets2012} for an introduction to the (Floquet) scattering theory and
to \cite{ Cookbook_Oleksii} for the numerical aspects. 
Refs.~\cite{Wang2003,Nikolic2012,Chen2013} discuss practical aspects associated
with developing schemes that preserve gauge invariance and charge conservation.
Recent developments on the numerical side may be found in \cite{Oriols2013}
while the emerging topic
of Floquet topological insulators is reviewed in \cite{Cayssol2013}. 

Time-resolved quantum transport is not, {\it a priori }, very different from AC
quantum transport. A series of seminal works on time-resolved quantum
electronics showed however that the current noise associated with voltage
pulses crucially depends on their actual shape (i.e. on the details of the
harmonics contents and of the relative phases of the various harmonics)
\cite{Levitov1996, Lorentzian_pulses}.  More precisely, Levitov and
collaborators found that pulses of Lorentzian shape can be noiseless while
other shapes are associated with extra electron-holes excitations that increase
the noise of the signal. It was predicted that Lorentzian pulses could form
noiseless single electron sources that could be used in quantum devices such as
flying Qubits. These predictions are the object of an intensive
experimental activity\cite{Glattli2013,Dubois2012}.  Meanwhile, other
experiments looked for various ways to construct coherent single electron
sources and reproduce known quantum optics experiments with electrons, a
program  sometimes referred to as ``electronic quantum optics''.
Ref. \cite{Single_e_source} used a small quantum dot to make such a source
\cite{Correlations_Single_e_source, Noise_Single_e_source,
Entanglement_single_particle,Partitioning_electrons,Degiovanni2011} which was
later used in \cite{Bocquillon2013} to perform  an electronic Hong-Ou-Mandel
experiment.  A similar source, yet working at much larger energy has been
recently demonstrated in \cite{Fletcher2012}. Another
route\cite{Electron_surfer, e_source_QD}  used SAW (Surface Acoustic Waves) to
generate a propagating confining potential that transports single electrons
through the sample. These experiments are mostly performed in two dimensional
gasses made in GaAs/GaAlAs heterostructures whose rather small velocities
(estimated around $10^4-10^5 m.s^{-1}$ in the quantum Hall regime) and large
sizes (usually several $\mu m$) make experiments in the GHz range possible.
Smaller devices, such as carbon nanotubes, require the use of frequencies in
the THz range. Although THz frequencies are still experimentally challenging,
detection schemes in these range have been reported recently \cite{McEuen2008}.

\subsection{Numerical simulations of time-resolved quantum transport}

While simulations of the time-dependent Schrödinger equation are almost as old
as quantum mechanics itself, time-resolved quantum transport requires that
two additional difficulties be dealt with: the statistical physics of the
many-body problem (whose minimum level is to include the Pauli principle and
the thermal equilibrium of the leads) and the fact that quantum transport takes
place in {\it infinite} systems. Early numerical simulations of time-resolved
quantum transport were based on a seminal paper by Caroli, Combescot, Nozières,
and Saint-James\cite{Caroli_and_co} which sets the basis of the NEGF formalism
(itself based on the Keldysh formalism \cite{Keldysh1964}) in a one dimensional situation. This
formalism was used in \cite{Patawski1992} to study resonant tunneling of a
single level. The formalism for a generic mesoscopic system was established by
Jauho, Wingreen and Meir\cite{Wingreen-Meir_1993, Wingreen-Meir_1994} extending
the stationary formalism put forward by Wingreen and
Meir\cite{Wingreen-Meir_1992}  which itself extends the original work of
\cite{Caroli_and_co}. The time-dependent NEGF approach described in these
papers is still the basis of most numerical works (while the scattering theory
is perhaps more favored for analytical calculations). Considering that the
formalism is 25 years old, the number of publications on the subject is rather
small. This is due in part to the fact that it only recently became possible to
perform experiments in the relevant regimes (i.e. GHz frequencies at dilution
fridge temperatures), and also to the extreme computational cost of a direct
integration of the NEGF equations. Many recent works describe various
strategies for integrating the integro-differential equation of the NEGF
formalism, including direct approaches
\cite{Guo_time_domain_analysis,Cuansing2011,Prociuk2008}, a semi analytical
approach \cite{Guo_square_steps_analytics}, a parametrization of the analytical structure of the equations\cite{Croy2009} and a recursive approach
\cite{square_pulse}.  The important issue of properly dealing with electron-electron interactions
has been discussed in \cite{Wei2009,Kienle2010,Wang2011,Wang2013}. Alternative approaches to NEGF include a direct treatment
of the one electron density matrix \cite{Xie2012} or the use of a
``stroboscopic'' wave packet basis \cite{Bokes2008,Bokes2012}. Perhaps the most advanced
alternative to NEGF is the partition-free approach introduced by Cini
\cite{Cini1980} in the early 80s. In this approach, instead of ``integrating
out'' the electrodes' degrees of freedom, as it is done in NEGF, one starts
with the exact density matrix at equilibrium at $t=0$ and follows the system
states as they are driven out of equilibrium by the time-dependent
perturbation. This approach can be followed with Green's functions
\cite{Stefanucci2004,Perfetto2010} or more conveniently directly at the wave
function level \cite{Gross_DFT, Gross_WF, Gross_bound_states}.

To the best of our knowledge, the best performance so far has been obtained
with the partition-free approach where around 100 sites could be studied
(the direct NEGF simulations are usually confined to 10 sites or fewer). The
wave function approach leverages the fact that calculations of the electric
current do not require all of the information contained within Green's
functions.  Nevertheless, all these techniques suffer from the fact that the
systems are intrinsically infinite which brings non local (in time) terms into
the dynamical equations.  An interesting approach followed in
Ref.\cite{Perfetto2010} consists of ignoring these non local terms and
considering a large finite system instead.

In the rest of this article, we will revisit in turn the three main approaches
discussed above. While our starting point is the NEGF formalism, we shall
construct systematically the two other approaches, thereby proving (both at the
mathematical and numerical level) the complete equivalence between time-dependent
NEGF, scattering theory and the partition-free approach.

%%%%%%%%%%%%%%%%%%%%%%%%%%%%%%%%%%%%%%%%%%%%%%%%%%%%%%%%%%%%%%%%%%%%%%%%%
\section{Generic model for time-dependent mesoscopic devices}
%%%%%%%%%%%%%%%%%%%%%%%%%%%%%%%%%%%%%%%%%%%%%%%%%%%%%%%%%%%%%%%%%%%%%%%%%
\label{sec: model}

We consider a quadratic discrete Hamiltonian for an open system
\begin{equation}
    \mathrm{\hat{\textbf{H}}}(t) =
    \sum_{i,j} \mathrm{\textbf{H}}_{ij}(t) c^{\dagger}_{i}c_{j}
\end{equation}
where $c^{\dagger}_{i}$ ($c_{j}$) are the usual Fermionic creation (annihilation)
operators of a one-particle state on site $i$. The site index $i$ includes all
the degrees of freedom present in
the system, i.e.  space but also spin, orbital (s,p,d,f) and/or electron/hole
(superconductivity),  so that a large number of situations can be modeled within
the same framework.  The system consists of a central region, referred to as
$\bar{0}$ connected to $M$ semi-infinite leads labeled $\bar{1}...\bar{M}$ as
depicted in Fig.~\ref{fig: system}. $\mathrm{\textbf{H}}(t)$ is formally an
infinite matrix and can
be viewed as consisting of sub-blocks $\mathrm{\textbf{H}}_{\bar{m}\bar{n}}$.
\be
\mathrm{\textbf{H}}=
\left(
\begin{array}{cccc} \mathrm{\textbf{H}}_{\bar{0}\bar{0}} &\mathrm{\textbf{H}}_{\bar{0}\bar{1}}&\mathrm{\textbf{H}}_{\bar{0}\bar{2}}& \dots \\
                             \mathrm{\textbf{H}}_{\bar{1}\bar{0}}     & \mathrm{\textbf{H}}_{\bar{1}\bar{1}}    & 0                                                     &\dots \\
                                \mathrm{\textbf{H}}_{\bar{2}\bar{0}}    & 0                                                      &       \mathrm{\textbf{H}}_{\bar{2}\bar{2}}    &\dots \\
                                                                                                     \dots & \dots & \dots & \dots
                             \end{array}
\right)
\ee
A semi-infinite lead $\bar m$ is itself a periodic system where a unit cell is
described by a Hamiltonian matrix $H_{\bar m}$ which is coupled to the
neighboring cells by  the coupling matrix $V_{\bar m}$:

\be
\label{Hlead}
\mathrm{\textbf{H}}_{\bar{m}\bar{m}}=
\left(
\begin{array}{ccccc} H_{\bar m} & V_{\bar m}  & 0  & 0 & \dots \\
                             V^\dagger_{\bar m} & H_{\bar m} & V_{\bar m} & 0 &\dots \\
                             0 & V^\dagger_{\bar m} & H_{\bar m} & V_{\bar m} &\dots \\
                             \dots & \dots & \dots & \dots & \dots
                             \end{array}
\right)
\ee

While the time dependence of the device region
$\mathrm{\textbf{H}}_{\bar{0}\bar{0}}(t)$ can (and will) be arbitrary, the
leads are only subject to homogeneous time-dependent voltages so that
$\mathrm{\textbf{H}}_{\bar{m}\bar{m}}(t)= w_{\bar m}(t) 1_{\bar m} +
\mathrm{\textbf{H}}_{\bar{m}\bar{m}}(t=0)$ ($1_{\bar m}$ is the identity matrix
in lead $m$). Following standard practice, we perform a  unitary gauge
transformation  $\hat{W} = \exp (-i \sum_{i \in \bar{m}} \phi_{\bar
m}(t)c^{\dagger}_{i}c_{i})$ on the Hamiltonian with $\phi_{\bar m}(t) =
\int_{-\infty}^t du\ w_{\bar m}(u)$ being the integral of the time-dependent
voltage. After the gauge transformation, we recover time-independent
Hamiltonians for the leads while the matrix elements that connect the lead to
the central part now acquire an time-varying phase:
\be
\mathrm{\textbf{H}}_{\bar{m}\bar{0}}\rightarrow e^{i\phi_{\bar m}(t)} \mathrm{\textbf{H}}_{\bar{m}\bar{0}}
\ee

The quantum mechanical aspects being properly defined, we are left to specify
the statistical physics; each lead  is supposed to remain at thermal
equilibrium with a chemical potential $\mu_{\bar m}$ and a temperature $T_{\bar
m}$. Note that the thermal equilibrium condition is most simply expressed for time-independent leads, i.e. after the gauge transformation. This particular choice of boundary condition is significant and its
physical meaning will be discussed in more depth later in the text (section
\ref{landauer}).

%%%%%%%%%%%%%%%%%%%%%%%%%%%%%%%%%
\section{Non-equilibrium Green's function (NEGF) approach}
%%%%%%%%%%%%%%%%%%%%%%%%%%%%%%%%%%
\label{sec: NEGF}
Here we summarize the basic equations of the time-dependent NEGF
formalism\cite{Wingreen-Meir_1992, Wingreen-Meir_1993} that constitutes the
starting point of our approach.  We refer to the original \cite{Rammer_review}
or more recent references \cite{Cookbook_Oleksii, Rammer_book} for a derivation
of these equations.  The basic objects under consideration are the Lesser
$\mathcal{G}^{<}(t,t')$ and Retarded $\mathcal{G}^{R}(t,t')$ Green's functions
of the system, \begin{align} \mathcal{G}_{ij}^{R}(t,t')
    &=-i\theta(t-t')\langle\{c_i(t),c^{\dagger}_j(t')\}\rangle\,,\label{Retarded_GF_definition}\\
\mathcal{G}_{ij}^{<}(t,t') &=i\langle c^{\dagger}_j(t')c_i(t)\rangle \ .
\label{Lesser_GF_definition} \end{align} where the operator $c_i(t)$
corresponds to $c_i$ in the Heisenberg representation and $\theta(t)$ is the
Heaviside function.  For a quadratic Hamiltonian, the Retarded Green's function
takes a simple form in terms of the ``first quantization'' evolution operator
of the system: \begin{equation} \mathcal{G}^{R}(t,t') = -i\theta(t-t')U(t,t')
\label{eq:evol} \end{equation} where the unitary evolution operator $U(t,t')$
verifies $i \partial_{t} U(t,t') = \mathrm{\textbf{H}}(t)U(t,t')$ and
$U(t,t)=1$.

The physical observables can be written simply in terms of the Lesser Green's
function. For instance the particle current between sites $i$ and $j$ reads,
\be
\label{current-lesser}
I_{ij}(t) = \mathrm{\textbf{H}}_{ij}(t)G^{<}_{ji}(t,t) - \mathrm{\textbf{H}}_{ji}(t) G^{<}_{ij}(t,t)
\ee
while local electron density is $\rho_{i}(t) = -i G^{<}_{ii}(t,t)$.

Suppose that one is interested in the quantum propagation of a wave packet
$\Psi(t)$ according to the  Schrödinger equation $i
\partial_t \Psi(t) = \mathrm{\textbf H} \Psi(t)$ with an initial condition
given by $\Psi(t=t_0)=\Psi_0$. Then one finds that $\Psi(t)$ is simply given by
$\Psi(t)=i\mathcal{G}^{R}(t,t_0)\Psi_0$.  In other words, the Retarded Green's
function encodes the quantum propagation of a wave packet. The Lesser Green's
function, on the other hand, captures the remaining many-body / statistical
physics aspects: the Pauli principle, the finite temperature properties of the
leads and the fact that the ``initial conditions'', say an electric voltage
pulse, are given in terms of macroscopic quantities (as opposed to an initial
microscopic wave packet) and spread over a finite time window.

\subsection{Equations of motion for the Retarded ($G^R$) and Lesser ($G^<$) Green's functions}
Introducing the projections of 
Green's functions on the central region $G^R(t,t')=\mathcal{G}^{R}_{\bar{0}\bar{0}}(t,t')$ and
$G^<(t,t')=\mathcal{G}^{<}_{\bar{0}\bar{0}}(t,t')$, one can obtain effective equations of
motion where the leads' degrees of freedom have been integrated out.
The equation of motion for $G^R$ reads\cite{Rammer_book},
\begin{align}
& i \partial_{t} G^{R}(t,t') =
 \mathrm{\bold{H}}_{\bar{0}\bar{0}}(t)G^{R}(t,t')
 + \int du\ \Sigma^{R}(t,u)G^{R}(u,t')
\label{eq:EOM}
\end{align}
or its symmetric counterpart
\begin{align}
& i  \partial_{t'} G^{R}(t,t') =
-G^{R}(t,t') \mathrm{\bold{H}}_{\bar{0}\bar{0}}(t') -
 \int du\ G^{R}(t,u)\Sigma^{R}(u,t')
 \label{eq:EOM2}
\end{align}
with the initial condition $\lim_{\tau\rightarrow 0} G^R(t+\tau,t) = -i$. The
self-energies encapsulate the effect of the leads:
\be
\Sigma^R(t,t') = \sum_{\bar m=1}^{\bar M} \Sigma^R_{\bar m}(t,t')
\ee
with
\be
 \Sigma^R_{\bar m}(t,t')= \mathrm{\textbf{H}}_{\bar{0}\bar{m}}(t)
    g^{R}_{\bar m}(t,t') \mathrm{\textbf{H}}_{\bar{m}\bar{0}}(t')
\ee
where $g^{R}_{\bar m}(t,t')$ is the isolated lead Retarded Green's function,
i.e. {\it the surface Green's function of the lead in the absence of coupling
to the central region}.

The equation of motion for the Lesser Green's function can be integrated
formally and reads\cite{Rammer_review, Rammer_book}
\be
G^< (t,t') = \int du \ dv \ G^{R}(t, u) \Sigma^{<}(u,v) [G^{R}(t',v)]^{\dagger}
     \label{eq:EOM<}
\ee
with $\Sigma^{<}(t,t')=\sum_{\bar m}\Sigma^<_{\bar m}(t,t')$ and $\Sigma^<_{\bar
m}(t,t')=\mathrm{\textbf{H}}_{\bar{0}\bar{m}}(t) g^{<}_{\bar m}(t,t')
\mathrm{\textbf{H}}_{\bar{m}\bar{0}}(t')$.
Equations (\ref{eq:EOM}) and (\ref{eq:EOM<}) form the starting point of this
paper.

\subsection{Equations of motion for lead self-energies}

To get a complete set of equations, we need to relate the self-energies of the
leads to the lead Hamiltonian matrices. While the corresponding calculation in
the energy domain is well developed, self-energies as a function of time have
been seldom calculated.  Here we use the following equation of motion,
\begin{align} i \partial_{t} g^{R}_{\bar m}(t,t') - H_{\bar{m}}(t) g^{R}_{\bar
    m}(t,t') = \int du\ V_{\bar{m}}(t)g^{R}_{\bar m}(t,u)
    V^\dagger_{\bar{m}}(u)g^{R}_{\bar m}(u,t') \label{eq:EOMlead}
\end{align}
This equation only provides the {\it surface } Green's function of the
lead, i.e. Green's function matrix elements for the last layer of the
semi-infinite periodic structure. For time-independent leads (the case
studied in this paper after the gauge transformation), $g^{R}_{\bar
m}(t-t')$ is a function of the time difference $t-t'$ only.  It is related
by a simple Fourier transform to the surface Green's function in energy:
\be g^{R}_{\bar m}(t-t')= \int \frac{dE}{2\pi} e^{-iE(t-t')} g^{R}_{\bar
m}(E).  \ee There are many techniques to calculate $g^{R}_{\bar m}(E)$ but
the presence of a cusp at $t=t'$ in the time domain and $1/\sqrt{E}$
singularities in the energy domain (whenever a new conducting channel
opens) renders a Fourier transform impractical and a direct use of
Eq.(\ref{eq:EOMlead}) much more convenient. The analogue of
Eq.(\ref{eq:EOMlead}) in the energy domain is a self-consistent equation
for $g^{R}_{\bar m}(E)$ \be \label{sce4lead} g^{R}_{\bar m}(E)
=1/[E-H_{\bar{m}}- V_{\bar{m}} g^{R}_{\bar m}(E) V^\dagger_{\bar{m}}] \ee
which is far less interesting than its time-dependent counterpart. Indeed, the
corresponding iterative solution converges poorly (each iteration
corresponds to adding one layer to the lead while other schemes allow to
double its size at each iteration) and it requires the use of a small
imaginary part in the self-energy.

As each lead is at thermal equilibrium, the Lesser surface Green's function for
the lead is obtained from the Retarded one through the use of the
fluctuation-dissipation theorem\cite{Datta, Rammer_book} :
\be g^{<}_{\bar
m}(E)= -  f_{\bar m}(E) \left(  g^{R}_{\bar m}(E) - [g^{R}_{\bar m}(E)]^\dagger
\right)
\ee
where $f_{\bar m}(E)=1/[1 + e^{(E-\mu_{\bar m})/k_{B}T_{\bar m}}]$
is the Fermi function of the lead.

%%%%%%%%%%%%%%%%%%%%%%%%%%%%%%%%%%%%%%%%%%%%%%%%%%%%%%%%%%%%%%%%%%
\section{Wave function (WF) approach}
%%%%%%%%%%%%%%%%%%%%%%%%%%%%%%%%%%%%%%%%%%%%%%%%%%%%%%%%%%%%%%%%%%
\label{sec: wf}
We now turn to the construction of our wave function approach. We seek
to explicitly construct  the wave function in terms of Green's
functions, relate the physical observables to the wave function and derive the
equations that this wave function satisfies.  Eventually, we arrive at a
closed set of equations where the original Green's function formalism has
disappeared entirely. The central object of the resulting theory lies halfway
between NEGF and the time-dependent scattering approach. Both Green's
functions and the (time-dependent) scattering matrix can be obtained directly
from the wave function.

In what follows we suppose that the voltage drop actually takes place {\it
inside } the central region ${\bar 0}$. This can be done without loss of
generality; if it is not the case then we simply change our definition of the
central region to include a few  layers of the leads. We always include at
least the first layer of each lead in our definition of the central region
$\bar 0$. This step is not necessary but somewhat simplifies the resulting
expressions.

\subsection{Construction of the wave function}

We start with a representation of the lead Lesser self-energy in the energy
domain: \be
\Sigma^{<}(t-t')=\sum_{\bar m} \int \frac{dE}{2\pi} if_{\bar m}(E) e^{-iE(t-t')} \Gamma_{\bar m}(E)
\ee
where $\Gamma_{\bar m}(E)= i  \mathrm{\textbf{H}}_{\bar{0}\bar{m}} \left(
g^{R}_{\bar m}(E) - [g^{R}_{\bar m}(E)]^\dagger \right)
\mathrm{\textbf{H}}_{\bar{m}\bar{0}}$ is the coupling matrix to the electrodes
(also known as the tunneling rate matrix in the context of weak coupling).
$\Gamma_{\bar m}(E)$ can be diagonalized into
\be
\Gamma_{\bar m}(E)=\sum_\alpha v_{\bar m\alpha} \xi_{\alpha E}\xi_{\alpha E}^\dagger
\ee
where the $\xi_{\alpha E}$ are the so-called dual transverse wave functions and
$v_\alpha (E)$ is the corresponding mode velocity\cite{Wimmer_thesis}. Note that the
$\xi_{\alpha E}$ are normalized but not necessarily orthogonal. They are related to the
transverse incoming modes $\xi^{in}_{\alpha E}$ to be introduced in the next
section by $\xi_{\alpha E}=\Gamma_{\bar m} \xi^{in}_{\alpha E}/v_{\bar m
\alpha}$.  Note that alternatively we could have used the fact that
$\Gamma_{\bar m}$ is a Hermitian matrix to justify its diagonalization into a
set of {\it orthonormal} vectors. However, by doing so we would have mixed
outgoing and incoming states and lost the connection with the scattering theory
described in the next section. We also note that all modes are in principle
included but the evanescent ones have vanishing velocities and will therefore
automatically drop out of the problem.

Eq. (\ref{eq:EOM<}) for the Lesser Green's function, hence the
observables, can be recast using the two above equations into,
\be
    G^<(t,t') =\sum_\alpha \int \frac{dE}{2\pi}\ if_\alpha(E)  \Psi_{\alpha E}(t) \Psi_{\alpha E}(t')^\dagger
    \label{eq:psi-less}
\ee
where we have used a unique index $\alpha$ to denote both the leads and the
channels inside the leads and introduced the wave function,
\begin{equation}
\Psi_{\alpha E}(t) = \sqrt{v_{ \alpha}} \int du \ G^{R}(t,u)  e^{-iEu} \xi_{\alpha E}.
\label{eq: wave_function}
\end{equation}
$\Psi_{\alpha E}(t)$ is the projection inside the device region of $\psi_{\alpha
E}(t)$ which is defined in the infinite system: $\Psi_{\alpha E}=[\psi_{\alpha
E}]_{\bar 0}$ with
\begin{equation}
\psi_{\alpha E}(t) = \sqrt{v_{\alpha}} \int du\  \mathcal{G}^{R}(t,u)  e^{-iEu} \xi_{\alpha E}.
\label{eq: wave_function2}
\end{equation}
$\Psi_{\alpha E}(t)$ and $\psi_{\alpha E}(t)$ are the basic objects that will
be discussed from now on.

We note that the Retarded Green's function
$G^{R}(t,t')=\theta(t-t')[G^>(t,t')-G^<(t,t')]$ can also be obtained from the
wave function,
\be
  G^{R}(t,t')=-i\theta(t-t') \int \frac{dE}{2\pi} \sum_{\alpha} \Psi_{\alpha,E}(t)\Psi_{\alpha,E}^{\dagger}(t')
\ee
from which we get the normalization condition,
\be
    \forall \ t \ \int \frac{dE}{2\pi} \sum_{\alpha} \Psi_{\alpha,E}(t)
    \Psi_{\alpha,E}^{\dagger}(t) = \mathds{1}_{\bar 0}
\ee

\subsection{Effective Schrödinger equation}

\begin{figure}[h]
    \centering
    \includegraphics[width=8cm]{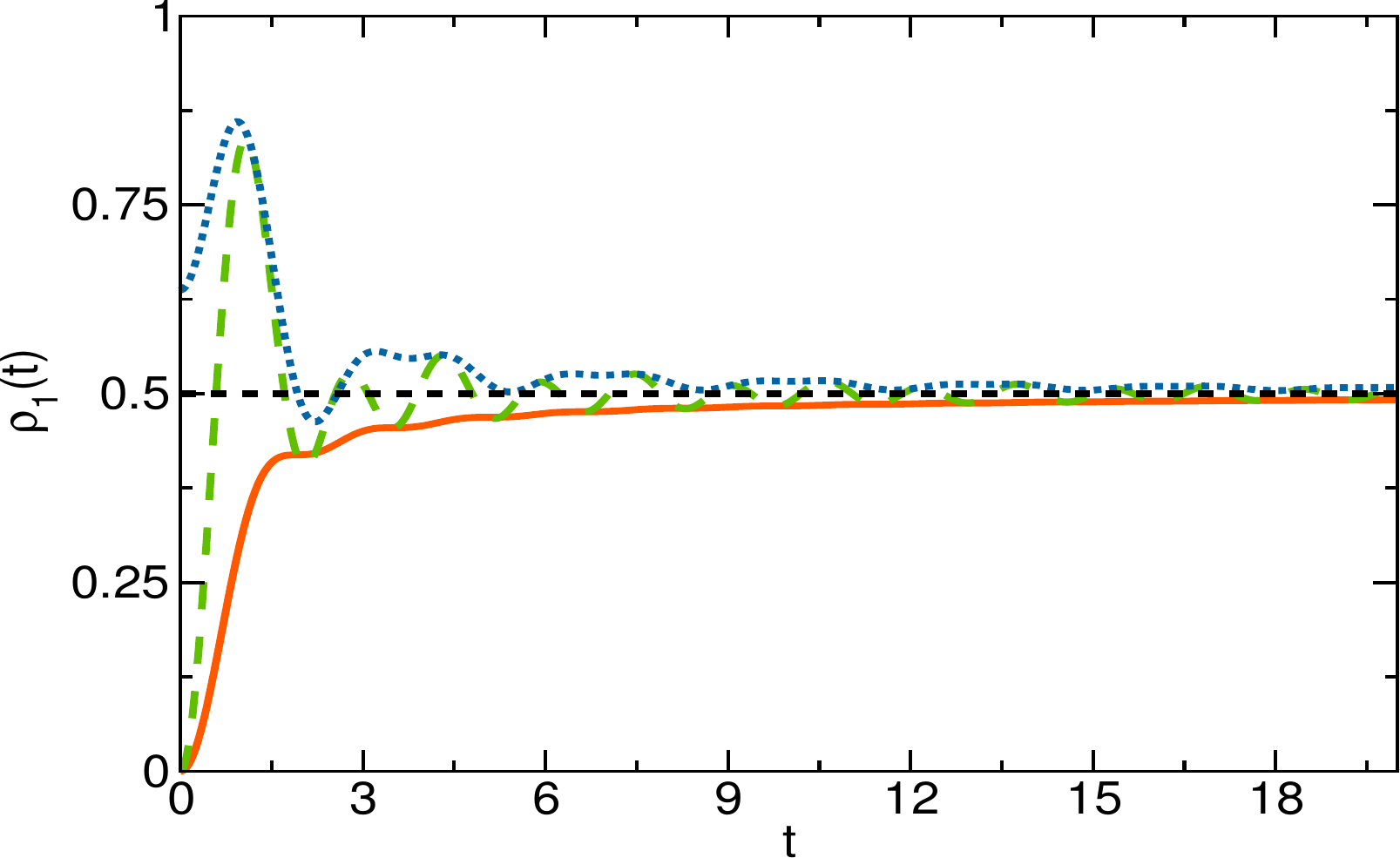}
    \caption{ Sensitivity of WF-A to initial conditions.  Local density of
    particle on site $1$ as a function of time within WF-A.  The calculations
    are done for $\Psi_{E,x} (t=0)=0$ (orange full line), $\Psi_{E,x}
    (t=0)=\delta_{x,1}$ (blue dotted line),
    $\Psi_{E,x} (t=0)=\delta_{x,2}$ (long green dashed line) and $\Psi_{E,x}
    (t=0)=\Psi^{st}_E$ (short black dashed
    line).  Except in the last case, we ignore the memory integral for negative
    times.}
    \label{fig: psi_convergence}
\end{figure}

The equations satisfied by the wave function derive directly from the equation
of motion for the Retarded Green's function. They read,
\be
    i \partial_{t} \Psi_{\alpha E}(t) = \mathrm{\textbf{H}}_{\bar{0}\bar{0}}(t) \Psi_{\alpha E}(t) +
       \int du\ \Sigma^{R}(t-u)\Psi_{\alpha E}(u) +
     \sqrt{v_\alpha}e^{-i E t}  \xi_{\alpha E} \label{eq:Psi}
\ee
and
\be
i \partial_{t} \psi_{\alpha E}(t) = \mathrm{\textbf{H}}(t) \psi_{\alpha E}(t) +
        \sqrt{v_\alpha}e^{-i E t}  \xi_{\alpha E} \label{eq:psi}
\ee
Remarkably, Eq.(\ref{eq:psi}) is almost the Schrödinger equation, up to the
source term $\sqrt{v_\alpha}e^{-i E t}  \xi_{\alpha E}$. Together,
Eq.(\ref{eq:psi-less}) and Eq.(\ref{eq:Psi}) (or alternatively
Eq.(\ref{eq:psi})) form a closed set of equations that permits the calculation
of the observables of the system. In particular, the Retarded Green's function
does not appear explicitly anymore. Note that the initial conditions for the
wave functions are not well defined. We shall find, however, that they are
essentially irrelevant and that after some relaxation time they are forgotten;
the source term controls the results (see Fig.~\ref{fig: psi_convergence}).

At this stage, several routes could be followed. If we suppose the
time-dependent perturbations to be periodic, we can make use of the Floquet
theorem to obtain a Floquet based wave function approach. Here, however, we
concentrate on the physics of pulses (perturbations of any sort but localized
in time). We suppose that the system is in a stationary state up to a time
$t=0$ and that the time-dependent perturbations (voltage pulses,
microwaves, etc.) are switched on at time $t>0$. We separate the problem into a
stationary part and a time-dependent perturbation
$\mathrm{\textbf{H}}_{\bar{0}\bar{0}}(t)=\mathrm{\textbf{H}}_{\bar{0}
st}+\mathrm{\textbf{H}}_{\bar{0} w}(t)$. The solution of the stationary problem
takes the form $e^{-iEt} \Psi^{st}_{\alpha E}$, where the stationary solution
can be obtained by solving the linear (sparse) equation,
\be
[E - \mathrm{\textbf{H}}_{\bar{0} st} - \Sigma^{R}(E)] \Psi^{st}_{\alpha E} = \sqrt{v_\alpha} \xi_{\alpha E}
    \label{eq: Psi_st}
\ee
$\Psi^{st}_{\alpha E}$ is a typical output of wave function based algorithms
for DC transport \cite{Kwant_preparation}. We now introduce a wave function measuring the {\it
deviation} with respect to the stationary solution,
\be
\Psi_{\alpha E}(t) = {\bar \Psi}_{\alpha E}(t) + e^{-iEt} \Psi_{\alpha E}^{st}.
\ee
 ${\bar \Psi}_{\alpha E}(t)$ satisfies,

\be
    i \partial_{t} \bar\Psi_{\alpha E}(t) = \mathrm{\textbf{H}}_{\bar{0}\bar{0}}(t) \bar\Psi_{\alpha E}(t) +
       \int_0^t du\ \Sigma^{R}(t-u)\bar\Psi_{\alpha E}(u) +
    \mathrm{\textbf{H}}_{\bar{0}w}(t)  e^{-iEt}\Psi_{\alpha E}^{st}
    \label{eq:Psibar2}
\ee
with the initial condition $\bar\Psi_{\alpha E}(t=0)=0$.  Eq.(\ref{eq:Psibar2})
is very similar to Eq.(\ref{eq:Psi}) but it has the advantage that the
equilibrium physics has been removed so that the memory kernel starts at $t=0$
(instead of $t=-\infty$). Also, the source term does not take place at the
system-leads interface anymore, but rather at the sites where a time-dependent
perturbation is applied.  A similar treatment can be done for $\psi_{\alpha
E}(t)$ and we obtain
\be
    i \partial_{t} \bar\psi_{\alpha E}(t) = \mathrm{\textbf{H}}(t) \bar\psi_{\alpha E}(t) +
    \mathrm{\textbf{H}}_{w}(t)  e^{-iEt}\psi_{\alpha E}^{st}
    \label{eq:Psibar3}
\ee
where $\psi^{st}_{\alpha E}$ satisfies $[E - \mathrm{\textbf{H}}_{st} ]
\psi^{st}_{\alpha E} = \sqrt{v_\alpha} \xi_{\alpha E}$ and
$\mathrm{\textbf{H}}(t)=\mathrm{\textbf{H}}_{st}+\mathrm{\textbf{H}}_{w}(t)$.
We shall find that Eq.(\ref{eq:Psibar2}) or Eq.(\ref{eq:Psibar3}) are much more
well suited for numerical simulations than the original NEGF equations
(see also \ref{appendix-src} for a simplified discussion in one dimension).

Finally, a common case of interest involves metallic electrodes coupled to
mesoscopic systems whose characteristic energy scales are much smaller than the
Fermi energy of the electrodes. In this limit (known as the wide band limit),
one can neglect the energy dependence of the electrode self-energy
$\Sigma^R(E+\epsilon)\approx \Sigma^R(E)$ and the self-energy memory kernel
becomes local in time resulting in,
\be\label{wbl}
   i \partial_{t} \bar\Psi_{\alpha E}(t) = [ \mathrm{\textbf{H}}_{\bar{0}\bar{0}}(t) +\Sigma^{R}(E)  ]\bar\Psi_{\alpha E}(t) +
    \mathrm{\textbf{H}}_{\bar{0}w}(t)  e^{-iEt}\Psi_{\alpha E}^{st}.
\ee

%%%%%%%%%%%%%%%%%%%%%%%%%%%%%%%%%%%%%%%%%%%%%%%%%%%%%%%%
\section{Time-dependent scattering theory}
%%%%%%%%%%%%%%%%%%%%%%%%%%%%%%%%%%%%%%%%%%%%%%%%%%%%%%%%
\label{sec:S}

So far our starting point has been the NEGF formalism from which we have
constructed the wave function $\Psi_{\alpha E}(t)$.  We now turn to a
``Landauer-Buttiker'' scattering approach of time-dependent quantum transport
in a mixed time-energy representation. We construct the time-dependent
scattering states of the system and find that their projection inside the
central region is in fact the wave function $\Psi_{\alpha E}(t)$. Hence, we
shall establish (as it is the case for DC transport) that the corresponding
scattering approach is rigorously equivalent to the NEGF formalism. Last, we
shall make the connection with the partition free approach thereby completing
the  formalism part of this article.  Note that the proofs below are a bit
technical.  \ref{appendix-src} illustrates them in the simplified case of a one dimensional model.

\subsection{Conducting modes in the leads}
We start by introducing the plane waves $\alpha$ inside a lead $\bar p$ which
take the form $\xi^{in}_{\bar p\alpha}(E) e^{-iEt-ik^{in}_\alpha(E) x}$ for the
incoming states and $\xi^{out}_{\bar p\alpha}(E) e^{-iEt+ik^{out}_\alpha(E) x}$
for the outgoing ones.  The integer $x$ labels the different layers of the lead ($x\in \{
1,2,3\cdots \}$) counted from the central system. The normalized vectors $\xi^{out}_{\bar p\beta}$
($\xi^{in}_{\bar p\beta}$) are the transverse part of the mode for the outgoing
(incoming) states, including the evanescent modes (although those will
eventually drop out for the incoming part). As the plane waves satisfy the
Schrödinger equation, we obtain
\be
\label{modes}
[H_{\bar p} - E + V_{\bar p}  \lambda_\alpha + V^\dagger_{\bar p} \lambda_\alpha^{-1}]\xi^{out}_{\bar p\alpha}(E)=0
\ee
with $\lambda_\alpha=e^{+ik^{out}_\alpha(E)}$. $\xi^{in}_{\bar p\alpha}(E)$
obeying the same equation with negative momenta. This (2$^{\rm{nd}}$ order) equation
can be recast in the form of a generalized eigenvalue problem,
\be
\label{modes2}
\left(\begin{array}{cc} H_{\bar p}-E  & V^\dagger_{\bar p} \\ 1 & 0  \end{array}\right)
\left(\begin{array}{c}  \xi_{\bar p\alpha}(E) \\  \chi_{\bar p\alpha}(E)  \end{array}\right)
= \lambda_\alpha
 \left(\begin{array}{cc} -V_{\bar p}  & 0 \\  0 & 1  \end{array}\right)
\left(\begin{array}{c}  \xi_{\bar p\alpha}(E) \\  \chi_{\bar p\alpha}(E)  \end{array}\right)
\ee
for which efficient techniques have now been developed \cite{Wimmer_thesis,Rungger2008}
($\chi_{\bar p\alpha}(E)$ is defined by the second line of Eq.(\ref{modes2})).
We note that solving Eq.(\ref{modes})  can be non trivial when $V$ is not
invertible, a common case when the lattice has more than one atom per unit cell
(e.g. graphene).  The corresponding mode velocity is given by $v^{out}_{\bar
p\alpha}= i(\xi^{out}_{\bar p\alpha})^\dagger [Ve^{+ik^{out}_\alpha(E)} - V^\dagger e^{-ik^{out}_\alpha(E)}]
\xi^{out}_{\bar p\alpha}$.
An interesting relation is obtained by observing that $\xi^{out}_{\bar
p\alpha}(E)$ ($\xi^{in}_{\bar p\alpha}(E)$) are the eigenvectors of the
Retarded (Advanced) Green's function of the lead,
\be
\label{eig-gr}
 g^{R}_{\bar p}(E) V^\dagger_{\bar p} \xi^{out}_{\bar p\alpha}(E)=  e^{+ik^{out}_\alpha(E)}\xi^{out}_{\bar p\alpha}(E)
\ee
\be
\label{eig-ga}
[g^R_{\bar p}(E)]^\dagger V^\dagger_{\bar p} \xi^{in}_{\bar p\alpha}(E)=  e^{-ik^{in}_\alpha(E)}\xi^{in}_{\bar p\alpha}(E)
\ee
as can be shown using Eq.(\ref{sce4lead}) and Eq.(\ref{modes}), see
Ref.\cite{Wimmer_thesis}.  Eq.(\ref{modes}) implies
 that for any two modes (incoming or outgoing) \cite{Wimmer_thesis},
\be
\label{modes3}
(\lambda_\alpha - [\lambda_\beta^*]^{-1})\xi^{in/out}_{\bar p\beta}(E) [V_{\bar
p}  \lambda_\alpha - V^\dagger_{\bar p} \lambda_\beta^{*}]\xi^{in/out}_{\bar
p\alpha}(E)=0.
\ee
It follows that, while in general different modes are not orthogonal, they
satisfy
\be
\label{ortho}
[\xi^{out}_{\bar p\alpha}(E)]^\dagger \Gamma_{\bar p}
\xi^{out}_{\bar m\beta}(E)=\delta_{\alpha\beta}\delta_{\bar m\bar p}v^{out}_{\bar
p\alpha}
\ee
with a similar expression for the incoming modes.

\subsection{Construction of the scattering states}
Our aim is to construct a wave function $\psi^{scat}_{\alpha E}(t) $ which (i)
is a solution of the Schrödinger equation and (ii) corresponds to an incoming
plane wave in mode $\alpha$ (belonging to lead $\bar m$) with energy $E$. This boundary condition
amounts to imposing the {\it incoming} part of the wave function, and leaving the outgoing
part free.  In particular, the system being time-dependent, the outgoing part can
contain many different energies.  In the rest of this section, we often drop
the indices $E$ and $\alpha$ when there is no risk of confusion. The value of
$\psi^{scat}_{\alpha E}(t)$ are noted $\psi^{scat}_{\bar 0}(t)$ in the central
region and $\psi^{scat}_{\bar px} (t)$ in the $x^{th}$ layer of lead $\bar p$.
In the leads, the wave function is formed by a superposition of plane waves,
\begin{eqnarray}
\label{scatteringstate}
\psi^{scat}_{\bar px}(t)&\equiv& \psi_{\bar px}^{in}(t) + \psi_{\bar px}^{out}(t) \ \ {\rm with}\\
\psi_{\bar px}^{in} (t) &=&\delta_{\bar p \bar m} \frac{\xi^{in}_{\bar p\alpha}(E)}{\sqrt{|v^{in}_{\bar m\alpha}|}} e^{-iEt-ik^{in}_\alpha(E) x} \nonumber \\
 \psi_{\bar px}^{out} (t) &=&\int \frac{dE'}{2\pi} \sum_\beta \frac{\xi^{out}_{\bar p\beta}(E')}{\sqrt{|v^{out}_{\bar p\beta}}|} e^{-iE't+ik^{out}_\beta(E') x} S_{\bar p\beta,\bar m\alpha}(E',E)\nonumber
\end{eqnarray}
$S_{\bar p\beta,\bar m\alpha}(E',E)$ is the central object of the scattering
theory, namely the probability amplitude for a mode $\alpha$ with energy $E$ to
be transmitted ($\bar p\ne\bar m$) or reflected ($\bar p= \bar m$) into mode
$\beta$ with energy $E'$. The formalism only differs from its time-independent
counterpart by the possibility to absorb or emit energy. The normalization has
been chosen so that the waves carry a current (per energy unit) unity.  As
Eq.(\ref{scatteringstate}) is made of a superposition of the eigenstates of the
leads, it satisfies the time-dependent Schrödinger equation in the lead by
construction.  Eq.(\ref{scatteringstate}) forms an ``incoming'' boundary
condition. One proceeds by writing the Schrödinger equation in the central
region and in the first layer of the leads (the ``matching conditions''):
\begin{eqnarray}
\label{match1}
i \partial_t  \psi^{scat}_{\bar 0}(t)&=& \mathrm{\textbf{H}}_{\bar{0}\bar{0}} \psi^{scat}_{\bar 0}(t) + \sum_{\bar p} V_{\bar p} \psi^{scat}_{\bar p1}(t)\\
\label{match2}
i \partial_t \psi^{scat}_{\bar p1} &=&  H_{\bar p}  \psi^{scat}_{\bar
p1}(t) + V^\dagger_{\bar p} P_{\bar p} \psi^{scat}_{\bar 0}(t) + V_{\bar p}
\psi^{scat}_{\bar p2}(t),
\end{eqnarray}
where the projector $P_{\bar p}$ projects the wave function of the central
region on the sites which are attached to the reservoir $\bar p$. The set of
the three above equations fully defines the scattering states as well as the
``scattering matrix'' $S_{\bar p\beta,\bar m\alpha}$ of the system.

\subsection{Connection to the wave function approach}
To proceed, we note that as $\psi^{scat}_{\bar px} (t)$ satisfies $i \partial_t
\psi^{scat}_{\bar p1} = H_{\bar p}  \psi^{scat}_{\bar p1}(t) + V^\dagger_{\bar p}
\psi^{scat}_{\bar p0}(t) + V_{\bar p} \psi^{scat}_{\bar p2}(t)$, Eq.(\ref{match2})
results in,
\be
\label{match3}
V^\dagger_{\bar p} P_{\bar p}  \psi^{scat}_{\bar 0}(t) = V^\dagger_{\bar p}  \psi^{scat}_{\bar p0}(t)
\ee
which relates the scattering matrix on the right to the wave function inside
the system on the left.

We now use the fact that  $\xi^{out}_{\bar p\alpha}(E)$ and $\xi^{in}_{\bar
p\alpha}(E)$ are the eigenvectors of the Retarded and Advanced surface Green's
function of the lead $\bar p$.  Equations (\ref{eig-gr}), (\ref{eig-ga}) and
Eq.(\ref{scatteringstate}) provide,
\be
\label{match4}
V_{\bar p}  \psi^{out}_{\bar p1}(t)=\int du \Sigma^R_{\bar p} (t-u)  \psi^{out}_{\bar p0}(u)
\ee
Finally, inserting the explicit decomposition Eq.(\ref{scatteringstate}) in terms
of incoming and outgoing waves inside Eq.(\ref{match1}) and using
Eq.(\ref{match3}) and Eq.(\ref{match4}), we obtain,
\be
\label{finalscatt}
i \partial_t \psi^{scat}_{\bar 0}(t)= \mathrm{\textbf{H}}_{\bar{0}\bar{0}}  \psi^{scat}_{\bar 0}(t) + \sum_{\bar p} \int_{-\infty}^t du \Sigma^R_{\bar p} (t-u) P_{\bar p} \psi^{scat}_{\bar 0}(u)
+i \Gamma_{\bar m} (E) \psi_{\bar m0}^{in}(t)
\ee
Eq. (\ref{finalscatt}) is identical to our main wave equation
(\ref{eq:Psi}) which completes the proof that \be
\psi^{scat}_{\bar 0}(t)= \Psi_{\alpha E}(t).
 \ee Hence
the equivalence between the scattering approach and the NEGF formalism can be
extended to time-dependent transport. We note however that
$\psi_{\alpha E} (t)$ and the Scattering state $\psi^{scat}_{\alpha E}(t)$ do
not match outside of the scattering region as the former only contains outgoing
modes (and no incoming ones).

\subsection{Generalization of the Fisher-Lee formula}
Besides proving the formal equivalence between the Scattering and NEGF
approaches
in this context, the above construction provides an explicit link between the
wave function and the scattering matrix.  Indeed, using the definition
Eq.(\ref{scatteringstate}) of the scattering matrix, one obtains after
integration over time,
\be
\label{fl1}
S_{\bar p\beta,\bar m\alpha}(E',E)=\int dt'  \ e^{iE' t'} \frac{[\xi^{out}_{\bar
p\beta}(E')]^\dagger}{\sqrt{|v^{out}_{\bar m\alpha}(E')|} } \Gamma_{\bar p}(E') [
\psi^{scat}_{\bar p0,\alpha E}(t') - \psi_{\bar p0, \alpha E}^{in}(t') ].
\ee
Eq. (\ref{fl1}) is a generalization of the Fisher Lee relation
\cite{Fisher1981} for time-dependent problems. As the numerical algorithms
described in the later sections allow one to compute the wave function
$\psi^{scat}_{\bar p0,\alpha E}(t')$ directly, they also provide means to
evaluate the scattering matrix through the above relation.  Equation
(\ref{fl1}) can be further simplified into,
\be
\label{fl2}
S_{\bar p\beta,\bar m\alpha}(E',E)=\frac{[\xi^{out}_{\bar p\beta}(E')]^\dagger}{\sqrt{|v^{out}_{\bar m\alpha}(E')|} } \Gamma_{\bar p}(E') \left[
\int dt'  \ e^{iE' t'}  \psi^{scat}_{\bar p0,\alpha E}(t')
-\frac{\xi^{in}_{\bar m\alpha}(E') }{\sqrt{|v^{in}_{\bar m\alpha}(E)|} } 2\pi \delta(E'-E)\right]
\ee
Inserting the definition of the wave function in term of the Retarded Green's
function inside Eq.(\ref{fl2}), one obtains another form, closer to the
original one of Ref. \cite{Fisher1981},
\be
\label{fl3}
S_{\bar p\beta,\bar m\alpha}(E',E)=\frac{[\xi^{out}_{\bar p\beta}(E')]^\dagger}{\sqrt{|v^{out}_{\bar m\alpha}(E')|} } \Gamma_{\bar p}(E') \left[
 \mathcal{G}^{R}(E',E) \Gamma_{\bar m}(E)
-  2\pi \delta(E'-E)\delta_{\bar m \bar p}\right]
\frac{\xi^{in}_{\bar m\alpha}(E)}{\sqrt{|v^{in}_{\bar m\alpha}(E)|} }
\ee
where we have introduced the (double) Fourier transform of the Retarded Green's
function,
\be
\mathcal{G}^{R}(E',E)= \int dt dt' \  \mathcal{G}^{R}(t',t)  e^{iE't'-iEt}.
\ee

\subsection{Link with the partition-free initial condition approach}
In the construction of the scattering states given above, we impose a boundary
condition where the form of the incoming modes is fixed  {\it for all times}
while the outgoing modes are free. Hence, this construction treats incoming
modes and outgoing ones on different footings. This might seem correct based on
physical arguments, yet we have seen in Section \ref{sec: wf} that the matrix
$\Gamma$ could be diagonalized in several different ways. In the rest of this
section, we follow a very simple route taken by Cini \cite{Cini1980} and
further developed in Refs. \cite{Gross_DFT, Gross_WF,Gross_bound_states,Stefanucci2013}
where such a distinction does not appear explicitly. The approach is
conceptually very simple. Let us suppose that the Hamiltonian is time-independent
up to $t=0$, then for $t<0$ we assume that the system is in an
incoherent superposition of all the eigenstates $e^{-iEt}\psi_{\alpha E}^{st}$
of the system with a filling factor $f_\alpha(E)$ (this may be thermal
equilibrium as in Ref.\cite{Gross_WF} or more generally a non-equilibrium
stationary state). At time $t>0$ the corresponding states
$\psi^{init}_{\alpha E} (t)$ simply evolve according to the Schrödinger equation $i
\partial_{t} \psi^{init}_{\alpha E} (t) = {\textbf{H}}(t) \psi^{init}_{\alpha
E} (t)$ with the initial condition $\psi^{init}_{\alpha E} (t=0)=\psi_{\alpha
E}^{st}$. Apparently, this is a different boundary condition from the one of
the scattering state above.

We now use the block structure of the Schrödinger equation (projected on lead
${\bar p}$) and obtain after integration between 0 and $t$ (momentarily
dropping the indices $E$ and $\alpha$),
\be
\psi^{init}_{\bar p} (t)+ i g_{\bar p}^R(t)\psi^{init}_{\bar p} (0)=  \int_0^t du g_{\bar p}^R(t-u) {\textbf{H}}_{\bar p\bar 0} \psi^{init}_{\bar 0} (u)
\ee
from which we get (after substitution inside the equation for
$\psi^{init}_{\bar 0}$),
\be
i \partial_{t} \psi^{init}_{\bar 0} (t) = {\textbf{H}}_{\bar 0\bar 0}(t) \psi^{init}_{\bar 0} (t) + \int_0^t du \Sigma^R(t-u) \psi^{init}_{\bar 0} (u)
-i \sum_{\bar p} {\textbf{H}}_{\bar 0\bar p}
g_{\bar p}^R(t) \psi^{init}_{\bar p} (0)
\label{gross}
\ee
Eq. (\ref{gross}) is essentially Eq.(4) of Ref.\cite{Gross_DFT}.
Eq. (\ref{gross}) is very similar to Eq.(\ref{eq:Psibar2}) with a crucial
practical difference: in the latter, the source term is present only at the
system's sites which are time-dependent while in the former it takes place at
the system-lead interfaces. Introducing $\bar \psi^{init}_{\alpha E} (t)\equiv
\psi^{init}_{\alpha E} (t) - e^{-iEt}\psi_{\alpha E}^{st}$, we find that $\bar
\psi^{init}_{\bar 0} (t)$ obeys Eq.(\ref{eq:Psibar2}) with $\bar
\psi^{init}_{\bar 0} (t=0)=0$. Hence, we have proved one more equivalence,
between the wave function $\bar \Psi_{\alpha E}(t)$ and $\bar
\psi^{init}_{\bar 0} (t)$:
\be
\psi^{init}_{\alpha E\bar 0} (t)= \Psi_{\alpha E}(t).
\ee
We note that the equivalence requires that the initial states at $t=0$ are the
scattering states $\psi_{\alpha E}^{st}$ of the stationary system. When the
system contains more than one channel, one finds that any choice of the initial
condition $\sum_{\alpha}U_{a\alpha}\psi_{\alpha E}^{st}$, where $U$ is a
unitary matrix, eventually gives the same total current and is therefore also
equivalent to the NEGF theory. However, the matrix $U$ must be unitary which
fixes the normalization of the initial states; they must carry a current unity.

\subsection{``Floquet wave function'' and link with the Floquet scattering theory}

Although this paper focuses on time-resolved electronics (typically transient
regimes or voltage pulses), the wave function formalism can also be used for
perturbations periodic in time. We refer to \cite{Moskalets2012} for an
introduction and bibliography on the subject. Let us briefly consider the
situation where
$\mathrm{\textbf{H}}_{\bar{0}\bar{0}}(t+T)=\mathrm{\textbf{H}}_{\bar{0}\bar{0}}(t)$
and introduce its decomposition in term of harmonics of $\omega=2\pi/T$,
\be
\mathrm{\textbf{H}}_{\bar{0}\bar{0}}(t)= \sum_{n=-\infty}^{\infty} H_n
e^{-i n\omega t}.
\ee
We also define the Fourier transform $\Psi_{\alpha E}(E')$ of $\Psi_{\alpha E}(t)$,
\be
\Psi_{\alpha E}(E')=\int dt' \ e^{i E' t'} \Psi_{\alpha E}(t')
\ee
from which we can express Eq.(\ref{eq:Psi}) as,
\be
E' \Psi_{\alpha E}(E')= \sum_n H_n \Psi_{\alpha E}(E'-n\omega) + \Sigma^R (E')
\Psi_{\alpha E}(E') + 2\pi \delta(E'-E) \sqrt{v_\alpha} \xi_{\alpha E}.
\ee
Introducing $\epsilon\in [-\omega/2,\omega/2]$ and $m$ such that $E'=E+\epsilon + m\omega$, one defines $\Psi_{m}(\epsilon)\equiv\Psi_{\alpha E}(E+\epsilon+m\omega)$ which verifies,
\be
\label{fl-s}
\epsilon \Psi_{m}(\epsilon) = \sum_n H_n \Psi_{m-n}(\epsilon) + [\Sigma^R
(E+\epsilon+m\omega) -m\omega -E]\Psi_{m}(\epsilon)+ 2\pi \delta(\epsilon)\delta_{m,0} \sqrt{v_\alpha} \xi_{\alpha E}
\ee
Last, we define
\be
\uppsi_{\alpha E\epsilon}(t)=\sum_m e^{-im\omega t} \Psi_{m}(\epsilon)
\ee
and obtain
\be
\label{floquet}
\Psi_{\alpha E}(t)=\int_{-\omega/2}^{\omega/2} \frac{d\epsilon}{2\pi} e^{-iEt-i\epsilon t} \uppsi_{\alpha E\epsilon}(t)
\ee
$\uppsi_{\alpha E\epsilon}(t)$ verifies $\uppsi_{\alpha
E\epsilon}(t+T)=\uppsi_{\alpha E\epsilon}(t)$ so that Eq.(\ref{floquet})
corresponds in fact to Floquet theorem.  We also note that the source term in
Eq.(\ref{fl-s}) is only present at $\epsilon =0$ so that the other energies do
not contribute to the scattering wave function.  Taking this last point into
account and computing (as an example) the current $I_{ij}(t)$ between site $i$
and site $j$, we arrive at,
\be
\label{floquet2}
I_{ij}(t) =-2\  {\rm Im } \sum_\alpha  \ \int \frac{dE}{2\pi}  f_\alpha (E)
\sum_{n,m,p} \Psi^*_{\alpha E,m}(i) [H_n]_{ij}\Psi_{\alpha
E,p}(j)e^{-i(n-m+p)\omega t}
\ee
where the wave function $\Psi_{\alpha E,n}(i)$ at site $i$  satisfies,
\be
\label{floquet3}
[E +m\omega-\Sigma^R (E+m\omega) ]\Psi_{\alpha E,m} - \sum_n H_n \Psi_{\alpha E,m-n}
= \delta_{m,0} \sqrt{v_\alpha} \xi_{\alpha E}
\ee
Eq.(\ref{floquet2}) and Eq.(\ref{floquet3}) provide a complete set of equations
to compute the current of the system. The corresponding ``Floquet wave function"
can be put in direct relation  to Floquet Scattering  theory using the link
with the Scattering matrix established at the beginning of this section. In
practice, the infinite set of equations defined by Eq.(\ref{floquet3}) needs to be truncated somehow~\cite{Nikolic2012} and one is left with solving a large, yet
finite, system of linear equations. Alternatively, a systematic perturbation theory can be constructed
taking the AC Hamiltonian as a small perturbation \cite{Cookbook_Oleksii}.

We have thus made explicit connections between various theoretical frameworks:
the NEGF, the scattering approach, the
partition-free initial condition approach and, for perturbations that are
periodic in time, the scattering Floquet approach.  This concludes the
formalism part of this paper. We now turn to a discussion of the strategies
that can be implemented to perform numerical simulations of the corresponding
theories. The formalism is also suitable for analytical approaches, and some
results will be given toward the end of this paper.

%%%%%%%%%%%%%%%%%%%%%%%%%%%%%%%%%%%%%%%%%%%%%%%%%%%%%%%%%%%%%%%%%%%%%%%%%%%%%%%%%%%
\section{Strategies for efficient numerical simulations}
%%%%%%%%%%%%%%%%%%%%%%%%%%%%%%%%%%%%%%%%%%%%%%%%%%%%%%%%%%%%%%%%%%%%%%%%%%%%%%%%%%%
\label{numerics}

We now turn to a discussion of various algorithms for simulating the formalism
introduced above. Here we provide a concrete example of an application to a
simple one dimensional chain, but the algorithms are general and apply to
arbitrary dimensions and geometries. In fact, we aim at developing a general
algorithm that would allow one to tackle a rather large class of problems, in
the same spirit as the existing numerical toolkits for the transport properties
of stationary devices\cite{Knit, Kwant_preparation}. Our Hamiltonian reads,
\begin{equation}
    \label{eq: 1d_Hamiltonian}
    \mathrm{\hat{\textbf{H}}}(t) = -\gamma \sum_{i=-\infty}^{+\infty} c^{\dagger}_{i+1}c_{i}  -\gamma  [e^{i\phi (t)} - 1] c^{\dagger}_{2}c_{1} + \sum_{i=1}^{N} \epsilon_i c^{\dagger}_{i}c_{i}+ h.c.
\end{equation}
where the system is subjected to a voltage pulse $w(t)$ with $\phi (t) =
\int_{-\infty}^t du\ w(u)$ and $\epsilon_i$ the potential inside the central
region ${\bar 0}=\{1,2,\dots N\}$.  The $\epsilon_i$ can in principle be time-dependent
but we restrict the examples to static cases; all the time dependence
comes from the voltage drop between site $1$ and site $2$. During the development of
the numerical techniques presented below, we used various analytical results
 to perform consistency checks of the validity of the numerics. They are summarized in \ref{benchmark}.

We denote $N$ the total number of sites of the
central region and $S$ the number of sites connected to the electrodes (for a
cubic system in $d$ dimensions we have $N\sim L^d$ and $S\sim L^{d-1}$). Let us
call $t_{\rm max}$ the maximum time of the simulations and $h_t$ the typical
discretization time step. In this section, we introduce the various algorithms
(three for NEGF labeled GF-A,B and C as well as four for the wave function
approach labeled WF-A, B, C and D) before turning to the numerics in the next
section. We emphasize that, although these algorithms have very different
computing efficiencies, they are all mathematically equivalent and -- as we have
checked explicitly -- give the same numerical results.

\subsection{GF-A: Brute-force integration of the NEGF equations}
The first technique consists in directly integrating the equations of motion of
the NEGF formalism treating the integro-differential equations as ordinary
differential equations. However,  the right hand sides of the equations contain
the self-energy integrals that need to be re-evaluated every time step. This
also means that some values of the Retarded Green's function in the past must
be kept in memory. The algorithm consists of 3 steps. (i) One starts with a
calculation of the leads' self-energy by a direct integration of
Eq.(\ref{eq:EOMlead}) for the $S\times S$ surface Green's function of the
leads. (ii) In the second step, one proceeds and integrates Eq.(\ref{eq:EOM})
which has a rather similar structure.  (iii) The last step is the calculation
of the Lesser Green's function using the double integration of
Eq.(\ref{eq:EOM<}). This last step is quite problematic as the integration over
times takes place over an infinite time window (as opposed to the calculation
of the Retarded Green's function where the self-energy terms only span a finite
window due to the causality of the Retarded Green's function). In practice, one
has to resort to using a cutoff within a large time window $\Delta_t$. We can
already note that the CPU cost of all these three steps scale as the square of
the total time, either $(t_{\rm max}/h_t)^2$ or $(\Delta_t/h_t)^2$ and that the
calculations of various observables (for different times for instance) involve
separate calculations for the last step.  For implementation purposes, we note
that the integrals containing the self-energy terms can be parallelized by
dividing the integral range into smaller pieces, which can be used to speed up
the calculations.  For integrating the equations of motion, we use either an
implicit linear multi-step scheme \cite{Delves} or an explicit $3^{\rm rd}$ order
Adams-Bashforth scheme (with slightly better performances for the latter).
Overall, the GF-A approach quickly becomes prohibitively expensive in CPU time.
This may explain why (to the best of our knowledge) the simulations
performed so far within this approach have been restricted to very small
systems and times.

\subsection{GF-B: Integrating out the time-independent subparts of the device}

\begin{figure}[h]
    \centering
    \includegraphics[width=8cm]{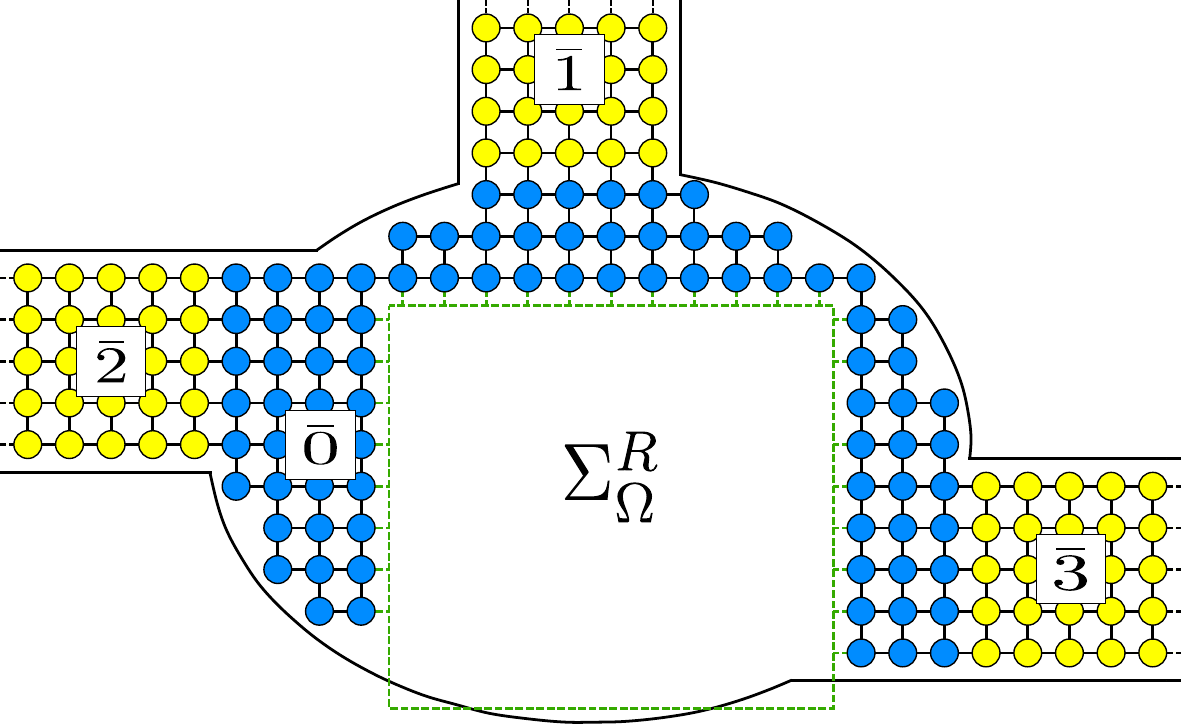}
    \caption{\label{fig: integrate_out}
     Sketch of the GF-B scheme. The degrees of freedom of the region $\Omega$
     inside the dashed green square are integrated out in a self-energy term
     denoted $\Sigma^R_{\Omega}$. This integration leads to an effective system
     containing a reduced number of sites.  }
\end{figure}

A first strategy to improve on the direct (naive) GF-A approach described above
is to integrate out the parts of the device region where we do not want to
compute observables. A typical example is shown in
Fig.~\ref{fig: integrate_out}.
 Suppose that a subset $\Omega$ of the sites in region $\bar 0$ has a
 ``sub''
 Hamiltonian matrix $\mathrm{\textbf{H}}_\Omega(t)$.  The Green's function for
 the {\it isolated } region $\Omega$ (i.e. when the coupling to the rest of
 region $\bar 0$ is zero) can be obtained by simply integrating the equation of
 motion of the finite region
 $i\partial_t g^{R}_{\Omega}(t,t')=\mathrm{\textbf{H}}_\Omega(t)
 g^{R}_{\Omega}(t,t')$. This is particularly simple when the
 region $\Omega$ is time-independent: diagonalizing the finite matrix
 $\mathrm{\textbf{H}}_\Omega \chi_\alpha = \epsilon_\alpha \chi_\alpha$, the
 Retarded Green's function simply reads,
 \be
 \label{eq:omega}
g^{R}_{\Omega}(t-t')=-i\theta(t-t') \sum_\alpha e^{-i\epsilon_\alpha (t-t')} \chi_\alpha \chi_\alpha^\dagger
\ee
Note that Eq.(\ref{eq:omega}) contrasts with its counterpart in the energy
domain: the Retarded Green's function as a function of energy of a finite region
is very ill defined numerically as it is essentially a sum of Dirac
distributions.
Noting $\mathrm{\textbf{H}}_{\bar 0\Omega}$ the matrix elements coupling the
$\Omega$ region to the rest of the device region $\bar 0$, we introduce the
self-energy due to the $\Omega$ region,
\be
 \Sigma^R_{\Omega}(t,t')= \mathrm{\textbf{H}}_{\bar{0}\Omega}(t)
    g^{R}_{\Omega}(t,t') \mathrm{\textbf{H}}_{\Omega\bar{0}}(t').
\ee
We can now proceed with solving Eq.(\ref{eq:EOM}) for the smaller region $\bar 0
\backslash \Omega$ with the added $\Sigma^R_{\Omega}$ in the self-energy,
\be
\Sigma^R(t,t')\rightarrow \Sigma^R(t,t')+\Sigma^R_{\Omega}(t,t').
\ee
Note however that the Lesser self-energy is {\it unchanged} as the $\Omega$
region is not a lead (i.e. is not at thermal equilibrium).  Using this
procedure, any region can be integrated out of the device region, effectively
reducing the effective total size $N$ of the simulation, but at the cost of
increasing the number of surface sites $S$.

When the size of the $\Omega$ region becomes large, a direct calculation of
$\Sigma^R_{\Omega}(t,t')$ becomes impractical. Fortunately, many schemes that
have been developed in the energy domain can be transposed to the time domain:
the original recursive Green's function algorithm, its variant the knitting
algorithm \cite{Knit} or the more involved nested dissection
algorithm\cite{Li20089408, Petersen20095020}. These schemes can be discussed
using the self-energy introduced above to ``decimate'' parts of the system, but
they are perhaps more transparent when discussed in the context of the Dyson
equation. Let $H_{ab}(t)$ be the Hamiltonian matrix of a system and let one
decompose it into the sum of two terms $H_{ab}=H_a+H_b$ (typically $H_a$ will
be the Hamiltonian matrix for two disconnected regions and $H_b$ connects these
two regions together) and we note $G^R_{ab}$ ($G^R_a$) the Retarded Green's function
associated with $H_{ab}$ ($H_a$).  In this context, the Dyson equation reads,
\be
G^R_{ab}(t,t')=G^R_a(t,t') + \int du \ G^R_a(t,u) H_b(u) G^R_{ab}(u,t')
\label{eq:dyson}
\ee
Eq.(\ref{eq:dyson}) allows the separated parts of the systems to be merged
(note that the structure in time of this equation is ``triangular'', i.e. one
can solve it for $t$ close to $t'$ and iteratively increase $t$). We refer to
Ref.\cite{Knit} for a detailed discussion of the procedure used for glueing
isolated parts together. Applying Eq.(\ref{eq:dyson}) recursively in a (quasi)
one dimensional geometry, one can add one slice of the system at each
iteration until the full system has been added (Recursive Green's function
algorithm).  Adding the sites one at a time, one obtains the knitting algorithm
which allows one to handle systems of arbitrary geometries. Both algorithms
have CPU times that scale as $S^2 N (\Delta_t/h_t)^2$ but memory footprints
much smaller than the direct method. In the last algorithm, nested dissection,
one cuts the system recursively into 2 (or more) pieces until the pieces are
small enough that their individual Green's functions may be calculated
directly. The gluing sequence is then applied backward to reconstruct the
Retarded Green's function of the full system. Note that the nested dissection
algorithm suffers from stability problems in the energy domain as some of the
pieces are
not in contact with the leads (and thus suffers from the problem discussed in
the beginning of this subsection). In the time domain, however, no such
limitation occurs.

\subsection{GF-C: Integration scheme that preserves unitarity}

In GF-A and GF-B, we use simple discretization schemes to integrate the
integro-differential equations for the Retarded Green's functions. However,
these schemes (as well as others, such as the Runge-Kutta method) do
not enforce unitarity of the evolution operator in Eq.(\ref{eq:evol}). The
scheme GF-C builds on GF-B but one replaces the discretization scheme by one
that preserves this important property of quantum propagation.

Eq.(\ref{eq:evol}) implies that for any intermediate time $u \ \in [t',t]$ we
have,
\begin{equation}
    \mathcal{G}^{R}(t,t') = i
    \mathcal{G}^{R}(t,u)\mathcal{G}^{R}(u,t').
    \label{eq: Gr_evol}
\end{equation}
which has a simple interpretation in terms of path integral: the propagator
between $t'$ and $t$ is a sum over all possible paths and this formula reflects
the fact that we keep track of the site where the particle is at time $u$. As
the particle may be in the central region or in one of the leads at time $u$ we
get, after integrating over the degrees of freedom of the leads (see
\ref{app: path_formula} for the derivation),
\be
G^{R}(t,t') = i  G^{R}(t,u)G^{R}(u,t') + \int_{u}^{t}
    dv \int_{t'}^{u} dv'\ G^{R}(t,v) \Sigma^{R}(v,v')
    G^{R}(v',t')
    \label{eq: path}
\ee
Eq.(\ref{eq: path}) is a sum of two terms which depend on the position of the
particle at time $u$. The first term corresponds to a particle which is in the
central region at time $u$ while the second term accounts for the paths
entering the leads at $v'<u$ and returning to the central region at a later
time $v>u$ (i.e. the particle is in the lead at time $u$). Eq.(\ref{eq: path})
encapsulates the unitarity of the evolution operator by construction. It can
be used to realize an efficient explicit integration scheme for the Retarded
Green's function. Applying Eq.(\ref{eq: path}) with $t\rightarrow t+h_t$ and
$u\rightarrow t$ we obtain:
\be
\label{unitaryscheme}
G^R(t+h_t,t') = i A_{h_t}(t) G^R(t,t')
+ \frac{h_t}{2} \int_{t'}^t dv [A_{h_t}(t) \Sigma^{R}(t,v)-i\Sigma^{R}(t+h_t,v) ] G^R(v,t')
\ee
where $A_{h_t}(t)$ is the short time propagator $A_{h_t}(t)=G^R(t+h_t,t)$.
Eq.(\ref{unitaryscheme}) provides an explicit scheme for integrating the
equation of motion which proves to be more stable than the naive ones. Note
that the Hamiltonian matrix has disappeared from Eq.(\ref{unitaryscheme}). It
is hidden in the short time propagator, $A_{h_t}(t)$, which can be obtained
``exactly'' from a direct integration of the equation of motion
Eq.(\ref{eq:EOM}) using a very small time step (much smaller than $h_t$).
The computing time to get this very precise estimate
is $\propto h_t^2$ and, $h_t$ being small, therefore negligible.

\subsection{WF-A: Direct integration of Eq. (\ref{eq:Psi})}
We now turn to the algorithms based on the wave function approach. We shall see
that they are much simpler and efficient than their NEGF counterparts. In the
first one, WF-A, we integrate directly  Eq.(\ref{eq:Psi}) using a $3^{\rm rd}$
order Adams-Bashforth scheme. The algorithm is intrinsically parallel as the
calculations for different energies are totally independent.  In a second step,
we calculate the energy integral of Eq.(\ref{eq:psi-less}) to obtain the
various observables. Note that this calculation can be done {\it on fly} so
that observables for all intermediate values of $t\le t_{\rm max}$ can be
obtained in a single run (in contrast to the GFs algorithms). A second level of
parallelism can be introduced with the calculation of the self-energy
``memory'' terms.  Note that in principle, the strategies developed for GF-B
and GF-C could be also used for the wave function approach. We shall take a
somewhat different route however.  A direct advantage of the WF approaches is
that the equations involved are on vectors rather than on matrices.
Sophisticated optimizations could be used in order {\it not} to calculate all
the matrix elements in the GF approaches (but only the relevant ones). However
in the WF approach, one naturally calculates the minimum information needed to
recover the observables.

\subsection{WF-B: Subtracting the stationary solution, integration of
Eq.(\ref{eq:Psibar2})} WF-B is very similar to WF-A except that we now use
Eq.(\ref{eq:Psibar2}) and therefore study the {\it deviation} from the
stationary case.  Being able to ``subtract'' the stationary physics from the
equations brings three distinct advantages compared to WF-A: (i) self-energy
``memory'' integrals start from $t=0$ (instead of $t=-\infty$) removing the
need for the large time cutoff $\Delta_t$ introduced earlier. In addition, the
initial condition is very well defined as the wave function vanishes. (ii) for
most practical physical systems, the characteristic energies involved are small
compared to the Fermi energy.  Subtracting the stationary physics allows one to
take advantage of these features to narrow down the integration of
Eq.(\ref{eq:psi-less}) to a region close to the Fermi energy. (iii) The source
terms in  Eq.(\ref{eq:Psibar2}) are present only at the sites where time-dependent
perturbations are present.

\subsection{WF-C: From integro-differential to differential equation.}
The scheme WF-B is already quite efficient and renders rather large problems
($N\sim 1000$) for rather long times ($t\sim 1000\gamma^{-1}$) tractable in a
reasonable CPU time (say, 1 hour).  Let us analyze its total CPU cost. We find,
$CPU(WF-B)\propto (t/h_t) [ N + S^2 (t/h_t) ] N_E$ where the first term comes
from the (sparse) matrix vector multiplication with the Hamiltonian matrix and
the second term accounts for the ``memory'' integral with the self-energy. The
factor $N_E$ accounts for the different energies and modes for which
Eq.(\ref{eq:Psibar3}) must be integrated. In general this $N_E$ is not an issue
as all these calculations can be done in parallel and for relevant regimes the
integral gets concentrated on a region close to the Fermi energy. The memory
footprint is $MEM(WF-B) \propto [ N + S (t/h_t) ]$ as we need
to keep in memory the wave function at time $t$ in the system plus its history
on the lead-system interfaces. The bottleneck of the calculation clearly comes
from the ``memory integral'' which itself comes from the information
corresponding to the wave function outside of the central region. The
computational time is essentially the same as if one had studied the time
evolution of a finite isolated system of $N + S^2 (t/h_t)$ sites. For the
typical values used here, $t=1000 \gamma^{-1}$ and $h_t=0.01$, we find that
WF-B's CPU is the same as if one was studying a finite system (i.e. no leads)
of size $N=100 000$. On the other hand we know that signal propagation in the
Schrodinger equation takes place at most at a speed $v=\partial E/\partial k$
with $E(k)=-2\gamma \cos k$ for a 1d chain. Hence at most $M\approx \gamma t$
layers of the lead can be probed by the propagation of the wave-function.  For
$t=1000 \gamma^{-1}$ this means at most $1000$ layers.

\begin{figure}[h]
    \centering
    \includegraphics[width=8cm]{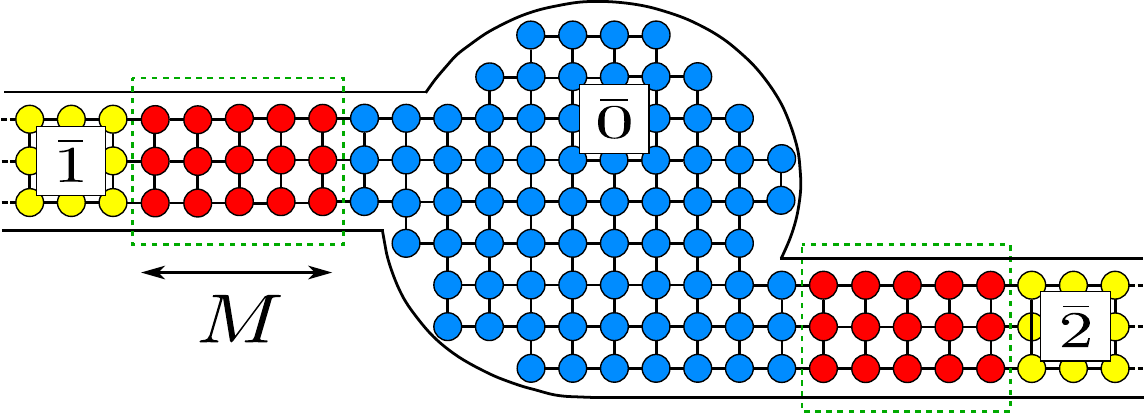}
    \caption{\label{fig:wfc}
 Sketch of the WF-C and WF-D schemes: $M$ layers of the leads (red) are added
 the central part $\bar{0}$ (blue circles) to constitute the effective central
 region. In WF-C the rest of the leads (yellow circles) are simply ignored
 while in WF-D, they are treated within the wide band approximation. }
 \end{figure}

The scheme WB-C is therefore very simple: instead of integrating the
integro-differential equation Eq.(\ref{eq:Psibar2}), one integrates the much
simpler differential equation Eq.(\ref{eq:Psibar3}).  As this cannot be done
for an infinite system, one simply truncates the system keeping the central
region plus $M$ layers of each leads (see Fig.\ref{fig:wfc}). The expected
correct value for $M$ is $M\approx v_{mx} t/2$ with the maximum speed being
$v_{mx}=\gamma \max_k |\partial E/\partial k|=\gamma z$. $z$ is the coordinance
of the system (number of neighbors per site) and the factor $1/2$ comes from
the fact that the signal has to travel up to the effective boundary (yellow-red
interface on Fig.\ref{fig:wfc}) {\it and} come back in order to disturb the
central region. Lower values of $M$ can be used if the Fermi energy is close to
the band edges and the system is therefore slower. According to the above
analysis, only $M\sim 1000\ll 100 000$ layers should be necessary, which should
lead to an important speed up compared to WF-B. It also considerably simplifies
the implementation and allows for very aggressive optimizations. The expected
gain is not a simple prefactor as $CPU(WF-C)\propto (t/h_t) [ N + S \gamma t ]
N_E$ is parametrically smaller than WB-B for 2D and 3D systems.

\subsection{WF-D: Faster convergence using the wide band limit.}
The drawback of WF-C is that hardwall boundary conditions are employed at the
yellow-red interface (see Fig.\ref{fig:wfc}). If one does not take a
large enough value of $M$, the particles will eventually bounce back toward the
central region. WF-D is a simple generalization of WF-C where the remaining
part of the leads (yellow sites in Fig.\ref{fig:wfc}) are treated within the
wide band limit Eq.(\ref{wbl}) so that we effectively have ``absorbing''
boundary conditions and faster convergence properties with respect to $M$.
Note that WF-D is an {\it exact} scheme, the (wide band limit) self-energy term is only used to
accelerate the convergence with respect to $M$ (as we shall see later in Fig.~\ref{fig:WBLtest}).

We shall see that WF-D will be by far the fastest of all the methods described
in this article.  We gather below the various steps associated with its
practical implementation (the equations that follow were given before and are
repeated here for convenience).
\begin{enumerate}
\item One starts with defining the Hamiltonian of the model, i.e. the two
    matrices $H_{\bar m}$ and $V_{\bar m}$ that define the Hamiltonian of each
    lead as well as the time-independent matrix
    $\mathrm{\textbf{H}}_{\bar{0}st}$ for the central part and the time-dependent
    counterpart $\mathrm{\textbf{H}}_{\bar{0} w}(t)$. In many cases
    (for instance for the voltage pulses discussed next), the time-dependent
    part of the Hamiltonian only connects a few subparts of the central region.
\item
\begin{enumerate}
\item One constructs the stationary modes of the leads, solving
    Eq.(\ref{modes2}). (There is a large literature on this topic which we
    refer to, see Ref.\cite{Knit} and references therein.)
\item One also computes the self-energy of the leads, defined by
    $\Sigma^R (E)= \sum_{\bar m} V_{\bar{m}} g^{R}_{\bar m}(E)
    V^\dagger_{\bar m}$ and Eq.(\ref{sce4lead}).
\end{enumerate}
\item Once the leads properties are known, one computes the stationary wave
    function of the system solving the following linear set of equations
\be [E
    - \mathrm{\textbf{H}}_{\bar{0} st} - \Sigma^{R}(E)] \Psi^{st}_{\alpha E} =
    \sqrt{v_\alpha} \xi_{\alpha E}.
\nonumber
\ee
Note that steps (2a), (2b) and (3) are
    standard steps of quantum transport calculations in wave function based
algorithms.
\item $M$ layers of the leads are now concatenated
    to the central region Hamiltonian matrix $\mathrm{\textbf{H}}_{\bar{0}st}$.
    Everything is now ready to form the main Eq. (\ref{wbl}) of the method
\be
   i \partial_{t} \bar\Psi_{\alpha E}(t) = [ \mathrm{\textbf{H}}_{\bar{0}st} + \mathrm{\textbf{H}}_{\bar{0} w}(t) +\Sigma^{R}(E)  ]\bar\Psi_{\alpha E}(t) +
    \mathrm{\textbf{H}}_{\bar{0}w}(t)  e^{-iEt}\Psi_{\alpha E}^{st}
\nonumber
\ee
which is integrated numerically using any standard integration scheme.
\item The full wave function of the system is then reconstructed,
\be
\Psi_{\alpha E}(t) = {\bar \Psi}_{\alpha E}(t) + e^{-iEt} \Psi_{\alpha E}^{st}.
\nonumber
\ee
\item The various observables (time-dependent current, electronic
    density...), which can be expressed in term of the Lesser Green's function,
    are obtained by the numerical integration (and sum over incoming modes)
    over the energy  of Eq.(\ref{eq:psi-less}).  For instance, the current
    between sites $i$ and $j$ reads,
\be
\label{current}
I_{ij}(t) =-2\  {\rm Im } \sum_\alpha  \ \int \frac{dE}{2\pi} f_\alpha (E)
\Psi^*_{\alpha E}(i,t)\mathrm{\textbf{H}}_{ij}(t)\Psi_{\alpha E}(j,t).
\ee
\end{enumerate}

%%%%%%%%%%%%%%%%%%%%%%%
\section{Numerical test of the different approaches.}
%%%%%%%%%%%%%%%%%%%%%%%
\label{sec: applications}

\subsection{Green's function based algorithms}
Let us start the numerical applications by sending a square voltage pulse
$w(t)=w_0 \theta(t-t_0)\theta(t_1-t)$ inside our quantum wire ($t_1>t_0$).  Fig.~\ref{fig:
IvsT} shows the pulse (dashed line) together with the calculation of the
current $I(t)$ using the GF-C technique (red line) and WF-B (black). Our first
finding is that both methods agree, which, given the fact that the two methods
are totally independent, is a strong check of the robustness of the approaches.
After relaxation, we expect the current to saturate to its DC value given by
the Landauer formula $I_{dc}=w_0$ (transmission is unity for a perfect 1d
chain), and indeed, it does.  Just after the abrupt rise of the potential, one
observes rapid oscillations of frequency $2\gamma/\pi$. These oscillations, often
observed in numerical simulations\cite{Wingreen-Meir_1994}, come from the fact
that the rise of the voltage is (infinitely) fast compared to the bandwidth of
the system, hence the band serves as a low-pass filter for the signal. Other large energy oscillations
of frequency $E_F/(2\pi)$ can also be observed. The
bandwidth usually corresponds to optical frequencies. For nanoelectronics
applications, therefore, one should stay in a regime where the characteristic
time scales of the time-dependent perturbations are large (say at least a
factor 10) compared to $\gamma^{-1}$.

\begin{figure}[h]
    \centering
    \includegraphics[width=8cm]{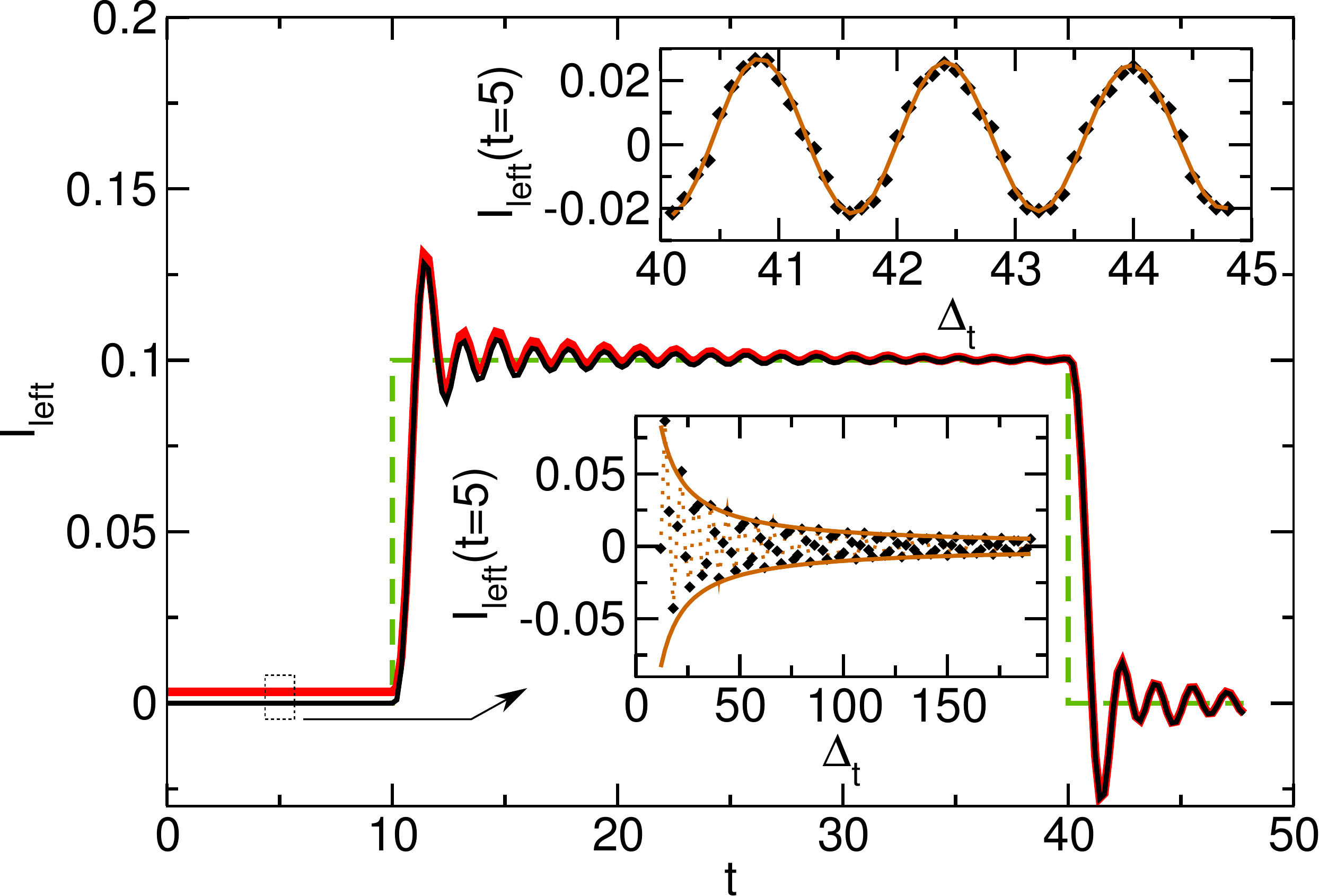}
    \caption{Current as a function of time for a square voltage pulse $w(t)=w_0
    \theta(t-t_0)\theta(t_1-t)$ with $w_0=0.1\gamma$,
    $t_0=10\gamma^{-1}$,  $t_1=40\gamma^{-1}$ and $E_F=0\gamma$. The lines show $w(t)$ (dashed), the GF-C result (red) and the WF-B result (black). Lower inset: current $I(t=5\gamma^{-1})$
    as a function of $\Delta_t$ for the GF-B scheme (symbols) together with the fit $1/\Delta_t$ (line). Upper inset: zoom of the lower inset with the fit $I=(0.1+\cos(4\Delta_{t}))/\Delta_{t}$.
    \label{fig: IvsT}
 }
\end{figure}

Before the pulse, the current vanishes. In the WF-B scheme, this is
automatically encoded as the system is in a stationary state. In the GF schemes
however, one needs a large value of the cut-off $\Delta_t$ to simply recover
this elementary fact. The lower inset of Fig.~\ref{fig: IvsT} shows the current
before the pulse as a function of the cut-off $\Delta_t$ together with a
$1/\Delta_t$ fit.  The data in the lower inset look noisy but upon closer
inspection (upper inset), one finds that the convergence shows fast
oscillations as $\cos(4\gamma\Delta_{t})/\Delta_{t}$. The slow convergence of
the GF schemes with respect to $\Delta_{t}$ is in itself a strong limitation.

\begin{figure}[h]
    \centering
    \includegraphics[width=8cm]{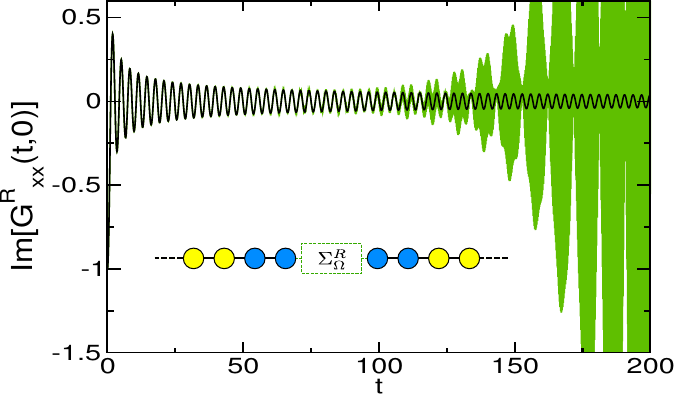}
    \caption{\label{fig: divergence}
    Comparison of GF-B (green, divergent) and GF-C (black, stable). We plot the
    imaginary part of the diagonal part of the Retarded Green's function as a
    function of time for $N=100$ (no time-dependent perturbation is applied).
    The $96$ central sites have been integrated out and an effective system of
    four sites remains.  $h_t=0.1$.}
\end{figure}

As Fig.~\ref{fig: IvsT} considers a perfect lead, it is enough to keep a small
($N\ge 2$) number of sites in the central region. If one is interested in, say,
the time it takes for a pulse to propagate, then a much larger system is
necessary and GF-A becomes impractical.  Fig.~\ref{fig: divergence} shows a
comparison between GF-B and GF-C for the calculation of the diagonal part of
the Retarded Green's function for a system with $N=100$ where the $96$ central
sites have been ``integrated out'' in order to reduce the effective size of the
system. We find that the naive discretization scheme (linear multi-steps in
this instance) used in GF-B fails and becomes unstable at large time while the
unitarity preserving scheme of GF-C restores the stability of the algorithm.
Further inspection showed that, indeed, extremely small values of $h_t$ were
needed in GF-B to enforce current conservation. GF-C is currently our best
Green's function based algorithm.

\subsection{Wave functions based algorithms}

We now turn to the wave function based algorithms. Fig.~\ref{fig:
psi_convergence} shows the local density of particles on site $1$ for a system
of two sites $N=2$ using WF-A  and various initial conditions.  We find that
the local density converges to its equilibrium value for any initial condition,
and rather faster than within Green's function algorithms.  More
importantly, by calculating the DC scattering wave function (a standard object
in computational DC transport), one can avoid the relaxation procedure and
automatically incorporate the equilibrium properties of the system (dashed
line).

\begin{figure}[h]
    \centering
    \includegraphics[width=15cm]{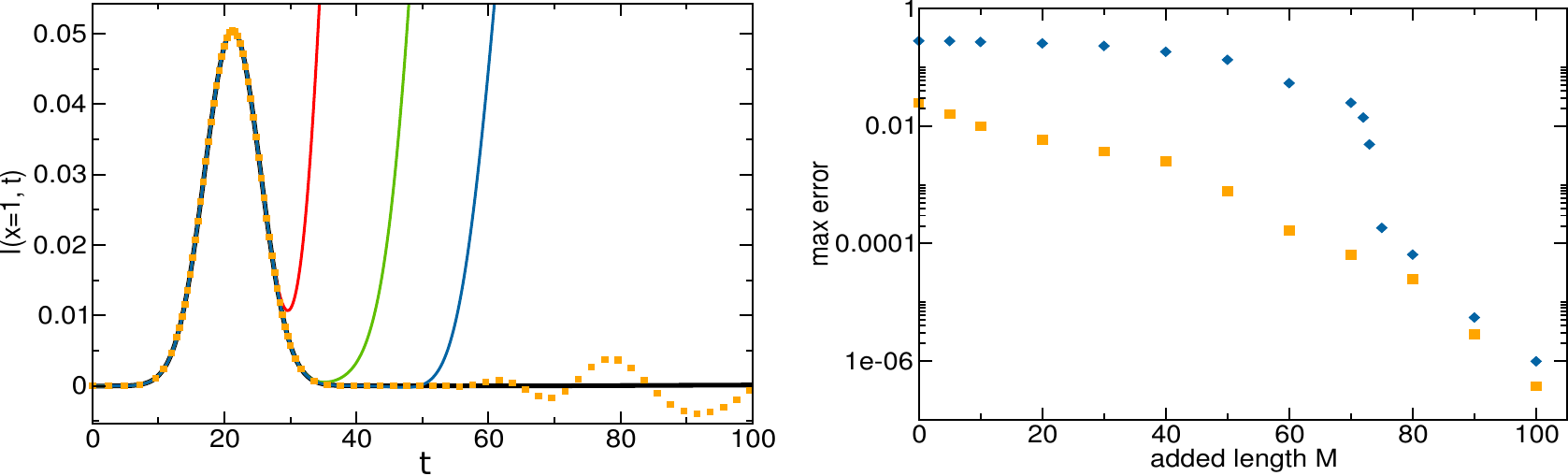}
    \caption{ Comparative study of WF-B, C and D for $N=100$.  $E_F=-1\gamma$
    and we send a Gaussian voltage pulse $w(t) = V_P
    e^{-4\log(2)t^2/\tau_P^2}$ with $V_P=0.05\gamma$ and
    $\tau_P=10\gamma^{-1}$ through the system.  Left plot: current as a function
    of time just after the voltage drop for WF-B (black), WF-C with (from left
    to right) $M=10$ (red), $M=20$ (green), $M=30$ (blue) and WF-D $M=30$ (orange
    squares).  Right graph: maximum error between $t=0\gamma^{-1}$ and
    $t=100\gamma^{-1}$ as a function of $M$ for WF-C (blue diamonds) and WF-D
    (orange squares).} \label{fig:WBLtest}
\end{figure}

WF-B which naturally captures the equilibrium conditions is a clear improvement
over WF-A. According to the arguments developed above, WF-C and D should permit
further improvements.  Upper Fig.~\ref{fig:WBLtest} shows current versus
time in presence of a Gaussian pulse for the three methods WF-B, C and D (and
various values of the number $M$ of added sites for the latter two).  In the
case of WF-C, one observes a very accurate regime until a time $t_0\propto M$
where the method abruptly becomes very inaccurate. This is due to the
finiteness of the effective system in WF-C. $t_0$ corresponds to the time it
takes for the signal to travel until the end of the sample and back again after
being reflected at the end. The wide band limit approximation used in WF-D
allows one to limit this abrupt failure and results in a much more robust (and
slightly faster) method. Lower Fig.~\ref{fig:WBLtest} shows the (maximum)
error made as a function of $M$. As surmised, very small values of $M$ are
needed for very accurate results. WF-D is our fastest and most robust method.

\subsection{Relative performance of the different approaches}
We end this section with Table~\ref{fig:benchmark} that compares
the relative performance of the various methods presented here.  We
find that WF-D is now fast enough to study two or three
dimensional systems with tens of thousand of sites (work station) or millions
of sites (supercomputers) with matching simulation times. More applications
will be shown later in the text and will show that WF-D essentially bridges the
gap between our simulation capabilities for stationary problems and time-dependent
ones.

Table~\ref{fig:benchmark} shows rather unambiguously the superiority of the
WF-D approach over all the others, especially the GF approaches. GF-B (not
stable for long times, otherwise similar to GF-C), WF-A (similar to WF-B but
much less robust) and WF-C (similar to WF-D but less robust and slightly
slower) are not shown. Note that the given times correspond to single core
calculations. WF-D can be further accelerated using two levels of parallelism:
a trivial one is the calculation of different energies on different cores
(allowing to drop the factor $N_E$). The second one is the sparse matrix -
dense vector multiplication in the evaluation of the product
$\mathrm{\textbf{H}}_{\bar{0}\bar{0}}(t) \bar\Psi_{\alpha E}(t)$ in
Eq.(\ref{wbl}). There are also two avenues for optimization which were not yet
explored in depth: the choice of the time integration scheme (e.g. an
adaptive time step) and the choice of the scheme for the integration over
energy (here again a combination of Gaussian quadrature scheme with an
adaptive energy mesh might be more effective than a naive approach).

\begin{table}[h]
%\smallskip
%\renewcommand{\arraystretch}{1.2}
\centering
\resizebox{8cm}{!} {
    \begin{tabular}{|c|c|c|c|}
        \hline
        Algorithm & CPU (1D) & Estimated CPU (2D) & Scaling of CPU \\
        \hline
        WF-D & $1$ & $10^4$ & $(t/h_t) N_E [N+\gamma t S]$ \\
         \hline
        WF-B & $40$ & $4.10^7$ & $(t/h_t) N_E [N+ (t/h_t) S^2]$ \\
         \hline
        GF-C & 10$^4$ & $10^{12}$ & $(t/h_t)^2 S^3$ (*)  \\
         \hline
        GF-A & 10$^5$ & $10^{14}$ & $(t/h_t)^2 S^2 N$ (*)  \\
          \hline
    \end{tabular}
}
\caption{Computation time in seconds for a calculation performed on a single
    computing core.  1D case: $N=20$ and $t_{max} = 10\gamma^{-1}$ (for GF-A
    the calculation has been done in parallel using 48 cores in order to obtain
    the results within a few hours). 2D case: $100\times 100$ sites hence,
    $S=100$, $N=10^4$ and $t_{max}=100\gamma^{-1}$. The CPU time is estimated
    from the scaling laws except for WF-D where calculations of similar sizes
    could be performed.  Third column: typical scaling of the computing time. A
    notable additional difference between the WF and GF methods is that (*) the
    GF methods only provide the observables at one given time per calculation
    while the WF methods give the full curve in one run. Typical values of
    $N_E$ needed for the integrations over energy are $20<N_E<100$.}
\label{fig:benchmark} \end{table}

%%%%%%%%%%%%%%%%%%%%%%%%%%%%%%%%%%%%%%%%%%%%%%%%%%%%%%%%%%%%%%%%%%%%%%%%
\section{A Landauer formula for voltage pulses}
%%%%%%%%%%%%%%%%%%%%%%%%%%%%%%%%%%%%%%%%%%%%%%%%%%%%%%%%%%%%%%%%%%%%%%%%
\label{landauer}

So far, the formalism and numerical techniques that have been presented are
applicable to arbitrary time-dependent perturbations. We now proceed with the
particular case where the perturbation is a  pulse of finite duration.

\subsection{Total number of injected particle}
We aim to define the generalization of the Landauer formula for pulse
physics. A natural extension would be to compute the time-dependent current
$I_{\bar p}(t)$ in lead $\bar p$. It is given by,
\be
I_{\bar p}(t)=\int \frac{dE}{2\pi} \sum_{\alpha} f_{\alpha}(E) I_{\alpha E, \bar p}(t)
\ee
with
\be
 I_{\alpha E, \bar p}(t) = 2 \ Im \ \Psi_{\alpha E, \bar px}^\dagger(t) V_{\bar p}^\dagger \Psi_{\alpha E, \bar p x-1}(t)
\ee
the notation corresponding to that introduced in the scattering matrix section \ref{sec:S}. We can now insert
Eq.(\ref{scatteringstate}) into the definition of $I_{\bar p}(t)$ and to
express it in term of the scattering matrix. The general formula involves a
triple integral over energy which is not very illuminating.  It also lacks the
basic properties of the Landauer-Buttiker approach which arise from current
conservation (time-dependent current is not conserved) and gauge invariance.
An important simplification occurs when one calculates the total number of
particles $n_{\bar p}=\int_0^{t_M} dt I_{\bar p}(t)$ received in lead $p$ in
the limit ${t_M\rightarrow\infty}$.  Of course, at this level of generality,
$n_{\bar p}$ can possibly diverge due to the presence of DC currents. Hence,
the following expressions assume a finite (large) value of the cutoff $t_M$.
Introducing $n_{\alpha E, \bar p}=\int_0^{t_M} dt\ I_{\alpha E,\bar p}(t)$, we
obtain
\be
\label{t-landauer}
n_{\alpha E, \bar p}=\sum_{\beta\in \bar p} \int \frac{dE'}{2\pi} P_{\bar p\beta,\bar m\alpha}(E',E)
 -\int_0^{t_M} dt \ \delta_{\alpha\beta} \delta_{\bar p\bar m}
\ee
\be
\lim_{t_M\rightarrow\infty} P_{\bar p\beta,\bar m\alpha}(E',E) =|S_{\bar p\beta,\bar m\alpha}(E',E)|^2
\ee

$P_{\bar p\beta,\bar m\alpha}(E',E)$ is thus interpreted as the
probability density to be scattered from channel $\alpha$ and energy $E$ to
channel $\beta$ and energy $E'$.  Equivalently, introducing the Fourier
transform $S_{\bar p\beta,\bar m\alpha}(t,E)\equiv \int \frac{dE'}{2\pi} e^{-i
E' t} S_{\bar p\beta,\bar m\alpha}(E',E)$ and using Parseval theorem, one
obtains
\be
\label{t-landauerB}
n_{\alpha E, \bar p}=\sum_{\beta\in \bar p}\int_0^{t_M} dt\  [ P_{\bar p\beta,\bar m\alpha}(t,E) -\delta_{\alpha\beta} ]
\ee
\be
\label{t-landauerC}
\lim_{t_M\rightarrow\infty} P_{\bar p\beta,\bar m\alpha}(t,E) =|S_{\bar p\beta,\bar m\alpha}(t,E)|^2
\ee

As the wave function $\Psi_{\alpha E}$ obeys the Schrödinger equation, one gets
a current conservation equation $\partial_t Q_{\alpha E,\bar 0} =\sum_{\bar p}
I_{\alpha E,\bar p}(t)$ where $Q_{\alpha E,\bar 0}= \Psi_{\alpha E}(t)^\dagger
\Psi_{\alpha E}(t)$ is the total number of particle inside the system
associated with mode $\alpha$ and energy $E$. Long after the pulse, the system
is back to equilibrium so that $Q_{\alpha E,\bar 0}(t_M)=Q_{\alpha E,\bar
0}(0)$ and the current conservation implies, \be \forall E\ ,\forall  \alpha
\sum_{\bar p} n_{\alpha E, \bar p} =0 \ee

Putting everything together, we obtain,
\be
n_{\bar p}=\sum_{\bar m}  \sum_{\alpha\in \bar m}\int \frac{dE}{2\pi} f_{\bar m}(E)  n_{\alpha E,\bar p}
\ee
To summarize, we find a formal analogy between the known rules of conventional
(DC) scattering theory and those of time-dependent transport. Summations over
channels are extended to a summation over channels {\it and} an integral over
energy (or time) while the current is replaced by the total number of
transmitted particles. In practice, the different terms contributing to
$n_{\bar p}$ should be grouped in such a way that the limit
${t_M\rightarrow\infty}$ can be taken without divergences (in the absence of DC current).

\subsection{Scattering matrix of a voltage pulse}

The theory above is rather general. We proceed with the particular case where
the perturbation is a voltage pulse applied to one electrode.
We consider an abrupt voltage drop across an infinite wire described by the
Hamiltonian matrix (\ref{Hlead}).  The voltage drop takes place between layers
$x=0$ and $x=1$. For this system, the Scattering matrix has a block structure
in term of the reflection $r$ and transmission $d$ amplitude,
\be
S_{\beta\alpha}(E',E)=
\left(
\begin{array}{cc}
r_{\beta\alpha}(E',E) & d_{\beta\alpha}(E',E) \\   d'_{\beta\alpha}(E',E) & r'_{\beta\alpha}(E',E)
\end{array}
\right)
\ee
which corresponds to the following form of the scattering wave function,
\be
x>0:\ \psi^{scatt}_x(t) = {\psi}^d_x(t), \ \  x\leq0:\  \psi^{scatt}_x(t) = {\psi}^r_x(t)
\ee
with
\begin{align}
    {\psi}^r_x(t) =& \frac{\xi_{\bar m \alpha}^+(E)}{\sqrt{|v^+_{\bar m \alpha}|}}
    e^{-iEt+ik^+_{\alpha}(E)x} + \sum_{\beta} \int \frac{dE'}{2\pi}
    \frac{\xi_{\bar m \beta}^-(E')}{\sqrt{|v^-_{\bar m \beta}|}} e^{-iE't-ik^-_{\beta}(E')x} r_{\beta
    \alpha}(E',E)
    \\
    {\psi}^d_x(t) = & \sum_{\beta} \int \frac{dE'}{2\pi}
    \frac{\xi_{\bar m \beta}^+(E')}{\sqrt{|v^+_{\bar m \beta}|}} e^{-iE't+ik^+_{\beta}(E')x} d_{\beta
    \alpha}(E',E)
\end{align}
where the subscript  $+$ ($-$) refers to right (left) going modes. $\psi^r_x(t)$ and $\psi^d_x(t)$
satisfy $i\partial_t \psi_x(t) = H_{\bar m} \psi_x(t) + V^{\dagger}_{\bar m}
\psi_{x-1}(t) +V_{\bar m} \psi_{x+1}(t)$ for all values of $x$ while
$\psi^{scatt}_x(t)$ satisfies the ``matching conditions'',
\be
i\partial_t \psi^{scatt}_0(t) = H_{\bar m} \psi^{scatt}_0(t) + V^{\dagger}_{\bar
m} \psi^{scatt}_{-1}(t) +V_{\bar m}e^{i\phi_{\bar m}(t)} \psi^{scatt}_{+1}(t)
\ee
\be
i\partial_t \psi^{scatt}_1(t) = H_{\bar m} \psi^{scatt}_1(t) + V^{\dagger}_{\bar
m}e^{-i\phi_{\bar m}(t)} \psi^{scatt}_{0}(t) +V_{\bar m} \psi^{scatt}_{+2}(t)
\ee
from which we directly get
\begin{align}
    V_{\bar m} \psi^r_1(t) =& V_{\bar m}e^{i\phi_{\bar m}(t)} \psi^d_1(t)
    \label{eq:match1}\\
   V_{\bar m}^\dagger \psi^r_0(t) =& V_{\bar m}^\dagger e^{i\phi_{\bar m}(t)} \psi^d_0(t)
    \label{eq:match2}
\end{align}
Inserting the explicit forms of $\psi^r_x(t)$ and $\psi^d_x(t)$ into
Eq.(\ref{eq:match1}) and Eq.(\ref{eq:match2}) (and making use of
Eq.(\ref{eig-gr}) and Eq.(\ref{eig-ga})), we obtain the equation satisfied by
the transmission matrix,
\begin{align}
\label{transmission}
    \sum_{\beta} \int &\frac{dE'}{2\pi} K_{\bar m}(\epsilon - E')
    \left[ \Sigma^R_{\bar m}(E') - \Sigma^R_{\bar m}(\epsilon)^{\dagger} \right]
    \frac{\xi_{\bar m \beta}^+(E')}{\sqrt{|v^+_{\bar m \beta}(E')|}} d_{\beta \alpha}(E',E)
    =\left[ \Sigma_{\bar m}^R(E) - \Sigma_{\bar m}^R(\epsilon)^{\dagger} \right]
    \frac{\xi_{\bar m \alpha}^+(E)}{\sqrt{|v^+_{\bar m \alpha}(E)|}} 2\pi \delta(\epsilon-E)
\end{align}
and similarly
\begin{align}
    \sum_{\beta} \int &\frac{dE'}{2\pi} K^*_{\bar m}(E'-\epsilon)
    \left[ \Sigma^R_{\bar m}(E') - \Sigma^R_{\bar m}(\epsilon)^{\dagger} \right]
    \frac{\xi_{\bar m \beta}^-(E')}{\sqrt{|v^-_{\bar m \beta}(E')|}} d'_{\beta \alpha}(E',E)
    =\left[ \Sigma_{\bar m}^R(E) - \Sigma_{\bar m}^R(\epsilon)^{\dagger} \right]
    \frac{\xi_{\bar m \alpha}^-(E)}{\sqrt{|v^-_{\bar m \alpha}(E)|}} 2\pi \delta(\epsilon-E)
\end{align}
where $K_{\bar m}(E)$ is the harmonic content of the voltage pulse,
\be
K_{\bar m}(E) = \int dt\ e^{i\phi_{\bar m}(t) +iEt}
\ee
In the situation where time-reversal symmetry is present
$\mathrm{\textbf{H}}_{\bar{m}\bar{m}}=\mathrm{\textbf{H}}^*_{\bar{m}\bar{m}}$
(no spin), one finds that to each right-going mode $\xi_{\bar m \alpha}^+$ is
associated a left-going one $(\xi_{\bar m \alpha}^+)^*$ with equal velocity. It
follows that
\be
\label{sym}
d'_{\beta \alpha}(E',E)=d_{\beta \alpha}(E,E')^*
\ee
The relation between left and right propagating modes is however more complex
in presence of magnetic field.  We continue with a physical assumption, namely
that the typical pulse height ($w_p$) is small compared to the Fermi energy
$w_P\ll E_F$.  We also suppose that its duration $\tau_P$ is rather long,
$\hbar/\tau_P\ll E_F$. This is in fact the typical situation in actual
experiments where the Fermi level $E_F\approx 1eV$ (metal) or $E_F\approx 10
meV$ (semi-conductor heterostructure) is much larger than the typical
characteristic energies of the pulses ($w_P< 1\mu eV$, $\tau_P\approx 1ns
\rightarrow \hbar/\tau_P\approx 1\mu eV$). As the kernel $K_{\bar m}(E)$
typically decays over  $\max (w_P,\hbar/\tau_P)$, we can therefore neglect the
energy dependence of the modes in Eq.(\ref{transmission}) (the so called wide
band limit) which are all taken to be at energy $E$.  The terms $\Sigma^R_{\bar
m}(E') - \Sigma^R_{\bar m}(\epsilon)^{\dagger}$ simplify into $\Sigma^R_{\bar
m}(E) - \Sigma^R_{\bar m}(E)^{\dagger} = -i\Gamma_{\bar m}(E)$ and
Eq.(\ref{ortho}) leads to

\be
    d_{\beta\alpha}(E',E) = \delta_{\alpha\beta} K_{\bar m}^*(E-E')
\label{d-wbl2}
\ee
or
\be
\label{d-wbl}
    d_{\beta\alpha}(t,E) =  \delta_{\alpha\beta} e^{-i\phi_{\bar m}(t) -iEt}
\ee
while $d'_{\beta\alpha}(E,E') = \delta_{\alpha\beta} K_{\bar m}(E'-E)$. We note
that in the wide band limit Eq.(\ref{sym}) holds even in the presence of
magnetic field.  Also, the reflection matrix $r_{\beta\alpha}(E',E)$ simply
vanishes in this limit. The role of the voltage drop is therefore purely to
redistribute the energy of the incoming electron into a larger energy window.

\subsection{Voltage pulses in multiterminal systems}

We now have all the ingredients to construct the theory of voltage pulses in
general multi-terminal systems. We assume that before the pulse, the system is
at equilibrium with no DC current flowing.  We also assume the wide band limit
of the above section, which implies that all the inelastic processes of the
scattering matrix take place at the position of the voltage drop. The
assumption that no reflection takes place at this place is important as each
electron experiences at most two inelastic events (upon entering and leaving
the sample) which considerably simplifies the theory. Introducing the DC
scattering matrix $S^0_{\bar p\beta,\bar m\alpha}(\epsilon)$ of the device {\it
in the absence of pulses}, we have
\be
S_{\bar p\beta,\bar m\alpha}(E',E)= \int \frac{d\epsilon}{2\pi} K_{\bar
p}(\epsilon-E')\  S^0_{\bar p\beta,\bar m\alpha}(\epsilon) \ K_{\bar m}^*(E - \epsilon)
\ee
Using $\int dE'/(2\pi) K_{\bar p}(\epsilon -E')K_{\bar p}^*(\bar\epsilon
-E')=2\pi\delta (\bar\epsilon - \epsilon)$, we find upon performing the integral
over $E'$ in Eq.(\ref{t-landauer}) \begin{align}
\label{final-Landauer}
n_{\bar p} =\sum_{\bar m}
\sum_{\beta\in \bar p} \sum_{\alpha\in \bar m} \int\frac{d\epsilon}{2\pi} \left[
\int \frac{dE}{2\pi}f(E) |S^0_{\bar p\beta,\bar m\alpha}(\epsilon)|^2 \ |K_{\bar
m}(E - \epsilon)|^2 -f(\epsilon) \int_0^{t_M} dt \ \delta_{\alpha\beta}
\delta_{\bar p\bar m}\right]
\end{align}
By using the unitarity of the device Scattering matrix $\sum_{\bar m\beta} |S^0_{\bar p\beta,\bar m\alpha}(\epsilon)|^2=\delta_{\alpha\beta} \delta_{\bar p\bar m}$ in the second part of Eq. (\ref{final-Landauer}), it can be rewritten in a more compact form where the limit $t_M\rightarrow\infty$ can be taken formally. It reads, 
\begin{align} 
n_{\bar p} &= \sum_{\bar m} N_{\bar p\bar m} \nonumber \\
N_{\bar p\bar m} &= \sum_{\beta \in \bar p} \sum_{\alpha \in \bar m}
                 \int\frac{d\epsilon}{2\pi}|S^0_{\bar p\beta,\bar m\alpha}(\epsilon)|^2 
                 \int\frac{dE}{2\pi}|K_{\bar m}(E - \epsilon)|^2 \left[ f(E) - f(\epsilon)
                 \right].
\label{finalfinal-Landauer}
\end{align}
Eq. (\ref{finalfinal-Landauer}) is the main result of this section. 
The ‘‘pulse conductance matrix'' $N_{\bar p\bar m}$ can be seen as the formal  generalization of the multiterminal DC conductance matrix~\cite{Buttiker1986} to voltage pulses.
In particular it shares two important properties of the DC conductance matrix: charge conservation and gauge invariance.
  Equations (\ref{final-Landauer}) and (\ref{finalfinal-Landauer}) call for a number of comments. In particular they consist
of the difference of two large terms so that some care is needed when performing practical calculations.

\begin{itemize}
\item First, Eq.(\ref{final-Landauer}) contains a diverging term on the right hand side which
    corresponds to the injected current from lead $\bar m$. Indeed, at
    equilibrium, although the net total currents coming from the different
    leads cancel, each lead injects a finite current, leading to a diverging
    number of injected particles. Therefore, to use Eq.
    (\ref{final-Landauer}) in practice, it is important to first sum the
    contribution from all leads before performing the integrals.  Also, one
    must add those contributions at fixed energy $\epsilon$ (i.e. the energy
    inside the device region, not $E$ the original energy of the injected
    particle) for those diverging terms to properly compensate.

\item Second, although Eq.(\ref{final-Landauer}) apparently contains
    contributions from the whole spectrum, one can show that the only
    non-compensating terms arise from a small region around the Fermi energy.
    Indeed, let us consider an energy $\epsilon$ well below $E_F$. The kernel
    $K_{\bar m}(E - \epsilon)$ vanishes when $E-\epsilon$  becomes larger than
    $\max (w_P,\hbar/\tau_P)$ so that the values of $E$ effectively
    contributing to the integral are also well below $E_F$, hence
    $f(E)=f(\epsilon)=1$.  The integral over the energy $E$ can now be
    performed and, using Parseval theorem, we get $\int dE |K_{\bar
    m}(E-\epsilon)|^2=\int_0^{t_M} dt$. We can now sum over the channel index
    $\alpha$ and lead index $\bar m$ using the unitarity condition
    $\sum_{\alpha\bar m} |S^0_{\bar p\beta,\bar m\alpha}(\epsilon)|^2=1$ and
    finally find that the first term of Eq.(\ref{final-Landauer}) compensates
    the second one for each energy $\epsilon$.  Again, to obtain this
    compensation it is important to {\it first} perform the integral over the
    injected energy $E$ at fixed energy $\epsilon$. The same point applies to Eq.(\ref{finalfinal-Landauer}): $E$ and $\epsilon$
must be close for $K_{\bar m}(E-\epsilon)$ to be non zero hence the term $f(E) - f(\epsilon)$ vanishes away from the Fermi level.
More discussion on this aspect can be found around Fig.\ref{blockade}.

\item Current conservation is one of the main features of the Landauer approach
    which is usually lost in non-interacting AC transport, as the electronic
    density varies in time inside the system \cite{Buttiker_capacitors}.
    However the total number of particles is a conserved quantity and
    \be
    \sum_{\bar p} N_{\bar p\bar m} =0
    \ee
    as can be seen directly on Eqs.(\ref{final-Landauer}), (\ref{finalfinal-Landauer}) or from the general
    argument at the beginning of this section.

\item Another equally important feature of the scattering
    approach is the gauge invariance -- raising the potential of all the leads
    simultaneously does not create any current -- which is also usually lost in
    the non-interacting AC theory. However Eq.(\ref{final-Landauer}) does
    satisfy gauge invariance. Indeed, suppose we send an identical voltage
    pulse on all the leads simultaneously. Then the term $|K_{\bar m}(E-\epsilon
    )|^2$ does not depend on $\bar m$ and one can immediately perform the sum
    over $\alpha$ and $\bar m$ and use $\sum_{\alpha\bar m} |S^0_{\bar
    p\beta,\bar m\alpha}(\epsilon)|^2=1$.  In a second step we perform the
    integral over $E$ of the first term of Eq.(\ref{final-Landauer}) using
    Parseval theorem and find again that it exactly matches and compensates the
    second term and $n_{\bar p}=0$.  Note that while the above statement is non
    trivial, there is a weaker form of gauge invariance which is always
    verified: the physics is entirely unaffected by a global change of the
    potentials of all the leads {\it and} the internal potential of the device
    (as such a global variation of the potential can be absorbed by a simple
    global phase in the wave function). The combination of both forms of gauge
    invariance (weak and strong) implies that a uniform voltage pulse applied to the
    central region $\bar 0$ (through a capacitive coupling to a gate) does not create
    any charge pumping, even in the non adiabatic limit.

\item One of the appealing aspects of Eq.(\ref{finalfinal-Landauer}) is that it
has a direct connection to
the DC conductance matrix in the adiabatic limit. Indeed the DC Landauer formula reads,
\be
I_{\bar p} = \frac{e^2}{h} \sum_{\bar m}T_{\bar p\bar m} V_{\bar m}
\ee
where $T_{\bar p\bar m}$ is the total transmission probability from lead $\bar m$ to $\bar p$. When the voltage pulse is
extremely slow (adiabatic limit) with respect to all the characteristic times of the device, one expects the
current to follow the voltage adiabatically, $I_{\bar p} (t) = (e^2/h) \sum_{\bar m}T_{\bar p\bar m} V_{\bar m}(t)$ and
\be
n_{\bar p} =  \sum_{\bar m} T_{\bar p\bar m} \bar n_{\bar m}
\ee
where $\bar n_{\bar m} =\int dt eV_{\bar m}(t)/h$ is the total number of  particles {\it injected} by the voltage pulse in lead $\bar m$.
Hence, in the adiabatic limit, $N_{\bar p\bar m} = T_{\bar p\bar m} \bar n_{\bar m}$ has a nice interpretation in term of the total transmission probability from $\bar m$ to $\bar p$
and the interesting question is how the physics deviates from this limit when the pulses get faster than the internal characteristic time scales of the device.
\end{itemize}

\subsection{A comment on the electrostatics}
\label{electrostatic}

\begin{figure}[h]
    \centering
    \includegraphics[width=12cm]{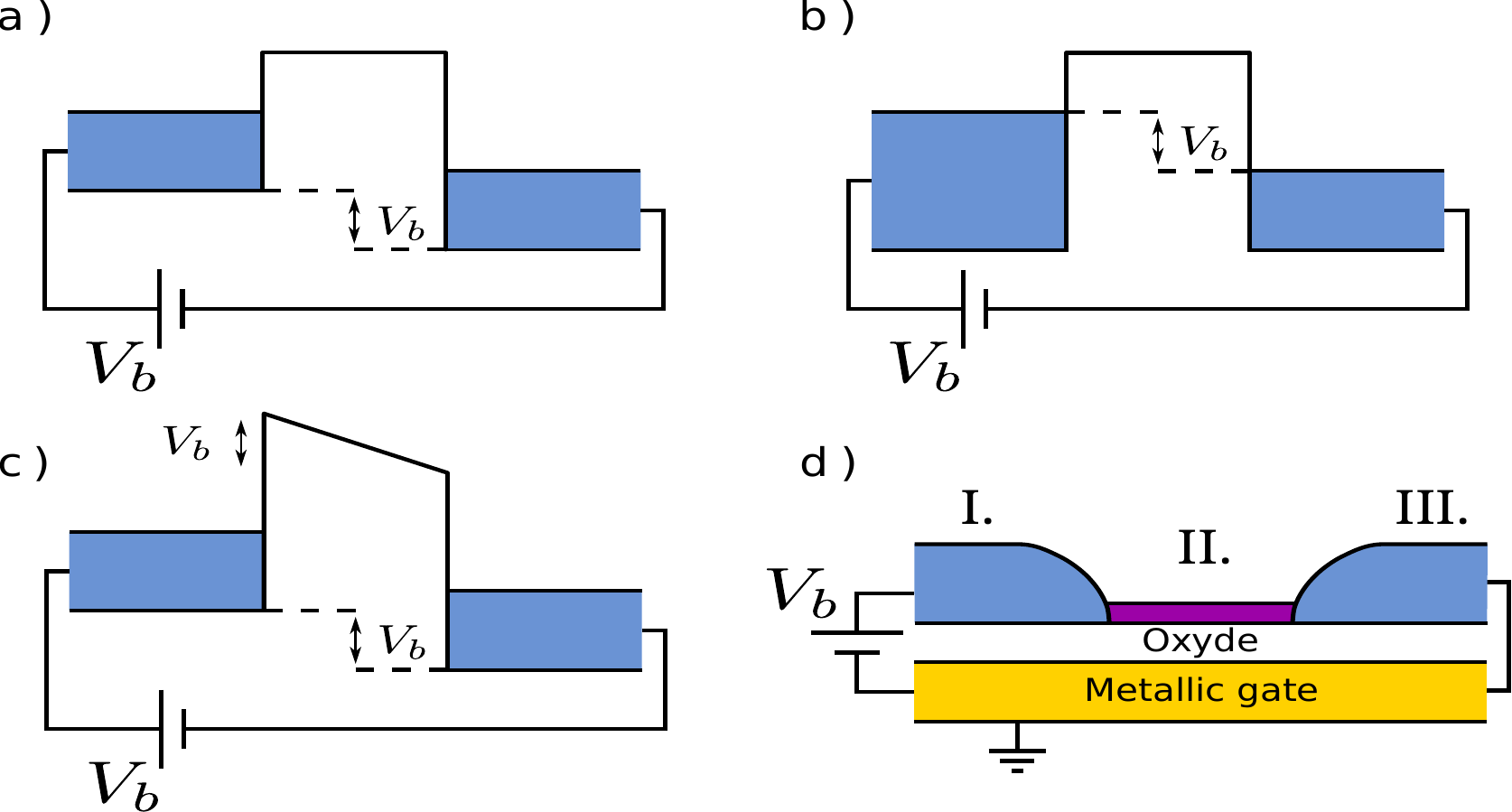}
    \caption{\label{fig:elec-chem}
  Sketch of different repartitions between chemical and electrical potential
  upon applying a difference of electrochemical potential $V_b$ between source
  and drain. a) Abrupt drop of purely electrical nature. b) The drop is purely
  of chemical nature. c) The purely electric drop takes place linearly over the
  sample (tunnel junction situation). d) Device corresponding to case a): the
  two electrodes I and III correspond to regions with high density of states
  while the central region II has a low density of states. A metallic gate, at
  a distance $d$ below the sample, screens the charges present in the sample.}
  \end{figure}

We end this section with a discussion of our choice of boundary conditions
in the electrodes and our model for an abrupt voltage drop. Following the usual
practice \cite{Wingreen-Meir_1994}, we have assumed (i) that the voltage drops
abruptly at the electrode -- system interface and (ii) that the electrodes remain
at thermal equilibrium (in the basis where the gauge
transformation has been performed so that the electrode Hamiltonian is time-independent).
Conditions (i) and (ii) correspond to case a) in Fig.\ref{fig:elec-chem}; an abrupt drop of the
electrical potential at the lead -- system interface. In an actual experiment,
however, a voltage source does not impose a difference of electric potential
but rather a difference of {\it electrochemical } potential. How the latter
is split between electric and chemical potential is a matter of the balance
between the electrostatic and chemical (i.e. kinetic) energy of the system
and is therefore extrinsic to the model discussed so far.
Fig.\ref{fig:elec-chem} b) and c) illustrate two possible ways of splitting
these contributions. In the former case the potential drop is of a purely
chemical nature, whereas in the latter the potential drop is purely electrical
and is not abrupt.

\begin{figure}[h]
    \centering
    \includegraphics[width=8cm]{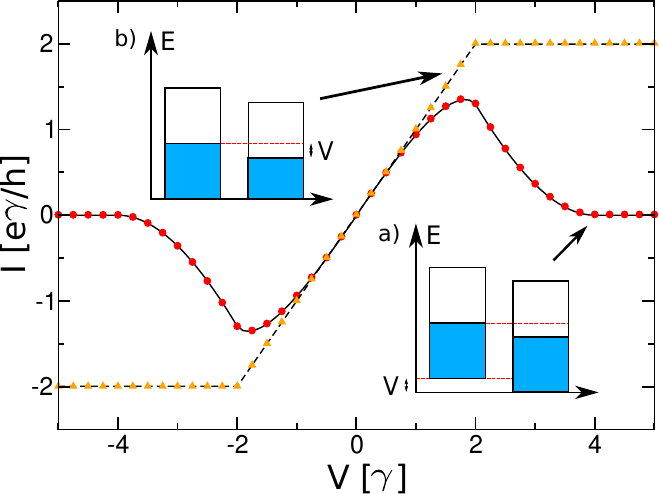}
    \caption{\label{fig: I-V}
    $I(V)$ characteristics of the 1D chain. Symbols: results obtained with GF-A
    after a fast voltage rise $w(t)=V \theta (t)$ and letting the system
    equilibrate for $t=100\gamma^{-1}$.  Lines: corresponding pure DC
    calculation.  We compare the case (a) where the drop of potential is purely
    electric (triangles, choice made everywhere else in this article) and (b)
    where it is purely chemical (circles). Inset, schematic of the
    corresponding adjustments of the band positions and Fermi levels. The
    shaded blue region corresponds to the filled states of the band.}
\end{figure}

Note that our model, Fig.\ref{fig:elec-chem} a), implies a small potential
mismatch at the electrode -- system interface which in turn induces a finite
reflection amplitude, which is not the case in Fig.\ref{fig:elec-chem} b). For
DC current with small bias both models coincide, but differences occur at
large biases. Fig.~\ref{fig: I-V} shows the stationary value of the current
after a fast increase of the voltage. We use a pulse of form $w(t)=V \theta
(t)$, wait for a long ($t=100$) time after the voltage has been established and
compute the corresponding stationary current (using any of the above equivalent
methods, in this case GF-A).  One can check from Fig.~\ref{fig: IvsT} that
$t=100$ is sufficient to achieve convergence toward the stationary value.  This
can be considered as a very elaborate (and ineffective) way to obtain the
$I(V)$ characteristics of the device. We also calculated directly the DC $I(V)$
characteristics using the stationary equations \cite{Knit} and checked that we
obtained matching results.  Fig.~\ref{fig: I-V} shows two curves. The first
curve, Fig.~\ref{fig: I-V} a): red circles, corresponds to the natural
condition in our formalism (case Fig.\ref{fig:elec-chem} a)). The voltage drop
is a drop of {\it electric} potential, hence there is a corresponding shift of
the band of the left lead with respect to the right one. The second
curve, Fig.~\ref{fig: I-V}b): triangles,  corresponds to a change of {\it
chemical} potential (Fig.\ref{fig:elec-chem} b), the bottom of the right and
left bands remain aligned). When $V$ becomes large compared to the Fermi energy
the two prescriptions differ; a drop of electric voltage implies backscattering
while in (b) the transmission probability is always unity. Also, a current in
(a) implies that the bands in the two leads overlap which does
not happen when $V$ is larger than the bandwidth of the system. At large $V$
the current therefore saturates to $2e\gamma /h$ in (b) while it vanishes in
(a).

While a full discussion of the electrostatics lies out of the scope of the
present paper, let us briefly discuss a simple situation which clarifies
which boundary condition is the most appropriate for a given situation.
A sketch of the system is given in Fig.\ref{fig:elec-chem} d). It consists of
two metallic electrodes I and III with a high electronic density of states (per
unit area) $\rho_I$ and $\rho_{III}$ connected to a central device region with
lower density of states $\rho_{II}$ (typically a GaAs/AlGaAs heterostructure or
a graphene sheet). Underneath the system, at a distance $d$, is a metallic gate
which is grounded. In a typical measurement setup the electrode III is
grounded while a voltage source $V_b$ is placed between the electrode I and the
metallic gate. Upon imposing the electrochemical potential $eV_b$ in region I,
a variation $U_I$ ($\mu_I$) of electric (chemical) potential takes place with
$eV_b=eU_I+\mu_I$. The variation of chemical potential corresponds to a
variation of electronic density $n_I= \rho_I \mu_I$ (quantum capacitance). On the other hand, the
presence of the underlying gate corresponds to an electric capacitance (per unit area)
$C=\epsilon /d$ and the electrostatic condition reads $n_I=C U_I/e$. Putting
everything together we arrive at
\be U=\frac{V_b}{1+C/(e^2\rho_I)}
\ee
Turning to concrete examples, we find that for typical transition metals (very
high density of states) $U\approx V_b$ as raising the chemical potential would
imply a huge increase in the density which in turn would induce a
correspondingly large increase in the electrostatic energy. Metallic electrodes
are thus typically associated with the cases of Fig.\ref{fig:elec-chem}  a) or
c). The behavior in region II depends acutely on $\rho_{II}$. If the density
of states in region II is high enough (say a 2D gas with a screening gate at
$d=100$nm) then the electric potential decays linearly from $eV_b$ (region I)
to 0 (region III), as shown in Fig.\ref{fig:elec-chem} c)]. If the density of
states in region II is small, however, (e.g. one dimensional systems such as
edge states in the Quantum Hall regime or a carbon nanotube) the ratio
$C/(e^2\rho_I)$ becomes large and $U_{II}$ vanishes  [case
Fig.\ref{fig:elec-chem} a)]. We conclude that while the situation depicted in
Fig.\ref{fig:elec-chem} b) is fairly rare (although possible using for instance
a graphene electrode coupled through a BN layer to an extremely close
underlying graphene gate), the situation of Fig.\ref{fig:elec-chem} a), which
is the focus of this paper, is typical of a mesoscopic system. In this case,
the drop of the electric potential will typically take place over a distance
$d$.  While the simulation of case (a) and (c) is straightforward within our
formalism, case (b) (fortunately often not realistic) would require additional
extrinsic inputs. While the electric potential adjusts itself instantaneously
(i.e. at the speed of light) inside the sample, the chemical potential
propagates at the group velocity of the electrons, and a proper model of the
inelastic relaxation inside the electrodes would be necessary.

%%%%%%%%%%%%%%%%%%%%%%%%%%%%%%%%%%%%%%%%%%%%%%%
\section{A  pedestrian example: propagation and spreading of a voltage pulse inside a one dimensional wire}
%%%%%%%%%%%%%%%%%%%%%%%%%%%%%%%%%%%%%%%%%%%%%%%
\label{sec:1d}

While most elementary courses on quantum mechanics concentrate on the
stationary limit, one aspect of the time-dependent theory stands out: the
spreading of a (mostly Gaussian) wave packet.  An initial wave packet with a
certain spatial width and average momentum experiences a ballistic motion of
its center of mass while its width spreads diffusively. The spreading of the
wave packet provides a simple illustration of a central concept of quantum
mechanics, the Heisenberg principle between time and energy. In this section,
we study a case which can be considered as the condensed matter analogue of the
spreading of the wave packet:  the propagation and spreading of an initial
condition which is given in term of a voltage pulse.  The voltage pulse shares
some similarities with the usual ``localized wave packet'', yet there are also
important differences. In particular, in an electronic system, there are
stationary delocalized waves which exist before the pulse. Hence, a voltage
pulse does not create a localized wave packet but a local deformation of
(mostly the phase of) an existing one.

The main result of this section is that the spreading of a voltage pulse is
accompanied by density (and current) oscillations that follow the propagation of
the pulse. The sort of wake which is formed by these oscillations is
unfortunately mainly of academic interest as its experimental observation
appear to be extremely difficult.

We start this section with a pedestrian construction of the scattering matrix
of a one dimensional chain. We then leave the discrete model for the
continuous limit which is more tractable analytically. We end this section
with an explicit calculation of the spreading of the wave packet and the above
mentioned wake that follows the ballistic propagation of the pulse.

\subsection{Scattering matrix: analytics}
Our starting point is the Schrödinger equation for the one dimensional chain
(i.e. the first quantization version of Hamiltonian (\ref{eq: 1d_Hamiltonian})
with a static potential $\epsilon_i=2\gamma$ over the entire infinite chain),
\begin{align}
i\partial_t \psi_x &= -\gamma \psi_{x-1} - \gamma \psi_{x+1} + 2 \gamma \psi_{x}, \ \forall x \ne 0,1 \\
\label{1d-match1} i\partial_t \psi_0 &= -\gamma \psi_{-1} - e^{i\phi (t)}\gamma^t \psi_{1} + 2 \gamma \psi_{0}, \\
\label{1d-match2} i\partial_t \psi_1 &= -\gamma \psi_{2} - e^{-i\phi (t)} \gamma^t \psi_{0} + 2 \gamma \psi_{1},
\end{align}
where the hopping element $\gamma_t$ between sites $0$ and $1$ can be different
from the hopping $\gamma$ of the rest of the system. As the time-dependent part
of the Hamiltonian concentrates on a single hopping term between sites $0$ and
$1$, we can build the states on either side with a linear combination of the
plane waves of the system,
\begin{align}
\psi_x &=  \frac{e^{-iEt+ik(E)x}}{\sqrt{|v (E)|}}  + \int \frac{dE'}{2\pi} \frac{e^{-iE't-ik(E')x}}{\sqrt{|v (E')|}} r(E',E)
   , \ \forall x \le 0 \\
   \label{1d-tr}
\psi_x &= \int \frac{dE'}{2\pi} \frac{e^{-iE't+ik(E')x}}{\sqrt{|v (E')|}} d(E',E)
   , \ \forall x \ge 1
\end{align}
with $E(k)=2\gamma (1 -\cos k)$ and $v=\partial E/\partial k$. The
``wave-matching'' conditions Eq.(\ref{1d-match1}) and (\ref{1d-match2})
translate, for our ansatz, into
\be
\label{rd-1d}
\frac{e^{-ik(E')}}{\sqrt{|v (E')|}} r(E',E) + 2\pi \frac{e^{ik(E)}}{\sqrt{|v (E)|}} \delta (E'-E)  = (\gamma^t/\gamma) \int \frac{d\epsilon}{2\pi} K(E'-\epsilon) \frac{e^{ik(\epsilon)}}{\sqrt{|v (\epsilon)|}} d(\epsilon,E)
\ee
\be
\label{dr-1d}
\frac{1}{\sqrt{|v (E')|}} d(E',E)  = (\gamma^t/\gamma)  \left[\frac{1}{\sqrt{|v (E)|}} K^* (E-E') + \int \frac{d\epsilon}{2\pi} K^*(\epsilon-E') \frac{1}{\sqrt{|v (\epsilon)|}} r(\epsilon,E) \right]
\ee
Equations (\ref{rd-1d}) and (\ref{dr-1d}) can be solved systematically, order
by order, in power of $\gamma^t/\gamma$. The first non vanishing term for the
transmission reads,
\be
d(E',E)=(\gamma^t/\gamma) \sqrt{\frac{v(E')}{v(E)}} [ 1 - e^{2ik(E)}] K^*(E-E') + O(\gamma^t/\gamma)^2
\ee
Of course, Equations (\ref{rd-1d}) and (\ref{dr-1d}) can also be solved in the
wide band limit, as in the previous section. The wide band limit leads to,
\be
r(t,E)  e^{-ik(E)}     + e^{ik(E)} e^{-iEt} =  (\gamma^t/\gamma)   e^{i\phi (t)} e^{ik(E)}  d(t,E)
\ee
\be
d(t,E)=(\gamma^t/\gamma) e^{-i\phi (t)} \left[  e^{-iEt} +   r(t,E)   \right]
\ee
from which we get,
\be
\label{wbl1d}
d(t,E)= (\gamma^t/\gamma)   e^{-i\phi (t)}e^{-iEt} \frac{e^{ik(E)}-e^{-ik(E)}}{(\gamma^t/\gamma) e^{ik(E)}- e^{-ik(E)}}
\ee
which is a simple generalization (for $\gamma^t\ne\gamma$) of  the
result  derived in the previous section. For $\gamma^t=\gamma$ one obtains
$d(E',E)=K^*(E-E')$.

Let us now briefly look at the shape of the transmitted wave that can be
reconstructed from the knowledge of $d(E',E)$ and Eq.(\ref{1d-tr}). In the wide
band limit $E(k')=E(k)$, it reads,
\be
\psi(x,t) = \frac{1}{\sqrt{v}} e^{-iEt} e^{ik x} e^{i\phi (t)}.
\ee
We find that in this solution the pulse does not propagate, which is to be
expected as the wide band limit neglects the system velocity. Using a linear
dispersion $E(k')=E(k)+v(k'-k)$ improves the situation as the corresponding wave
function,
\be
\psi(x,t) = \frac{1}{\sqrt{v}}  e^{ik x-iEx/v} d(t-x/v)
\ee
shows the ballistic propagation of the pulse. In the limit where the velocity
of the wave is slow (with respect to the typical scales of the voltage pulse)
one can use $d(t)=e^{-i\phi (t)-iEt}$,
\be
\psi(x,t) = \frac{1}{\sqrt{v}}  e^{ik x-iEt} e^{-i\phi (t-x/v)}
\ee
At this level of approximation the voltage pulse can be considered as a
``phase domain wall'' which propagates ballistically inside the wire. The
spreading of the voltage pulse is associated with the mass of the particle, i.e.
to the curvature of the dispersion relation, and therefore is beyond the linear
dispersion considered here.
Also, the expression $d(t-x/v)=e^{-i\phi (t-x/v)}$ is slightly ill-defined as
it does not fulfill particle conservation (it corresponds to a uniform density
yet a non uniform current). This reflects the fact  the transmission matrix
itself was calculated in the wide band limit, i.e. without taking the
electronic propagation into account.

We continue by taking the continuum limit of the problem, i.e. we introduce a
small discretization step $a$, set $\gamma=\hbar^2/(2m a^2)$ and $k\rightarrow
k a$. The limit $a\rightarrow 0$ provides the usual quadratic dispersion of the
Schrödinger equation, $E(k)=\hbar^2 k^2/(2m)$. In this limit, we can solve
equations  (\ref{rd-1d}) and (\ref{dr-1d}) for a linear spectrum, beyond the
wide band limit. We obtain,
\be
\label{lin-1d-1}
\dot r- iE r +2iEe^{-iEt} = (\gamma^t/\gamma)
e^{i\phi (t)} [iE d- \dot d]
\ee
\be
\label{lin1d-2}
e^{i\phi (t)} d=(\gamma^t/\gamma)  \left[  e^{-iEt} +   r   \right]
\ee
where we have used the notation $\dot r=\partial_t r(t,E)$. This set of linear
equations can be formally integrated and one obtains the correction to the wide
band limit. For $\gamma_t=\gamma$, we get,
\be
\label{toto}
\dot r -i (E +w(t)/2) r = i e^{-iEt} w(t)/2
\ee
Assuming  that the voltage is small compared to $E$, we can neglect $w(t)$ in
the left hand side of Eq.(\ref{toto}) and obtain
\be
r(E',E)=-\frac{w(E'-E)}{E'+E}+ O[w(E)/E]^2
\ee
where $w(E)$ is the Fourier transform of the voltage pulse $w(t)$, or
equivalently,
\be
r(t,E)=\frac{i}{2}\int_{-\infty}^t du e^{-i2E u+iEt} w(u)
\ee
and
\be
\label{azerty}
d(t,E)=e^{-i\phi (t)-iEt} + \frac{i}{2}e^{-i\phi (t)} \int_{-\infty}^t du e^{-i2E u+iEt} w(u)
\ee
It is interesting to look at Eq.(\ref{azerty}) for a time larger than the total
duration of the pulse, so that the integral of the right hand side is simply
$w(-2E)$. We get,
\be
d(t,E)=e^{-i\phi (t)-iEt} [ 1 + \frac{i}{2} w(-2E) e^{i2Et} ] + O[w(E)/E]^2
\label{eq:2E}
\ee
We find that the first correction to the wide band limit corresponds to a
beating of frequency $2E$. The corresponding term is, however, very small as
$w(\epsilon)$ vanishes when $\epsilon$ is larger than $\max(V_P,\hbar/\tau_P)$
which, under the assumptions of the wide band limit, is much smaller than
$E_F$.

\subsection{Scattering matrix: numerics}
\begin{figure}[h]
    \centering
    \includegraphics[width=10cm]{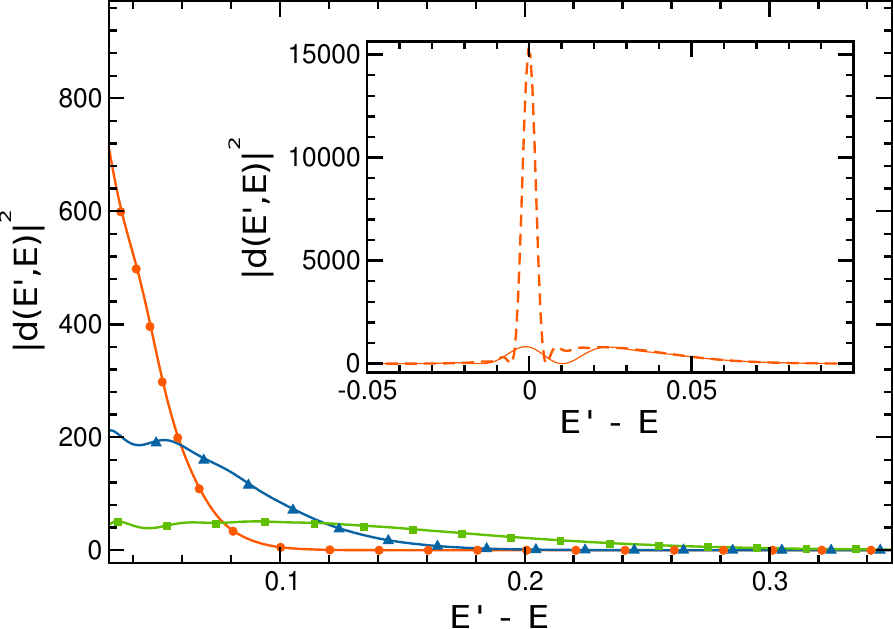}
    \caption{\label{dransmission}
    Transmission probability of an incoming particle at energy $E=-1\gamma$ for
    a Gaussian voltage pulse $w(t) = V_P  e^{-4\log(2)t^2/\tau_P^2}$ with
    amplitude $V_P$, width $\tau_P$ and fixed product $V_P\tau_P=5.9$. Full
    lines corresponds to Eq.(\ref{d-wbl2}) while symbols are numerical results.
    Orange circles : $V_P=0.059\gamma$, $\tau_P=100\gamma^{-1}$, blue
    triangles: $V_P=0.118\gamma$, $\tau_P=50\gamma^{-1}$, green squares:
    $V_P=0.236\gamma$, $\tau_P=25\gamma^{-1}$. Inset: convergence of the
    discrete Fourier transform for two different values of $t_M$ (same
    parameters as the orange circles).}
\end{figure}

As a test of the consistency of our different approaches,
Fig.\ref{dransmission} shows the transmission probability $d(E',E)$ of the one
dimensional chain as obtained from a numerical calculation [WF-D method
followed by the generalized Fisher-Lee formula Eq.(\ref{fl1})] and the
analytical result $d(E',E)=K^*(E-E')$ in the wide band limit
[Eq.(\ref{d-wbl2}), the Fourier transform was performed numerically]. First, we
find that the wide band limit gives excellent results; the analytics match the
numerical results even for pulses that are quite large in energy ($V_P$ up to
20\% of the injected energy $E$). Second, we find that, as expected, the
characteristic energy for the decay of $d(E',E)$ is indeed given by
$\max(V_P,\hbar/\tau_P)$. Last, we find (inset) a large peak of width
$\hbar/t_M$ and height $t_M/\hbar$ around $E'=E$.  This peak, which converges
to $\delta (E'-E)$ when $t_M\rightarrow\infty$ corresponds to the fact that for
most of the time  there is no time-varying voltage in the system which is therefore elastic. This can also be seen from
the analytical expression of $K(E)$, which can be obtained in the case of a
Lorentzian pulse \cite{Lorentzian_pulses}. Indeed for $w(t) =
\frac{2\tau_P}{\tau_P^2 + t^2}$, one obtains,
\be
\label{lorentzian}
e^{i\phi(t)} = \frac{t - i\tau_P}{t + i\tau_P} \ \ {\rm and } \ \ \
K(E)=2\pi\delta(E) - 4\pi \tau_P e^{E \tau_P}\theta(-E)
\ee

\subsection{Spreading of a voltage pulse inside a one dimensional wire: analytics}
Going beyond the linear dispersion to study the spreading of the voltage pulse
is not straightforward using the above wave matching method; we now take
a different approach. We consider a pulse whose duration $\tau_P$ is short with
respect to the total propagation time that will be considered, yet long with
respect to $\hbar/E$. At a small time $t_0$ just after the pulse we can safely
ignore the spreading of the pulse and the wave function is given by
\be
\label{init}
\psi (x,t_0)= \frac{1}{\sqrt{v}} e^{-i\phi (-x/v)} e^{-iEt_0} e^{ik x}
\ee
Eq.(\ref{init}) will be used as our initial condition. As noticed before, the
voltage pulse takes the form of a phase domain wall that modifies the existing
plane wave, as the function $\phi (-x/v)$ is constant except within a small
window of size $v\tau_P$. We now introduce explicitly the modulation of the
plane wave $Y(x,t)$,
\be
\label{hello}
    \psi(x,t) = \frac{1}{\sqrt{v}}Y(x,t) e^{-iEt +ik x}
\ee
$Y(x,t)$ verifies $Y(x,t_0)=e^{-i\phi (-x/v)}$. To obtain the evolution of
$Y(x,t)$ for times $t>t_0$, we inject the definition  of the wave function
Eq.(\ref{hello}) into the (free) Schrödinger equation and obtain,
\be
    i\partial_t Y(X,t) = -\frac{1}{2m^*} \Delta_{X} Y(X,t)
\ee
where the Laplacian $\Delta_{X}=\partial_{XX}$ acts on the coordinate $X=x-vt$
which follows the ballistic motion of the pulse. Solving this free Schrödinger
equation is now straightforward and one proceeds as for a ``regular'' wave
packet. In momentum space we have
\be
\label{wavepacket}
Y(X,t)=\int \frac{dQ}{2\pi} e^{-iQ X} e^{-iQ^2 t/(2m^*) } Y(Q,t=0)
\ee
with
\be
Y(Q,t=0)=v K^*(Qv)
\ee
In a few cases one knows $K(E)$ explicitly and one an explicit formula for the
wave function can be obtained. In the case of a Lorentzian pulse $K(E)$ is
given by Eq.(\ref{lorentzian}) and the integration in Eq.(\ref{wavepacket})
provides an explicit expression,
\be
Y(X,t) = 1 - v\tau_P \sqrt{\frac{2m^*\pi}{it}} \exp\left( \frac{m^* (iX -
v\tau_P)^2}{2it}\right)  \left[ 1 + {\rm Erf}
\left(\frac{iX-v\tau_P}{2\sqrt{it/(2m^*)}}\right) \right]
\label{eq:Ylor}
\ee
with the usual definition of the error function ${\rm Erf}(x) =
(2/\sqrt{\pi})\int_0^x e^{-x^2}dx$.

%%%%%%%%%%%%%%%%%%%%%%%%%%%
\subsection{Spreading of a voltage pulse inside a one dimensional wire: numerics}
%%%%%%%%%%%%%%%%%%%%%%%%%%%%%%%%%%%%%%%

\begin{figure}[h]
    \centering
    \includegraphics[width=12cm]{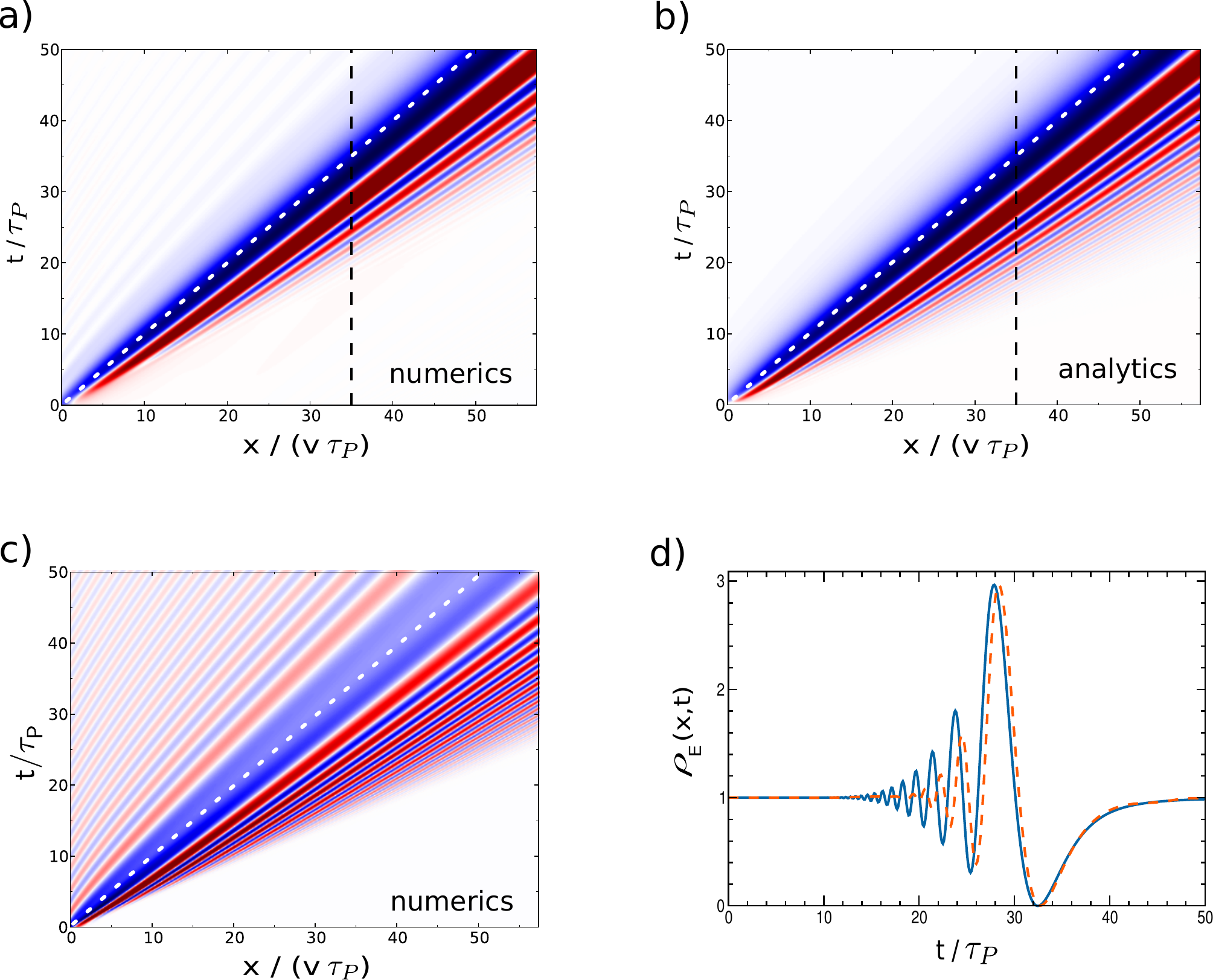}
    \caption{\label{wakes}
    Color plot of the local charge density $\rho_{E}(x,t)/\rho_E(x,t=0) $ as a
    function of space (in unit of $v\tau_P$) and time (in unit of $\tau_P$) at
    energy $E=-1.8\gamma$ and $\tau_P=10\gamma^{-1}$.  Levels of red (blue)
    correspond to local densities higher (lower) than one.  The white dashed
    lines indicates the ballistic propagation $x=v t$.  Panels a) and b)
    correspond to a  Lorentzian pulse $w(t)=2\tau_P/(\tau_P^2+t^2)$ calculated
    analytically [right,  Eq.(\ref{eq:Ylor})] and numerically [left].  Panel c)
    shows the numerical result for a Gaussian pulse
    $w(t)=V_Pe^{-4log(2)t^2/\tau_P^2}$ with $V_P=0.59\gamma$.  Panel d) shows a
    cut at $x=35 v \tau_P$ of the results of panel a) (orange dashed line) and
    panel b) (full blue line). }
\end{figure}

The previous form of $Y(x,t)$ is the voltage pulse analogue of the spreading of
a wave packet.  It can be recast as a function of the dimensionless position
$\bar X = X/(v \tau_P)$ and time $\bar t = t / [m^*(v\tau_P)^2]$. The typical
spreading takes place ``diffusively'', i.e. $\Delta \bar X \propto \sqrt{\bar
t}$, as for a regular wave packet. However, the peculiarity of the voltage
pulse (i.e. it is merely a localized deformation of the {\it phase} of an
existing stationary wave rather than the modulation of its amplitude) manifests
itself in the presence of oscillations in the charge density. Fig.\ref{wakes}
shows the calculation of the local charge density
\be
    \rho_E(x,t) = |\Psi_E(x,t)|^2
\ee
obtained from numerical calculations (left panels) and from Eq.(\ref{eq:Ylor})
(upper right panel).  The two upper color plots provide the same quantity as
calculated numerically (left) and analytically (right). We find that the
analytical description is fairly accurate despite various possible sources of
discrepancy. The numerics are performed  with our tight-binding model which
slightly deviates from the continuum and the analytics neglect the quadratic
dispersion at small times. A more detailed comparison is shown in
Fig.\ref{wakes} d) where we have plotted a cut at fixed $x$ of the local charge
density. The lower left plot corresponds to a different (Gaussian) form of the
pulse from which a close analytical expression could not be obtained.

The most striking feature of the ``spreading of the voltage pulse'' is the
appearance of density oscillations which are reminiscent of a wake. Although we
could only analyze these oscillations analytically for the Lorentzian pulse, we
actually found them for other shapes, the specificity of the Lorentzian pulse
being that these oscillations always travel faster than the Fermi group
velocity (the electrons' energy can only increase with a Lorentzian pulse, see
Eq.(\ref{lorentzian})).  Indeed for a Gaussian pulse (Fig.\ref{wakes}c)), the
oscillations also take place \emph{after} the passage of the pulse.

\begin{figure}[h]
    \centering
    \includegraphics[width=16cm]{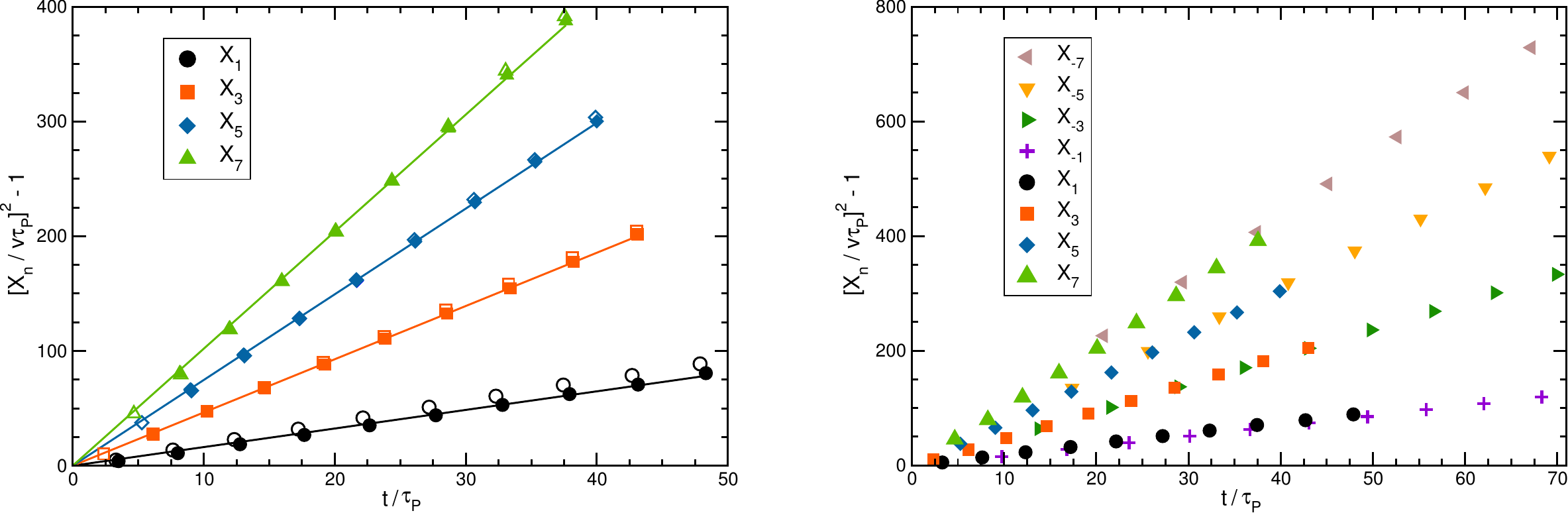}
    \caption{\label{wakes2}
    Maxima of the oscillations appearing in Fig. \ref{wakes}a) and Fig.
    \ref{wakes}c) as a function of time. Left graph: full (empty) symbols
    correspond to the Lorentzian (Gaussian) pulse. Both cases are hardly
    distinguishable.  Lines are linear fits of the numerical data obtained for
    the Lorentzian case. Right graph: all symbols correspond to the Gaussian
    pulse, negative (positive) values of $n$ refer to maxima  appearing before
    (after) the pulse.  }
\end{figure}

At large time,  Eq.(\ref{eq:Ylor}) indicates that the amplitude of
$\rho_E(x,t)$ scales as $1/\sqrt{\bar t}$ while the ``period'' of the
oscillations increases as $\sqrt{\bar t}$.  More
precisely the $n^{th}$ extremum $X_n$ of these oscillations obeys the relation,
\be
    X_n^2  = \frac{2\pi}{m^*} \Big(n +
    \frac{1}{4}\Big) \ t +(v\tau_P)^2
\label{eq:maxima}
\ee
In other words the positions $X_n$ of the extrema increase diffusively with the
quantum diffusion constant $D=h/m^*$.  Fig. \ref{wakes2} shows the values of
$X_n$ as obtained numerically for a Gaussian or a Lorentzian pulse. We find (i)
that the positions of the peaks in front of the pulse is not affected by the
shape of the pulse (Lorentzian or Gaussian). Also (ii) the peaks behind the
pulse (negative $n$, not present in the Lorentzian case)  are positioned
symmetrically with respect to the peaks with positive $n$.

In order to be able
to observe these oscillations, one would need $Dt$ to be larger than the
original size of the pulse $v \tau_P$ which unfortunately happens to be very
difficult. Indeed, one finds $D \approx 10^{-4}-10^{-2} m^2. s^{-1}$ which translates into
$X_1\approx 1 nm$ for a large propagation time $t=10 ns$ that would require,
assuming $v\approx 10^4 m.s^{-1}$, a $100\mu m$ long coherent sample and
$\tau_P <100 fs$. This is clearly beyond available technology.
In addition, the numerics and expressions obtained so far in this section refer to
the contribution  to the electronic density $\rho (x,t)$ at a given energy $E$. This contribution
corresponds to the derivative of the corresponding density with respect to
Fermi energy $d\rho (x,t)/dE_F=\rho_{E_F}(x,t)$. It can therefore be, in principle,  directly measured by
modulating the system with a uniform electrostatic gate, but its main interest lies in the physical insights it conveys.
\begin{figure}[h]
    \centering
    \includegraphics[width=14cm]{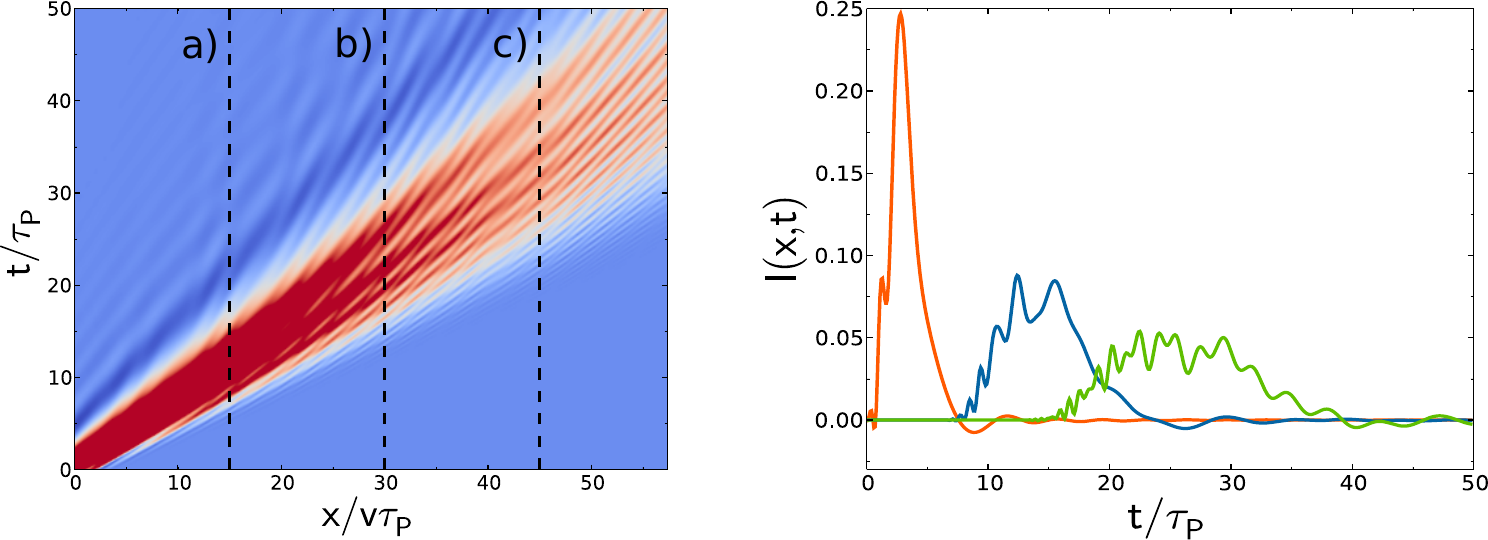}
    \caption{\label{wakes_current}
    Current density as a function of space (in unit of $v\tau_P$) and time (in
    unit of $\tau_P$) for the Gaussian pulse of Fig. \ref{wakes}c). Fermi level
    is set at $E_F=-1.8\gamma$. Left panel:
    the color map goes from zero values (blue) to $0.6$ (red). Right panel:
    cut of the left panel  at three positions in space a), b) and c) corresponding to the three dashed lines  shown on the left panel.
    Orange: $x=15v\tau_P$, blue: $x=30v\tau_P$, green: $x=45v\tau_P$.}
\end{figure}
Fig.\ref{wakes_current} shows full current (integrated over energy) as a function of
space and time corresponding to the Gaussian pulse of Fig. \ref{wakes}c).
Beside the ballistic propagation of the pulse (at the Fermi velocity), one indeed observes
that the oscillating tail survives the integration over energies.
Note that these oscillations are reminiscent of other oscillations, associated with shock waves,
that  were predicted in\cite{Bettelheim2006,Bettelheim2012,Protopopov2013}.
In the latter case,  a quantum wire was perturbed with a local density perturbation
(as opposed to the voltage pulse studied here). However,  as those oscillations
also appear for a non-interacting gas and a finite curvature is needed to obtain them, they might be related to the present case.

\begin{figure}[h]
    \centering
    \includegraphics[width=12cm]{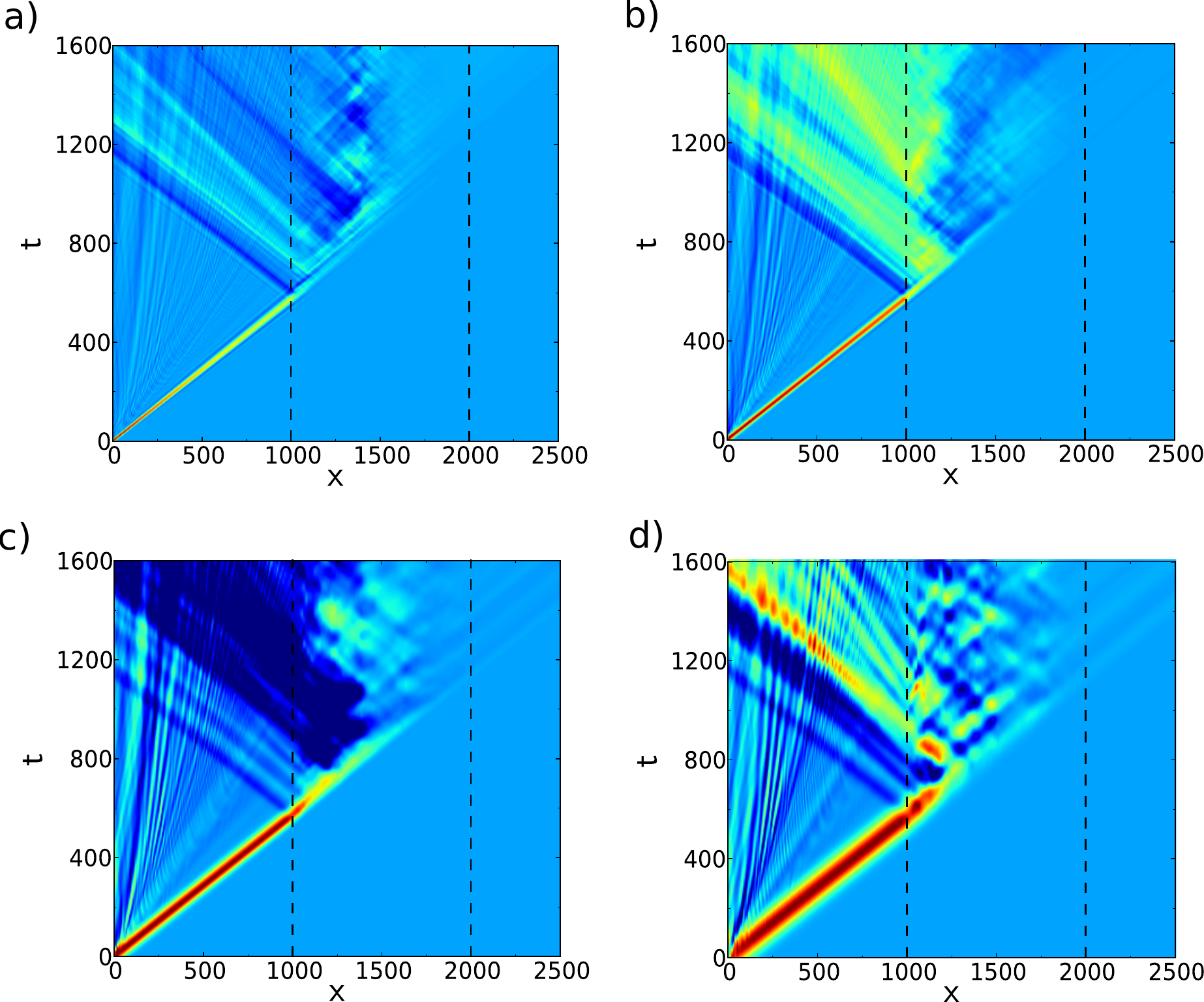}
    \caption{\label{fig:disorder} Propagation of a Gaussian voltage pulse in a
        1D quantum wire up to $t_{max}=1600\gamma^{-1}$. $E_F=-1\gamma$,  $w(t)
        = V_P e^{-4\log(2)t^2/\tau_P^2}$ with $V_P=0.05\gamma$ and $N=2500$. A
        disordered region ($W=0.5$) has been included between the two dashed
        lines (see text).  The different color plots correspond to
        $\tau_P=10\gamma^{-1}$ (a), $\tau_P=20\gamma^{-1}$ (b),
        $\tau_P=50\gamma^{-1}$ (c) and  $\tau_P=100\gamma^{-1}$ (d).  }
\end{figure}

The last figure of this section illustrates that our method is in no way limited
to the simplest case of a ballistic one dimensional wire: higher dimensions,
other lattices (e.g. graphene), or perturbations (polarized light, electrostatic
gates) can be studied as well.  Fig. \ref{fig:disorder} shows again a one
dimensional wire, but a disordered region (Anderson model) has been placed
between the sites $i=1000$ and $i=2000$ (dashed line).  In this region, the
on site energies $\epsilon_i$ are given by static random variables uniformly
chosen between $[-W/2, W/2]$. These preliminary results show the propagation of
a voltage pulse for different values of the width $\tau_P$. The Anderson
localization length $\xi$ for this system is roughly equal to $400$ sites ($\xi
\approx 96/W^2$ at the center of the band) and we indeed find that after a set
of multiple reflections, the transmitted current essentially vanish after the
middle of the disordered region. Further analysis would allow one to discuss the
interplay between the duration of the pulse and the phenomena of Anderson
localization. We differ such a step to a future publication.

%%%%%%%%%%%%%%%%%%%%%%%%%%%%%%%%%%%%%%%%%%%%%%%%%%%%%%%%%%%%%%%%%%%%%%%
\section{A two dimensional application to a flying Qubit}
%%%%%%%%%%%%%%%%%%%%%%%%%%%%%%%%%%%%%%%%%%%%%%%%%%%%%%%%%%%%%%%%%%%%%%%
\label{sec:2d}
We end this article by a simulation that goes beyond the one dimensional case
studied so far.  We will discuss an implementation of a solid state quantum bit
known as a ``flying Qubit'' and directly inspired from recent experiments
\cite{Yamamoto2012}.  The results below are similar in spirit to those obtained
earlier in \cite{Bertoni1999}, but the addition of the Fermi-Dirac statistic
(taken into account here but not in\cite{Bertoni1999}) allows one to make
actual predictions for transport experiments.  Our model system is sketched in
Fig.  \ref{qubitsystem}. It consists of a large quasi-one dimensional wire of
(width $2W$) which is split in the middle by a top gate to which a depleting
voltage $V_T$ is applied. One effectively has two (weakly coupled) wires which
form the two states of the flying Qubit (the up and down ``states'' correspond
to the upper and lower wire respectively). The system has four terminals and we will compute the
effective transmission probability of the wire into the up and down channel, i.e. the ratio between
 the total number of transmitted particle and the total number particles injected in the wire.

\subsection{Integral over energies}
Before going to the simulations of this device, let us briefly discuss the last
technical difficulty that one is faced with when performing such
simulations: the integral over incident energies. We have seen in Section
\ref{landauer} that only a small energy window around the Fermi level
contributes to the transport properties and we would like now to understand how
this fact manifests itself in the numerical calculations.

\begin{figure}[h]
    \centering
    \includegraphics[width=12cm]{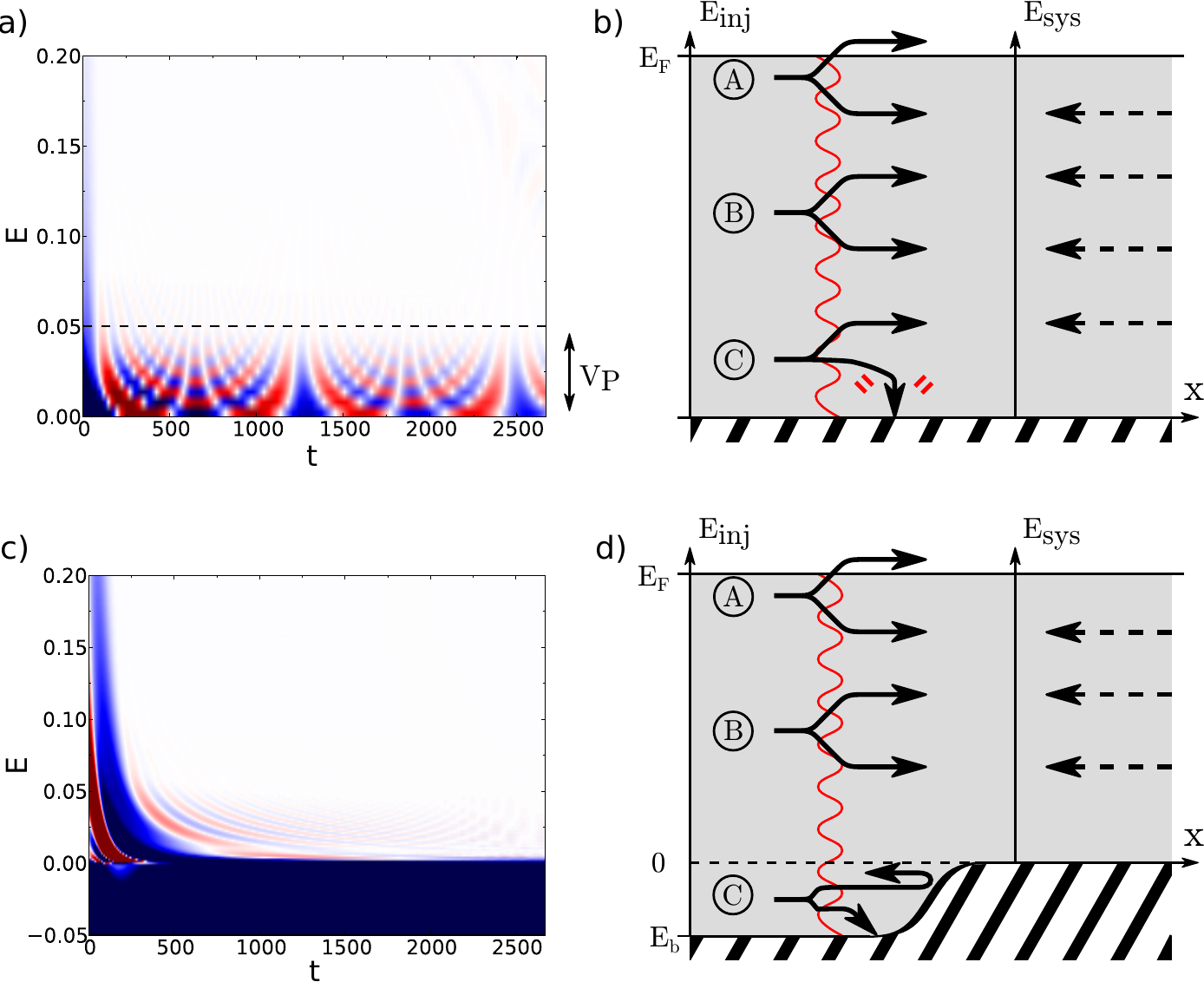}
    \caption{\label{blockade}
     Left panels: contribution $I(E,t)$ to the current $I(t)$ as a function of the injected energy $E$ and time $t$. The system is a one dimensional wire where one send a Gaussian pulse,
    $V(t)=V_Pe^{-4log(2)t^2/\tau_P^2}$, with width $\tau_P=100\gamma^{-1}$ and
    amplitude $V_P=0.05\gamma$. Red (blue) indicates values above
    (below) one.
     Right panels:  Schematic of the various contributions coming from different energies:
Case A:  the injected energy $E_{inj}$ is close to the Fermi
    energy $E_F$. Case B: the injected energy $E_{inj}$ is well below $E_F$
(these terms eventually give a vanishing contribution). Case C: the injected energy $E_{inj}$ is
close to the bottom of the band. These terms also give a vanishing contribution but they relax extremely slowly with time.
Lower panels: same as the upper panels but including our energy filtering scheme which removes the
contributions from case C.}
\end{figure}

The central technical issue when performing the energy integral numerically is
that within the WF method, one integrates over the injection energy $E_{inj}$
(see Fig. \ref{blockade} for a schematic). On the other hand, we have seen in
Section \ref{landauer} that in order to understand the various compensations
that take place between the currents coming from different leads, we must add
the contributions at a given energy $E_{sys}$ (energy of the electron inside
the mesoscopic region, i.e. after the pulse). This is illustrated in the upper
right panel of Fig. \ref{blockade}: in case A) the injected energy
$E_{inj}<E_F$ is close enough to the Fermi energy that the voltage pulse can
bring it to an energy $E_{sys}>E_F$ large enough for this contribution not to
be compensated by electrons coming from the other side. In case B) however,
$E_{inj}\ll E_F$ so that $E_{sys}<E_F$ and all contributions are compensated by
electrons injected from the right (at energy $E_{sys}$).  Unfortunately, in the
numerics we only control $E_{inj}$ so that we cannot differentiate between case
A) and B) and need to integrate over the whole energy range. This is not a real
issue, however, as several tens of energy points are usually enough and these
calculations can be performed in parallel. A real difficulty comes from case C)
where the injected energy $E_{inj}$ is close to the bottom of the band so that
after the pulse the electron can end up at a vanishing energy $E_{sys}=0$
which results in a vanishing velocity. As a result these contributions get
stuck at the place where the voltage drop takes place and cannot relax. This is
illustrated on the left panel of Fig. \ref{blockade} where we have plotted the
current flowing through the device as a function of $t$ and $E_{inj}$. We find
indeed that contributions that are too close to the bottom of the band relax
extremely slowly (by too close we mean closer than $\max (V_P, \hbar/\tau_P)$).
This makes numerical convergence difficult as one needs very long
simulation time to recover particle conservation.

Our strategy to remove the effect of those contributions is to improve our model
of the electrodes. In actual experimental setups the electrodes are
essentially metallic (high Fermi energy) so that the contributions
corresponding to case C) are essentially negligible. We therefore add an
external potential which is vanishing in the mesoscopic system and negative in
the electrode, as seen in the lower part of Fig. \ref{blockade}. As the current is measured
in the region where this potential vanishes (i.e. on the right in Fig. \ref{blockade}), the very low injected
energies (case C) will not contribute to the current any more and one recovers
particle conservation even for rather small simulation times.

\subsection{Model}
\begin{figure}[h]
    \centering
    \includegraphics[width=13cm]{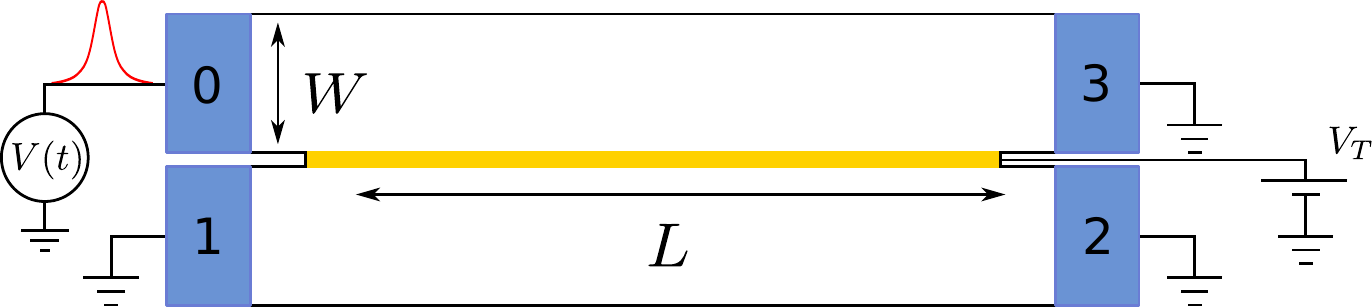}
    \caption{\label{qubitsystem}
    Sketch of our flying Qubit consisting of two wires of width $W$ connected
    to four leads (blue). The wires are coupled via a tunneling gate $V_T$ with
    length $L$ (yellow). Particles are injected by means of a
    voltage pulse which acts as the initial condition of the problem.
    }
\end{figure}

We consider the device sketched Fig. \ref{qubitsystem} for an electronic density
of  $n_s=0.3\ 10^{10} cm^{-2}$ which corresponds, for a GaAs/GaAlAs
heterostructure ($m^*=0.069 m_e$), to $E_F=108 \mu eV$ and $\lambda_F=457 nm$
($E_F=h^2/(2m^* \lambda_F^2) =\hbar^2 \pi n_s/m^*$  with $n_s$ being the full
electron density including spins). The device half width is $W=360 nm$ and the
length is $L=10 \mu m$. We use Gaussian pulses $w(t)=e
V_Pe^{-4log(2)t^2/\tau_P^2}$ with a width $\tau_P= 37 ps$ and $V_P=18 \mu V$.
The total simulation time was $t_{max}=1.5 ns$ with a time step $h_t=3.7 fs$.

We use a simple one band Schrodinger equation that includes the confining
potential $V(x,y,t)$ (due to the mesa and the gates),
\be
i\hbar\partial_t \psi(x,y,t)= -\frac{\hbar^2}{2m^*}\Delta \psi(x,y,t) +
V(x,y,t)\psi (x,y,t).
\ee
We rescale time in unit of the inverse of the Fermi energy $\tilde t =t
E_F/\hbar$, and space in unit of the Fermi wave length $\tilde x = 2\pi
x/\lambda_F$. The Fermi energy is rescaled to $\tilde{E}_F=1$, and we get
the dimensionless Schrodinger equation,
\be
i\partial_{\tilde t} \psi(\tilde x,\tilde y,t)= -\tilde \Delta \psi(\tilde
x,\tilde y,\tilde t) + [V(\tilde x,\tilde y,\tilde t)/E_F] \psi (\tilde x,\tilde
y,\tilde t).
\ee

The confining potential includes, in particular, the contribution from the
tunneling gate $V_T(x,y,t)=V_T(t) \delta (y)$, the other gates being always
static.  For actual simulations, the model is discretized on a square lattice
with lattice constant $a$ and we introduce $\psi_{n_x,n_y}(\tilde t)\equiv \psi
(n_x a,n_y a,\tilde t)$.  The discretized Schrödinger equation reads,
\begin{align}
   i\partial_{\tilde t} \psi_{n_x,n_y} = -\gamma [\psi_{n_x+1,n_y}+\psi_{n_x-1,n_y} +
\psi_{n_x,n_y+1}+\psi_{n_x,n_y-1} - 4\psi_{n_x,n_y}] + V_{n_x,n_y}(\tilde t) \psi_{n_x,n_y}
\end{align}
where $\gamma=1/a^2$. Note that $V_T(\tilde x,\tilde y,\tilde t)$ is
discretized into $[V_T]_{n_x,n_y}(\tilde t)=(V_T(\tilde t)/a)\delta_{n_y,0}$.

\subsection{Time-resolved simulations}
\begin{figure}[h]
    \centering
    \includegraphics[width=0.6\textwidth]{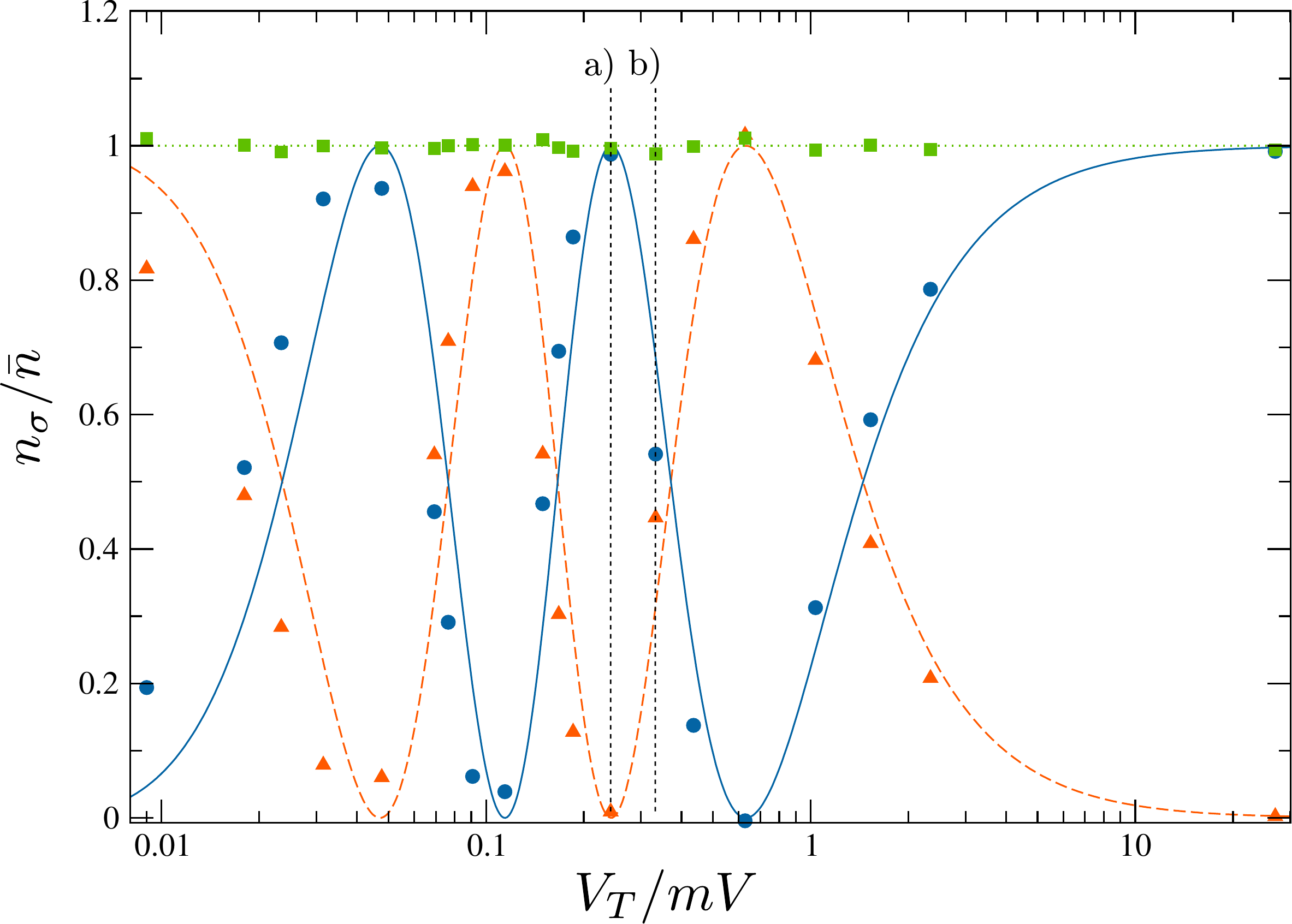}
    \caption{\label{qubit_nsigma}
        The number of transmitted particles through lead 2 (orange triangles)
        and lead 3 (blue circles) ($n_{\downarrow}$ and $n_{\uparrow}$,
        collectively $n_{\sigma}$) normalized by the number of injected
        particles in lead 0 ($\bar{n}$) as a function of the tunneling gate
        potential, $V_T$. The green squares are the sum
        $n_{\downarrow}+n_{\uparrow}$. The corresponding lines are the DC
        transmission probabilities of the system between leads 0\&2 (dashed
        orange) and 0\&3 (full blue) and their sum (dotted green). The values
        of $V_T$ indicated by the black dashed lines labeled a) and b)
        correspond to the values of $V_T$ used to produce figure
        \ref{qubit_propagation}. These results were produced with a Gaussian
        voltage pulse as described in the main text.
    }
\end{figure}

\begin{figure}[h]
    \centering
    \includegraphics[width=0.8\textwidth]{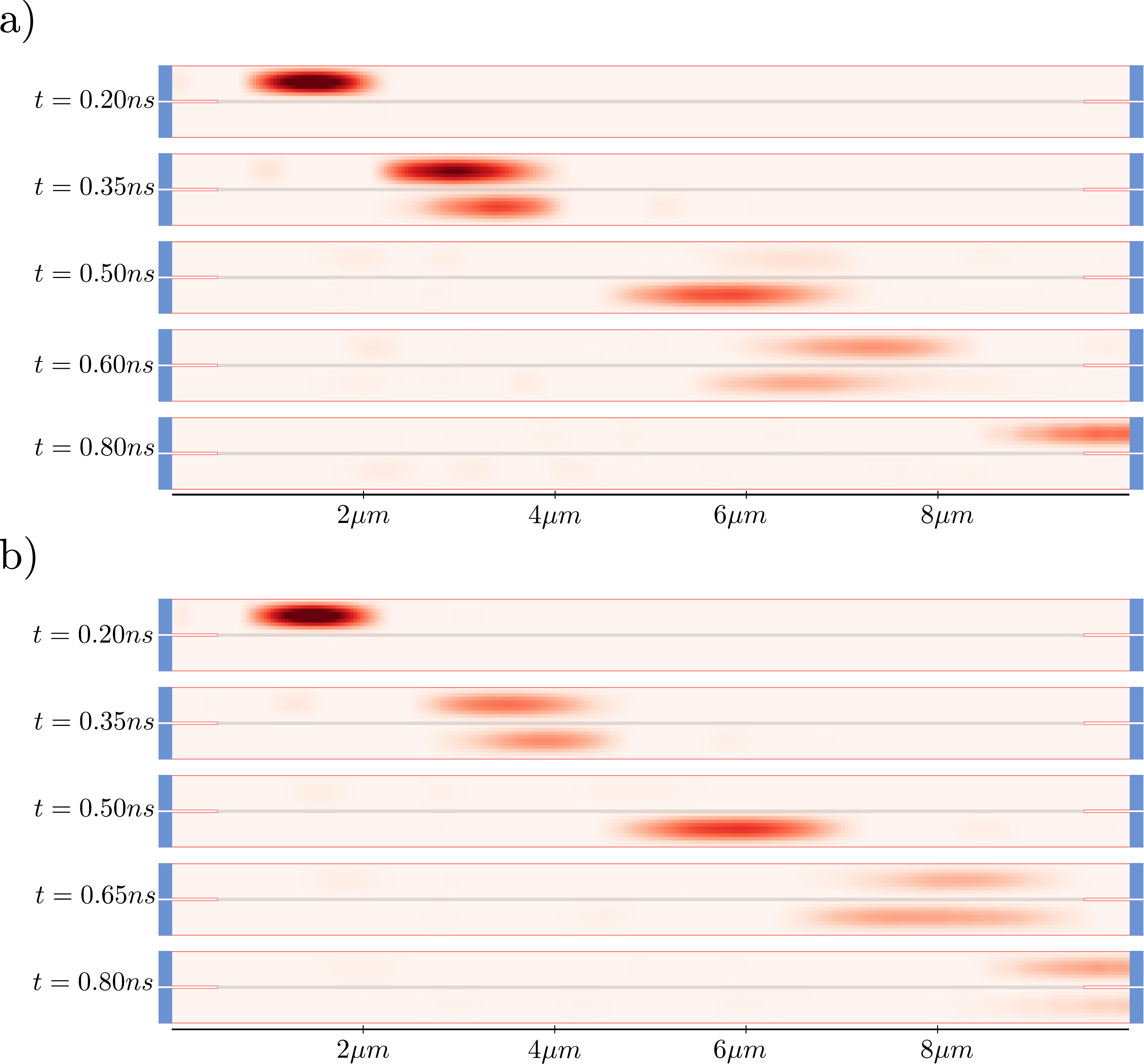}
    \caption{\label{qubit_propagation}
        Propagation of a voltage pulse within the coupled wire system. The
        figures show snapshots of the difference of the local charge density
        from equilibrium at different points in time. Figures a) and b)
        correspond to two different values of the tunneling gate voltage,
        $0.24 mV$ and $0.34 mV$ respectively. These results were produced with a
        Gaussian voltage pulse as described in the main text. Each of these two
        runs corresponds
to a computing time per energy and per channel of 30 minutes on one computing core
($a=0.5$, $7250$ sites and $400000$ time steps). }
\end{figure}

\begin{figure}[h]
    \centering
    \includegraphics[width=0.6\textwidth]{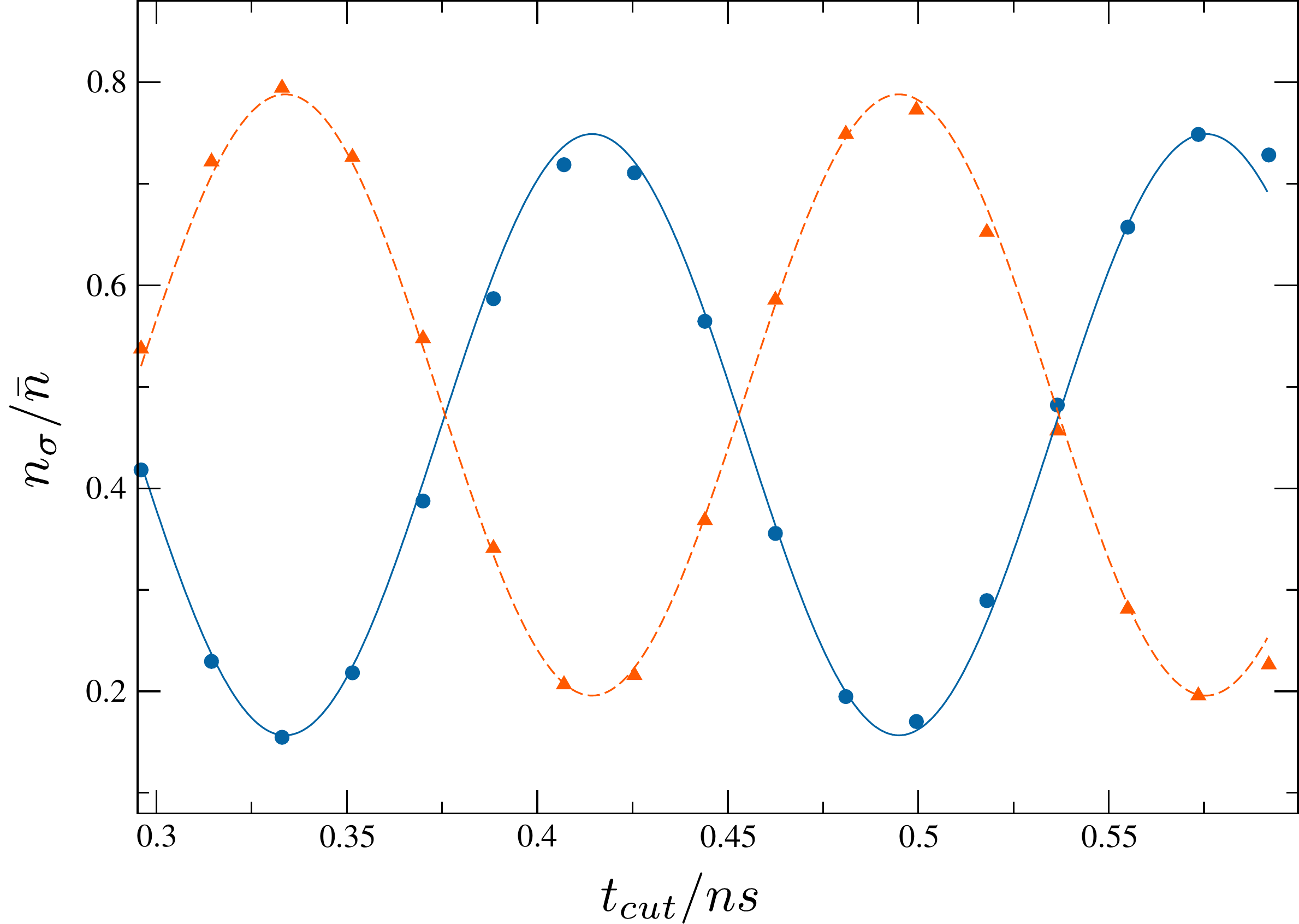}
    \caption{\label{qubit_slicer}
        The number of transmitted particles through lead 2 (orange triangles)
        and lead 3 (blue circles) ($n_{\downarrow}$ and $n_{\uparrow}$,
        collectively $n_{\sigma}$) normalized by the number of injected
        particles in lead 0 ($\bar{n}$) as a function of the time at which
        the coupling between the two wires of the flying Qubit system are cut,
        $t_{cut}$. The lines are cosine fits to the results from the numerics
        to guide the eye. These results were produced with a Gaussian voltage
        pulse as described in the main text.  }
\end{figure}

Let us now discuss the results of the simulations. Fig. \ref{qubit_nsigma} shows
the total number of transmitted particles in the upper (lead 3, $n_{\uparrow}$)
and lower (lead 2, $n_{\downarrow}$) channels as a result of the voltage pulse
sent in the upper electrode (lead 0).  We find that these numbers oscillate as
a function of the tunneling gate voltage $V_T$ which demonstrates that it is
possible to dynamically control the superposition of the wave function into the
upper and lower part of the leads. In the DC limit, one expects $n_{\uparrow} =
\bar n \cos^2 [(k_A-k_S)L]$ ($n_{\downarrow} = \bar n \sin^2 [(k_A-k_S)L]$)
where $\bar n$ is the total number of electrons sent by the pulse and $k_A$
($k_S$) is the momentum of the antisymmetric (symmetric) mode in the wire.  We
find a good agreement between the simulations and the DC results, which is not
trivial as we have worked in a fast regime where $\tau_P$ is smaller than the
characteristic energy of the wire, $\hbar v_F (k_A-k_S)$ (in the adiabatic
limit where $\tau_P$ is longer than all the characteristic times of the system,
one should always recover the DC result). Indeed, Fig. \ref{qubit_propagation}
shows two snapshots of Fig. \ref{qubit_nsigma} for $V_T=0.24 mV$ and $V_T=0.34
mV$. We find that the electronic density does oscillate, as a function of time,
between the two arms of the flying Qubit.  We find (case a) that at $t=0.35 ns$
the ``pulse'' is in a superposition of the up and down state while slightly later
($t=0.5 ns$) the ``pulse'' is almost entirely in the lower arm. In the last
simulation, shown in Fig. \ref{qubit_slicer}, we send the same pulse, wait for
some time and abruptly raise the value of $V_T$ to infinity (therefore
effectively slicing the wire in two) at a time $t_{cut}$.  From Fig.
\ref{qubit_propagation}, one expects to observe oscillations of $n_{\downarrow}$
and $n_{\uparrow}$ as a function of $t_{cut}$ and indeed Fig. \ref{qubit_slicer}
shows them. Note however that, in addition to ``freezing'' the system in one
arm,  the ``slicing'' operation also repels all the electrons beneath the
tunneling gate. As a result several electrons (around 6 in our simulations) get
expelled from the system. Fig. \ref{qubit_slicer} is obtained by performing two
simulations, one with the pulse and one without, and subtracting the two
results in order to cancel out this spurious effect.

The device of Fig.\ref{qubitsystem} contains, in fact, quite rich physics but
we shall end our discussion here. The purpose was to show
that the formalism introduced in this article allows one to perform
simulations on models and time scales large enough to be meaningful for
mesoscopic physics. Further discussions of the physics involved in actual
devices will be conducted elsewhere.

%%%%%%%%%%%%%%%%%%%%%%%%%%%%%%%%%%%%%%%%%%%%%%%%%%%%%%%%%%%%%%%%%%%%%%%%
\section{Conclusion}
%%%%%%%%%%%%%%%%%%%%%%%%%%%%%%%%%%%%%%%%%%%%%%%%%%%%%%%%%%%%%%%%%%%%%%%%

In the first part of this paper we have shown the equivalence between three
different theoretical approaches of time-dependent nanoelectronics: the NEGF
formalism, the Scattering approach and the partition-free initial condition
approach.  Building on these different theories, we have developed various
strategies to perform numerical simulations for those systems. We eventually
converged to a very simple algorithm (WF-D) whose performance is many orders of
magnitudes better than a brute force integration of the NEGF equations. Systems
with more than $10^5$ sites with times long enough to probe their ballistic or
even diffusive dynamics are  now  accessible to direct simulations. In the last
part of this article, we have specialized the formalism to the particular case
of voltage pulses. We found that the total number of transmitted particles is
an observable that satisfies the basic requirements of a well behaved theory:
particle conservation and gauge invariance. The article ends with two
practical examples that illustrate our approach: a solid state equivalent of
the spreading of the wave packet and an implementation of a ``flying Qubit''.

A strong emphasis was put on the technical aspects of time-dependent transport (and the corresponding
simulations) with little room left for discussing the actual physics involved.
We believe however that conceptually new physics will soon emerge from fast quantum electronics and that simulation methods, such as the one presented in this article, will play an important role in this development.

\section*{Acknowledgements}
The DC numerical simulations were performed with the Kwant software package,
developed by C. W. Groth, M. Wimmer, A. R. Akhmerov and X. Waintal \cite{Kwant_preparation}. The simulations of section \ref{sec:2d} were performed with an implementation of WF-D which is itself based on Kwant. Hence it shares all its general aspects
(arbitrary dimension (1D, 2D, 3D), lattices (including say graphene), geometries (multiterminal) or internal structure (spins, superconductivity, spin-orbit, magnetic fields, etc.).

This work was supported by the ERC grant MesoQMC. MH also acknowledges the support from the French
ANR grant (ANR-11-JS04-003-01).

\clearpage
\bibliographystyle{model1-num-names}
\bibliography{references}

%%%%%%%%%%%%%%%%%%%%%%%%%%%%%%%%%%%%%%%%%%%%
\clearpage
\appendix

%%%%%%%%%%%%%%%%%%%%%%%%%%%%%%%%%%%%%%%%%%%%%%%%%%%%%%%%%%%%%%%%%%%%%%%%
\section{Understanding the origin of the ``source'' term and ``memory'' kernel}
%%%%%%%%%%%%%%%%%%%%%%%%%%%%%%%%%%%%%%%%%%%%%%%%%%%%%%%%%%%%%%%%%%%%%%%%
\label{appendix-src}
As we have seen, there exist general connections between the various approaches
used for (time-resolved) quantum transport (NEGF, scattering theory or the
partition-free approach).  As these connections were proved for a rather
general class of problems (allowing for non trivial electrodes such as graphene
as well as arbitrary time-dependent perturbations), the basic mathematical
structure of these connections might be somewhat obscured. In this appendix, we
consider the simplest situation where one can understand the origin of the
``memory kernel'' (term involving the self-energy) and ``source'' term that play a central role in our
formalism: a simple one dimensional chain with just one electrode and no time-dependent
perturbations.

Our starting point is the Schrödinger equation for the 1D chain in the energy
domain,
\be
\Psi_{x-1} + \Psi_{x+1} = E \Psi_x
\ee
We suppose that the ``system'' corresponds to $x\ge 1$ (where one can possibly
add terms such as $V_x \psi_x$ but those will be irrelevant for the present
discussion) and the ``electrode'' corresponds to $x\le 0$. As a boundary
condition in the electrode, we impose the incoming part of the wave: for
$x\le 0$,
\be
\Psi_x = e^{ikx} + r e^{-ikx}
\ee
which in turn implies that $E= 2\cos k$. At this stage, we could proceed with
``wave matching'' and try to obtain an expression for the reflection amplitude
$r$. Another possibility involves deriving an effective equation where $r$ has
disappeared, which amounts to finding the effective boundary condition imposed
on the system due to the presence of the electrode. Writing the Schrödinger
equation for $x=0$ and $x=1$ we get,
\be
1+r + \Psi_{2} = E \Psi_1
\ee
\be
e^{-ik} + r e^{ik} + \Psi_{1} = E (1+r)
\ee
Now using $E=e^{ik} + e^{-ik}$ we find,
\be
\label{src-kernel}
[\Sigma^R \Psi_1 +i \Sigma^R v ] + \Psi_2 = E \Psi_1
\ee
where we have introduced the self-energy $\Sigma^R=e^{ik}$ and the velocity
$v=\partial E/\partial k$.  Eq.(\ref{src-kernel}) is reminiscent of the original
equation $\Psi_{0} + \Psi_{2} = E \Psi_1$. The value of the wave function in the
electrode $\Psi_0$
has been replaced by an effective boundary condition and the electrodes
effectively drop out of the problem. This effective boundary condition [first
two terms in Eq.(\ref{src-kernel})] contains a self-energy term
(proportional to $\Psi_1$) and a source term. This is in fact a generic
consequence of the peculiar sort of boundary condition where we impose the
incoming waves (as opposed to more conventional Dirichlet or Neumann
boundary conditions). Upon transforming into the time domain, the self-energy
term transforms into a convolution which gives rise to the memory kernel.

Let us now introduce an equation for a new wave function $\psi_x$ which is
defined for an infinite system,
\be
\psi_{x-1} + \psi_{x+1}  +\delta_{x,1} i \Sigma^R v = E \psi_x
\ee
such that for $x\le 0$ one imposes the presence of outgoing modes only, i.e.
$\psi_x$ takes the form $\psi_x = r e^{-ikx}$. Then, upon performing the same
algebra as above, one finds that for $x\ge 1$, $\psi_x$ satisfies
Eq.(\ref{src-kernel}). In other words, $\Psi_x$ and $\psi_x$ are equal for
$x\ge 1$ even though the latter lacks the incoming part for $x\le 0$. The
generalization of these two wave functions corresponds directly to their
equivalent as defined in the core of the text.

%%%%%%%%%%%%%%%%%%%%%%%%%%%%%%%%%%%%%%%%%%%%%%%%%%%%%%%%%%%%%%%%%%%%%%%%
\section{Derivation of the path integral formula (\ref{eq: path})}
%%%%%%%%%%%%%%%%%%%%%%%%%%%%%%%%%%%%%%%%%%%%%%%%%%%%%%%%%%%%%%%%%%%%%%%%
\label{app: path_formula}
The projection of Eq. (\ref{eq: Gr_evol}) onto the central region $\bar 0$
yields,
\be
    \label{eq: Gr_projected}
    \forall \  u \ \in [t',t],\  \mathcal{G}^{R}_{\bar{0}\bar{0}}(t,t') =
    i
    \mathcal{G}^{R}_{\bar{0}\bar{0}}(t,u)\mathcal{G}^{R}_{\bar{0}\bar{0}}(u,t')
    + i
    \sum_{i=1}^{M} \mathcal{G}^{R}_{\bar{0}\bar{i}}(t,u)\mathcal{G}^{R}_{\bar{i}\bar{0}}(u,t').
\ee
We use the Dyson equation to rewrite $\mathcal{G}^{R}_{\bar{0}\bar{i}}(t,u)$
and
$\mathcal{G}^{R}_{\bar{i}\bar{0}}(u,t')$ as follows:
\be
    \mathcal{G}^{R}_{\bar{0}\bar{i}}(t,u)  = \int_{u}^{t} dv \
    \mathcal{G}^{R}_{\bar{0}\bar{0}}(t,v)\mathrm{\bold{H}}_{\bar{0}\bar{i}}(v)g^{R}_{\bar{i}}(v,u)
    \mathcal{G}^{R}_{\bar{i}\bar{0}}(u,t')  = \int_{t'}^{u} dv \
    g^{R}_{\bar{i}}(u,v)\mathrm{\bold{H}}_{\bar{i}\bar{0}}(v)\mathcal{G}^{R}_{\bar{0}\bar{0}}(v,t')
\ee
Lastly, we substitute these above relations into Eq.(\ref{eq: Gr_projected})
and obtain,
\be
    \forall \  u \ \in [t',t],\  G^R(t,t') =  i G^{R}(t,u)G^{R}(u,t') +
    \sum_{i=1}^{M} \int_{u}^{t} dv\ G^{R}(t,v) \int_{t'}^{u} du\
    \mathrm{\bold{H}}_{\bar{0}\bar{i}}(v)g^{R}_{\bar{i}}(v,v')\mathrm{\bold{H}}_{\bar{i}\bar{0}}(v')G^{R}(v',t')
    \label{eq: rewrite}
\ee
which is essentially Eq.(\ref{eq: path}).
%%%%%%%%%%%%%%%%%%%%%%%%%%%%%%%
\section{Various analytical results for the Green's functions of the 1d chain}
%%%%%%%%%%%%%%%%%%%%%%%%%%%%%%%
\label{benchmark}
We gather here a few analytical results for the 1d chain that were used to
benchmark the numerical results shown in this work. Given an analytic function
$f$ our convention for Fourier transforms is
\begin{align}
    &f(t) = \int \frac{dE}{2\pi}\ f(E) e^{-iEt}\\
    &f(E) = \int dt\ f(t) e^{iEt}
\end{align}
The expressions below correspond to the Hamiltonian (\ref{eq: 1d_Hamiltonian})
for the perfect one dimensional chain ($\epsilon_i=0$).  The Lesser Green's
functions were computed at zero temperature with $E_F=0$. Energies are written
in units of the hopping parameter $\gamma$, and times are in units of
$\gamma^{-1}$.

\begin{itemize}
\item We begin with self-energies in energy for a semi-infinite lead,
\begin{align}
    \Sigma^R(E) &= \left\{
    \begin{array}{lll}
        \frac{E}{2} - i \sqrt{1 - (\tfrac{E}{2})^{2}} & \mbox{if } |E| \leq 2\\
        \frac{E}{2} -\sqrt{(\tfrac{E}{2})^{2} - 1} & \mbox{if } E > 2 \\
        \frac{E}{2} + \sqrt{(\tfrac{E}{2})^{2} - 1} & \mbox{if } E < -2
    \end{array}
    \right.\\
    \Sigma^<(E) &= \left\{
    \begin{array}{ll}
    2i \sqrt{1 - (\frac{E}{2})^2} & \mbox{if } -2 \leq E \leq E_F\\
        0 & \mbox{else}
    \end{array}
    \right.
\end{align}
\item The corresponding Fourier transforms in time yields,
\begin{align}
    &\Sigma^{R}(t) = -i \frac{J_{1}(2t)}{t}\theta(t)  \\
    &\Sigma^{<}(t) = i \frac{J_{1}(2t)}{2t} -
    \frac{H_{1}(2t)}{2t}
\end{align}
where $J_{n}$ is the Bessel function of the first kind, and $H_{n}$ is the
Struve function of order $n$.

\item We also computed Green's functions for the infinite 1d chain at
    equilibrium.  The diagonal elements of the Retarded and Lesser Green's
    functions in energy read,
\begin{align}
    G^{R}_{xx}(E) &= \left\{
    \begin{array}{lll}
        \frac{1}{2i \sqrt{1-(\frac{E}{2})^{2}}} & \mbox{if } |E| \leq
        2\\
        \frac{1}{2 \sqrt{(\frac{E}{2})^{2}-1}} & \mbox{if } E > 2 \\
        \frac{1}{-2\sqrt{(\frac{E}{2})^{2}-1}} & \mbox{if } E < -2
    \end{array}
    \right.\\
    G^{<}_{xx}(E) &= \left\{
    \begin{array}{lll}
        \frac{i}{\sqrt{1 - (\frac{E}{2})^2}} & \mbox{if } -2 \leq E \leq E_F \\
        0 & \mbox{else}
    \end{array}
    \right.
\end{align}
\item and their counterparts in the time domain,
\begin{align}
    &G^R_{xx}(t) = -iJ_{0}(2 t)\theta(t) \\
    &G^<_{xx}(t) = \frac{i}{2} J_{0}(2 t) - \frac{H_{0}(2 t)}{2}
\end{align}
\item The off diagonal element $G^<_{x,x+1}$ in energy and time domains read,
\begin{align}
    G^{<}_{x,x+1}(E) &= \left\{
    \begin{array}{lll}
        \frac{iE/2}{\sqrt{1 - (\frac{E}{2})^2}} & \mbox{if } -2 \leq E \leq E_F\\
        0 & \mbox{else} \\
    \end{array}
    \right.\\
    G^{<}_{x,x+1}(t) &= \frac{J_{1}(2t)}{2}  - \frac{i}{2} H_{-1}(2t).
\end{align}
\end{itemize}

\end{document}